\documentclass[aps,pra,preprint,showpacs,amsmath,amssymb]{revtex4}
\input{epsf}

\usepackage{graphicx}
\usepackage{array}
\usepackage{longtable}
\usepackage{rotating,booktabs}
\usepackage{booktabs,threeparttable}
\usepackage{amsmath}
\allowdisplaybreaks[4]

\usepackage{rotating}

\begin{document}

\newcommand{\ket}[1]{\ensuremath{\left|#1\right\rangle}} % Dirac Kets
\newcommand{\bra}[1]{\ensuremath{\left\langle #1 \right|}} % Dirac Bras
\newcommand{\braket}[2]{\ensuremath{\left\langle #1 | #2 \right\rangle}} % Dirac Braket

\title{Calculations of long-range three-body interactions for He($n_0\,^{\lambda}S$)-He($n_0\,^{\lambda}S$)-He($n_0^{\prime}\,^{\lambda}L$)}

\author{Pei-Gen Yan$^{1,2}$, Li-Yan Tang$^{2,*}$~\footnotetext{*Email: lytang@wipm.ac.cn}, Zong-Chao Yan$^{1,2,3}$, and James F Babb$^{4}$}

\affiliation {$^1$Department of Physics, University of New Brunswick, Fredericton, New Brunswick, E3B5A3, Canada\\
$^2$State Key Laboratory of Magnetic Resonance and Atomic and Molecular Physics, Wuhan Institute of Physics and Mathematics, Chinese Academy of Sciences, Wuhan 430071, People's Republic of China\\
$^3$Center for Cold Atom Physics, Chinese Academy of Sciences, Wuhan 430071, People's Republic of China\\
$^4$Harvard-Smithsonian Center for Astrophysics, ITAMP, Cambridge, Massachusetts 02138, USA}

\date{\today}

\begin{abstract}

We theoretically investigate long-range interactions between an excited $L$ state He atom and two identical $S$ state He atoms,
for the cases of the three atoms
all in spin singlet states or all in spin triplet states,
denoted by He($n_0\,^{\lambda}S$)-He($n_0\,^{\lambda}S$)-He($n_0^{\prime}\,^{\lambda}L$),
with $n_0$ and $n_0'$ principal quantum numbers, $\lambda=1$ or 3 the spin multiplicity, and $L$
the orbital angular momentum of a He atom.
Using degenerate perturbation theory for the energies up to second-order, we evaluate the coefficients $C_3$ of the first order dipolar interactions and the coefficients $C_6$ and $C_8$ of the second order additive and nonadditive interactions. Both the dipolar and dispersion interaction coefficients, for these three-body degenerate systems, show  dependences on the geometrical configurations of the three atoms.
The nonadditive interactions start to appear in second-order.
To demonstrate the results and for applications, the obtained coefficients $C_n$
are evaluated with highly accurate variationally-generated nonrelativistic wave functions in Hylleraas coordinates
for
He($1\,^{1}S$)-He($1\,^{1}S$)-He($2\,^{1}S$),
He${(1\,^{1}S)}$-He${(1\,^{1}S)}$-He${(2\,^{1}P)}$,
He${(2\,^{1}S)}$-He${(2\,^{1}S)}$-He${(2\,^{1}P)}$, and
He${(2\,^{3}S)}$-He${(2\,^{3}S)}$-He${(2\,^{3}P)}$.
The calculations are given for three like-nuclei
for the cases of  hypothetical infinite mass He nuclei,
and of real finite mass $^4{}$He or $^3{}$He nuclei.
The special cases of the three atoms in equilateral triangle configurations are
explored in detail, and for
the cases where one of the atoms
is in a $P$ state, we also present
results for the  atoms in an  isosceles right triangle configuration or
in an equally spaced co-linear configuration.
The results can be applied to construct potential energy surfaces for three helium atom systems.
\end{abstract}

\maketitle

\section{Introduction}\label{Int}

We recently demonstrated for the case of three Li atoms with two atoms
in their ground states and one
atom in the first excited $P$ state
that their long-range
(\textit{i.e.} atoms sufficiently separated that electron exchange is negligible)
interactions exhibit a
first order interaction potential dependent on
the geometrical configuration of the atoms and that in second order
additive and non-additive dispersion interactions appear~\cite{yan16} (subsequently referred
to as Paper~I).
These long-range interactions are in sharp contrast to the case of three ground state atoms, where
geometric-configuration non-additive dispersion interactions (sometimes
called Axilrod-Teller-Muto terms) appear in third order, cf. Ref.~\cite{bell70}.
Our results provide accurate long-range potential energies
of electronically excited Li trimers and the results may be useful in the analysis of processes
such as optical blockade effects~\cite{LukHem00} and cooperative spontaneous emission~\cite{FenLiZhu13}
where long-range dipole-dipole interactions amongst several atoms appear.

In the present paper, we extend the theory of Paper~I to the case
of three He atoms.
We investigate the long-range interactions
between an excited $L$ state He atom and two identical $S$ state He atoms,
for the cases of the three atoms
all in spin singlet states or all in spin triplet states,
denoted by He($n_0\,^{\lambda}S$)-He($n_0\,^{\lambda}S$)-He($n_0^{\prime}\,^{\lambda}L$),
with $n_0$ and $n_0'$ principal quantum numbers, $\lambda=1$ or 3 the spin multiplicity, and $L$
the orbital angular momentum of a He atom.

Interactions between three {ground state} He atoms were studied extensively,
and elaborate calculations of three-body interactions, including dispersion interactions contributions,
are available (see, for example, Ref.~\cite{CenJezAki07} and references therein).
The three-ground state He atom calculations are valuable for modeling liquid ${}^4$He and solid He \cite{MurBar71}, for recombination and scattering studies~\cite{SunEsr08},
investigations of Efimov states~\cite{SunEsr08,Sun14}, and thermophysics~\cite{CenJezAki07}.

Rare gas metastable helium atoms have been widely used in many studies~\cite{VasCohTan12}
such as photoassociation spectroscopy~\cite{Herschbach00}, metastable loss in magneto-optical traps~\cite{Weiner99}, Penning ionization~\cite{Leo01,Mastwijk98,Kumakura99,Tol99}, and associative ionization~\cite{Weiner99}.
Indeed, the three-body loss rate in a $^4\textrm{He}(2\,^{3}S$) Bose-Einstein condensate
was measured providing evidence from the trimer system $^4\textrm{He}_3(2\,^{3}S$) for a universal three-body parameter~\cite{KnoBorVas12}.
For the case of two metastable He atoms, with one atom in
excited $P$-state level, the first-order resonant dipole-dipole interaction produces an interatomic potential varying as $C_3/R^3$, where the corresponding coefficients $C_3$ for weakly bound dimers of helium atoms
were given in many papers~\cite{zhang06,zhang06-1}
(and references therein).
However, due to the degeneracy, these coefficients may not be used in the study of helium clusters or the study of atom-molecule and molecule-molecule collision that involve a $P$-state atom.
This is because, when a helium excited dimer turns into an excited trimer or a cluster involving excited atoms,
the interactions between atoms are changed due to  quantum many-body effects.
So to proceed with
calculations of molecular He$_n^{*}$ excimer potential energy surfaces,
investigations of the long-range multi-body interactions are warranted.
For example, early work by Phelps~\cite{Phe56} studying the destruction of
He($2\,^{3}S$) atoms using absorption measurements in a discharge found evidence for
for the formation of metastable helium dimers
through the reaction He($2\,^{3}S$) +  He($1\,^{1}S$) +   He($1\,^{1}S$)
and evidence that the three-body interaction between
a bound He$_2$($2\,^{3}\Sigma)$ molecule and a He($1\,^{1}S$)
atom may be ``repulsive or, at most, weakly attractive.''
However,  trimer excited potential energy calculations were not available at the time,
and, to date, may still be unavailable, making interpretations
based on collisional dynamics uncertain.
Later studies, however, confirm the importance
and complexity, see Ref.~\cite{PouKhaSte88} and references therein,
of the kinetics between the He($2\,^{3}S$) atoms and the
vibrationally excited He$_2$($a\,^{3}\Sigma_u)$ molecules.
Further work,
such as we  present here, concerning the long-range interactions of combinations
of three He atoms when at least one is metastable is desirable, and may lead to understanding of
various recombination and scattering processes.

In this work, we present our research on long-range interactions for three like helium atoms involving
at least one atom in an excited state.
Using variationally optimized nonrelativistic atomic helium wave functions in  Hylleraas coordinates, we present our theoretical calculations of long-range interaction coefficients for  He($n_0\,^{\lambda}S$)-He($n_0\,^{\lambda}S$)-He($n_0^{\prime}\,^{\lambda}L$) considering the energetically lowest five states: He($1\,^{1}S$), He($2\,^{3}S$), He($2\,^{1}S$), He($2\,^{3}P$) and He($2\,^{1}P$).
We present the additive ``dipolar'' interactions coefficients $C_3$ and additive dispersion interactions coefficients $C_6$, $C_8$ that enter, respectively, in first- and second-order perturbation theory. We also evaluate the second-order nonadditive dispersion interactions coefficients $C_6$, $C_8$ that contain a dependence on the geometrical configuration of the three atoms.  In addition, the coefficients are given explicitly and as numerical values for the three basic geometrical configurations of the nuclei in an equilateral triangle, in an isosceles right triangle or equally spaced collinearly. Finally, long-range potentials in the sum of first and second order energies for these three geometrical configurations are given.

Due to the quantum three-body effect, both the dipolar and dispersion interaction coefficients, for the degenerate He($n_0\,^{\lambda}S$)-He($n_0\,^{\lambda}S$)-He($n_0^{\prime}\,^{\lambda}L$) system, show a dependence on the geometrical configurations of the three atoms. Currently, the obtained coefficients are given in three very common configurations. But in this paper, we would also talk about the connection of this degenerate three-body system with nondegenerate three-body system and three-body atom-molecule system, which could be used in the study of three-body recombination or ultracold atom-molecule collision.

%-------new

\section{Theoretical Formulation}\label{The}

In this paper, atomic units are used throughout. The three atoms are labeled by $I$, $J$ and $K$, with, respectively, internal coordinates $\boldsymbol{\sigma}$, $\boldsymbol{\rho}$, and $\boldsymbol{\varsigma}$.
When the labels $I$, $J$, or $K$ appear, it is understood that cyclic permutation
can be used.

In the present work, we take the mutual electrostatic interactions $V_{IJ}$ between pairs of atoms for the He($n_0\,^{\lambda}S$)-He($n_0\,^{\lambda}S$)-He($n_0'\,^{\lambda}L$) system as a perturbation. According to degenerate perturbation theory, the zeroth-order wave function of the unperturbed system can be written as,
\begin{equation}\label{e111}
\ket{\Psi^{(0)}}=a\ket{\phi_1}+b\ket{\phi_2}+c\ket{\phi_3} \,.
\end{equation}
where $\phi_1$, $\phi_2$, $\phi_3$ are three orthogonal eigenvectors corresponding to the same energy eigenvalue $E_{n_0n_0n_0'}^{(0)}=2E_{n_0S}^{(0)}+E_{n_0'L}^{(0)}$,
\begin{eqnarray}
\ket{\phi_1}&=&\ket{\varphi_{n_0'}(LM;\boldsymbol{\sigma})\varphi_{n_0}(0;\boldsymbol{\rho})\varphi_{n_0}(0;\boldsymbol{\varsigma})}\,, \label{e6a_a} \\
\ket{\phi_2}&=&\ket{\varphi_{n_0}(0;\boldsymbol{\sigma})\varphi_{n_0'}(LM;\boldsymbol{\rho})\varphi_{n_0}(0;\boldsymbol{\varsigma})}\,, \label{e6b_b}
\\
\ket{\phi_3}&=&\ket{\varphi_{n_0}(0;\boldsymbol{\sigma})\varphi_{n_0}(0;\boldsymbol{\rho})\varphi_{n_0'}(LM;\boldsymbol{\varsigma})}\,.  \label{e6c_c}
\end{eqnarray}
The expansion coefficients $a$, $b$, $c$ are determined by diagonalizing the perturbation in the basis set \{$\phi_1$, $\phi_2$, $\phi_3$\}, which depends on the geometrical configuration formed by the three atoms. In the following, we show that all the dispersion interaction coefficients contain some of or all of these three expansion coefficients $a$, $b$, $c$
leading to  dependences on the configuration of the three atoms.

For three well-separated helium atoms, the mutual interaction energy $V_{IJ}$ can be expanded according to Refs.~\cite{bell70, fontana61}
\begin{equation}\label{e2}
V_{IJ}=\sum_{l_Il_J}\sum_{m_Im_J}T_\text{$l_I-m_I$}(\boldsymbol{\sigma})T_\text{$l_Jm_J$}(\boldsymbol{\rho})W_{l_Il_J}^{m_I-m_J}(IJ) \,.
\end{equation}
 In Eq.(\ref{e2}), the multipole transition operators are
\begin{eqnarray}
T_{l_I-m_I}(\boldsymbol{\sigma})&=&\sum_{i}Q_i\sigma ^{l_I}_{i}Y_{l_I-m_I}(\hat{\boldsymbol{\sigma}_i}) \,,  \\ \label{e3a}
T_{l_Jm_J}(\boldsymbol{\rho})&=&\sum_{j}q_j\rho ^{l_J}_{j}Y_{l_Jm_J}(\hat{\boldsymbol{\rho}_j}) \,,  \label{e3b}
\end{eqnarray}
and the geometry factor is
\begin{eqnarray}\label{e4}
W_{l_Il_J}^{m_I-m_J}(IJ)&=&\frac{4\pi(-1)^{l_J}}{R_{IJ}^{l_I+l_J+1}}\frac{(l_I+l_J-m_I+m_J)!(l_I,l_J)^{-1/2}}{[(l_I+m_I)!
(l_I-m_I)!(l_J+m_J)!(l_J-m_J)!]^{1/2}}
P_{l_I+l_J}^{m_I-m_J}(\cos\theta_{IJ})\nonumber\\
&\times & \exp[{i(m_I-m_J)\Phi_{IJ}}] \,,
\end{eqnarray}
where ${\bf R}_{IJ}={\bf R}_J-{\bf R}_I$ is the relative position vector from atom $I$ to atom $J$, the notation
$(a,b,\ldots)=(2a+1)(2b+1)\ldots$, and $P_{l_I+l_J}^{m_I-m_J}(\cos\theta_{IJ})$ is the associated Legendre function with $\theta_{IJ}$ representing the angle between ${\bf R}_{IJ}$ and the $z$-axis. If we now choose the $z$ axis to be normal to the plane of the three atoms, giving $\theta_{12}=\theta_{23}=\theta_{31}=\pi/2$, the associated Legendre functions can be  simplified as
\begin{eqnarray}
P_{l}^{m}(0)&=&\frac{1}{2^{l+1}}[1+(-1)^{l+m}](-1)^{\frac{l+m}{2}}(l+m)!
\bigg[\bigg(\frac{l+m}{2}\bigg)!\bigg]^{-1}\bigg[\bigg(\frac{l-m}{2}\bigg)!\bigg]^{-1} \,.
\end{eqnarray}
$\Phi_{IJ}$ denotes the angle between ${\bf R}_{IJ}$ and the $x$-axis. It shows the dependence of the mutual dipole-dipole interaction between two atoms on the orientation of the interacting dipoles relative to the line connecting them \cite{axilrod43}. Similar expressions result for $V_{JK}$ and $V_{KI}$. For simplicity, in this work, we  transfer all these $\Phi$
into interior angles ($\alpha$, $\beta$, $\gamma$) of the triangle formed by the three helium atoms with the same method
as used in Paper~I.
%shown in Ref. \cite{yan16}.
%

\subsection{The formulas for He($1\,^{1}S$)-He($1\,^{1}S$)-He($2\,^{1}S$)}\label{A1}

The second-order energy correction for the He($1\,^{1}S$)-He($1\,^{1}S$)-He($2\,^{1}S$) system can be written as
\begin{eqnarray}\label{eee2}
\Delta E^{(2)}&=&-\sum_{n_sn_tn_u}\sum_{L_sL_tL_u}\sum_{M_sM_tM_u}
\frac{|\langle\Psi^{(0)}|V_{123}|\chi_{n_s}(L_sM_s;\boldsymbol{\sigma})\chi_{n_t}(L_tM_t;\boldsymbol{\rho})
\chi_{n_u}(L_uM_u;\boldsymbol{\varsigma})\rangle |^{2}}{E_{n_sL_s;n_tL_t;n_uL_u}-E_{n_0S;n_0S;n_0'S}^{(0)}} \nonumber \\
\nonumber \\
&=&-\sum_{n\geq 3}\bigg(\frac{C_{2n}^{(12)}}{R_{12}^{2n}}+\frac{C_{2n}^{(23)}}{R_{23}^{2n}}+\frac{C_{2n}^{(31)}}{R_{31}^{2n}}\bigg) \,,
\end{eqnarray}
where $\chi_{n_s}(L_sM_s;\boldsymbol{\sigma})\chi_{n_t}(L_tM_t;\boldsymbol{\rho})
\chi_{n_u}(L_uM_u;\boldsymbol{\varsigma})$ is an intermediate state of the system with the energy eigenvalue $E_{n_sL_s;n_tL_t;n_uL_u}=E_{n_sL_s}+E_{n_tL_t}+E_{n_uL_u}$. It is noted that the above summations should exclude terms
with $E_{n_sL_s;n_tL_t;n_uL_u}=E_{n_0S;n_0S;n_0'S}^{(0)}$. $C_{2n}^{(12)}$, $C_{2n}^{(23)}$, and $C_{2n}^{(31)}$ are the additive dispersion coefficients. In this work we are only concerned with $n=3$ and 4 in Eq.~(\ref{eee2}). The corresponding dispersion coefficients are
\begin{eqnarray}
C_{6}^{(IJ)}&=&(|A_{I}|^2 + |A_{J}|^2)\mathbb{T}_1 +|A_{K}|^2\mathbb{T}_2+(A_{I}^{*}A_{J}+ A_{J}^{*}A_{I})\mathbb{T}_3  \,, \label{C6a1}
\end{eqnarray}
\begin{eqnarray}
C_{8}^{(IJ)}&=&(|A_{I}|^2+|A_{J}|^2)\mathbb{R}_1+|A_{K}|^2\mathbb{R}_2+(A_{I}^{*}A_{J}+ A_{J}^{*}A_{I})\mathbb{R}_3 \,, \label{C8a1}
\end{eqnarray}

\begin{eqnarray}\label{eta}
A_{1}=a, A_{2}=b, A_{3}=c\,,
\end{eqnarray}
where $a$, $b$ and $c$ are defined in Eq. (\ref{e111}). The other terms in Eqs. (\ref{C6a1}) and (\ref{C8a1}) are given by

\begin{eqnarray}
\mathbb{T}_1&=&\sum_{n_sn_t} K_1(n_s,n_t,1,1) \, ,
\end{eqnarray}

\begin{eqnarray}
\mathbb{T}_2&=&\sum_{n_sn_t} K_2(n_s,n_t,1,1) \, ,
\end{eqnarray}

\begin{eqnarray}
\mathbb{T}_3&=&\sum_{n_sn_t} K_3(n_s,n_t,1,1) \, ,
\end{eqnarray}

\begin{eqnarray}
\mathbb{R}_1&=&\sum_{n_sn_t} [K_1(n_s,n_t,1,2)+K_1(n_s,n_t,2,1)] \, ,
\end{eqnarray}

\begin{eqnarray}
\mathbb{R}_2&=&\sum_{n_sn_t} [K_2(n_s,n_t,1,2)+K_2(n_s,n_t,2,1)] \, ,
\end{eqnarray}

and

\begin{eqnarray}
\mathbb{R}_3&=&\sum_{n_sn_t} [K_3(n_s,n_t,1,2)+K_3(n_s,n_t,2,1)] \, .
\end{eqnarray}

The $K_i$-Functions are defined by Eqs. (\ref{AK1})-(\ref{AK3}) in the Appendix.

\subsection{The formulas for He($n_0\,^{\lambda}S$)-He($n_0\,^{\lambda}S$)-He($n_0^{\prime}\,^{\lambda}P$)}\label{B}

\subsubsection{The first-order energy}\label{B1}
 According to the perturbation theory, the first-order energy correction for the He($n_0\,^{\lambda}S$)-He($n_0\,^{\lambda}S$)-He($n_0^{\prime}\,^{\lambda}P$) system is
\begin{equation}\label{e14}
\Delta E^{(1)}=\langle \Psi^{(0)}|V_{123}|\Psi^{(0)}\rangle=-\frac{C_3^{(12)}(1,M)}{R_{12}^3}-\frac{C_3^{(23)}(1,M)}{R_{23}^3}-\frac{C_3^{(31)}(1,M)}{R_{31}^3} \,,
\end{equation}
where
\begin{eqnarray}\label{e15}
C_{3}^{(IJ)}(1,M)&=& (A_{I}^{*}A_{J}+A_{J}^{*}A_{I}) \mathbb{D}_0(M) \,, \label{C3a}
\end{eqnarray}
\begin{eqnarray}\label{AI}
A_{1}=a, A_{2}=b, A_{3}=c\,,
\end{eqnarray}
and
\begin{eqnarray}\label{D0}
\mathbb{D}_0(M)&=&\frac{4\pi(-1)^{1+M}}{9(1-M)!(1+M)!}
|\langle\varphi_{n_0}(0;\boldsymbol{\sigma})\|T_1(\boldsymbol{\sigma})
\|\varphi_{n_0'}(1;\boldsymbol{\sigma})\rangle|^2\, ,
\end{eqnarray}
where $a$, $b$, $c$ are defined in Eq. (\ref{e111}). It should be mentioned that there exist only additive long-range interaction terms at this order of perturbation.

\subsubsection{The second-order energy}\label{B2}

The second-order energy correction for the He($n_0\,^{\lambda}S$)-He($n_0\,^{\lambda}S$)-He($n_0^{\prime}\,^{\lambda}P$) system can be written as
\begin{eqnarray}\label{ee2}
\Delta E^{(2)}&=&-\sum_{n_sn_tn_u}\sum_{L_sL_tL_u}\sum_{M_sM_tM_u}
\frac{|\langle\Psi^{(0)}|V_{123}|\chi_{n_s}(L_sM_s;\boldsymbol{\sigma})\chi_{n_t}(L_tM_t;\boldsymbol{\rho})
\chi_{n_u}(L_uM_u;\boldsymbol{\varsigma})\rangle |^{2}}{E_{n_sL_s;n_tL_t;n_uL_u}-E_{n_0S;n_0S;n_0'L}^{(0)}} \nonumber \\
\nonumber \\
&=&-\sum_{n\geq 3}\bigg(\frac{C_{2n}^{(12)}(1,M)}{R_{12}^{2n}}+\frac{C_{2n}^{(23)}(1,M)}{R_{23}^{2n}}+\frac{C_{2n}^{(31)}(1,M)}{R_{31}^{2n}}
\nonumber \\
&+&\frac{C_{2n}^{(12,23)}(1,M)}
{R_{12}^nR_{23}^n}+\frac{C_{2n}^{(23,31)}(1,M)}{R_{23}^nR_{31}^n}+\frac{C_{2n}^{(31,12)}(1,M)}{R_{31}^nR_{12}^n}\bigg) \,,
\end{eqnarray}
where $\chi_{n_s}(L_sM_s;\boldsymbol{\sigma})\chi_{n_t}(L_tM_t;\boldsymbol{\rho})
\chi_{n_u}(L_uM_u;\boldsymbol{\varsigma})$ is an intermediate state of the system with the energy eigenvalue $E_{n_sL_s;n_tL_t;n_uL_u}=E_{n_sL_s}+E_{n_tL_t}+E_{n_uL_u}$. It is noted that the above summations should exclude terms
with $E_{n_sL_s;n_tL_t;n_uL_u}=E_{n_0S;n_0S;n_0'L}^{(0)}$. $C_{2n}^{(12)}(1,M)$, $C_{2n}^{(23)}(1,M)$, and $C_{2n}^{(31)}(1,M)$ are the additive dispersion coefficients, and $C_{2n}^{(12,23)}(1,M)$, $C_{2n}^{(23,31)}(1,M)$, and $C_{2n}^{(31,12)}(1,M)$ are the nonadditive dispersion coefficients. In this work we are only concerned with $n=3$ and 4 in Eq.~(\ref{ee2}). The corresponding dispersion coefficients are
\begin{eqnarray}
C_{6}^{(IJ)}(1,M)&=&(|A_{I}|^2 + |A_{J}|^2)\mathbb{D}_1(M) +|A_{K}|^2\mathbb{D}_2  \,, \label{C6a}
\end{eqnarray}
\begin{eqnarray}
C_{8}^{(IJ)}(1,M)&=&(|A_{I}|^2+|A_{J}|^2)\mathbb{Q}_1(M)+|A_{K}|^2\mathbb{Q}_2+(A_{I}^{*}A_{J}+ A_{J}^{*}A_{I})\mathbb{Q}_3(M) \,, \label{C8a}
\end{eqnarray}
\begin{eqnarray}
C_{6}^{(IJ,JK)}(1,M)&=&\mathbb{Q}_4(A_{K},A_{I},1,M,\eta_{J})  \,, \label{C6d}
\end{eqnarray}
\begin{eqnarray}
C_{8}^{(IJ,JK)}(1,M)&=&\mathbb{Q}_4(A_{K},A_{I},2,M,\eta_{J}) \,, \label{C8d}
\end{eqnarray}
with
\begin{eqnarray}\label{eta}
A_{1}=a, A_{2}=b, A_{3}=c, \eta_{1}=\alpha, \eta_{2}=\beta, \eta_{3}=\gamma\,,
\end{eqnarray}
where $a$, $b$ and $c$ are defined in Eq. (\ref{e111}) and $\alpha$, $\beta$ and $\gamma$ are the interior angles,
and the other terms in Eqs. (\ref{C6a})-(\ref{C8d}) are given by
\begin{eqnarray}
\mathbb{D}_1(M)&=&\sum_{n_sn_t L_s}F_1(n_s,n_t,L_s,1;1,1;1,M) \,, \\
\mathbb{D}_2&=&\sum_{n_sn_t} F_2(n_s,n_t,1,1) \,,\\
\mathbb{Q}_1(M)&=& \sum_{n_sn_t L_s} [F_1(n_s,n_t,L_s,1;1,3;1,M) + F_1(n_s,n_t,L_s,1;2,2;1,M) \nonumber \\
&+&F_1(n_s,n_t,L_s,1;3,1;1,M) + F_1(n_s,n_t,L_s,2;1,1;1,M)] \,, \\
\mathbb{Q}_2&=&\sum_{n_sn_t}[F_2(n_s,n_t,1,2)+F_2(n_s,n_t,2,1)]  \,, \\
\mathbb{Q}_3(M)&=& \sum_{n_sn_t}[F_3(n_s,n_t,1,1;2,2;1,M)+F_3(n_s,n_t,1,2;2,1;1,M) \nonumber \\
&+&F_3(n_s,n_t,2,1;1,2;1,M)+F_3(n_s,n_t,2,2;1,1;1,M)] \,,\\
\end{eqnarray}
and
\begin{eqnarray}
\mathbb{Q}_4(A_K,A_I,\lambda,M,\eta_J)&=&2\sum_{n_tM_t}\operatorname{Re}[
A_K^*A_I\,e^{i(M_{t}-M)\eta_J}]F_4(n_t,\lambda,M_t;1,M) \,.\label{QQ4}
\end{eqnarray}
In the above $(I,J,K)$ forms a permutation of (1,2,3).
%Many of the details of the perturbation theory were given in Ref.~\cite{yan16}, which we refer to as Paper I.
Many of the details of the perturbation theory were given in Paper I.
The expressions of the $F_i$ appearing above
are given
%in Ref.~\cite{yan16} are preceded
by Eqs. (A10)-(A13) of Paper~I.

\section{Results and Discussion}\label{Res}

In the present work, the atomic wave functions of helium were constructed variationally using Hylleraas basis sets
and the intermediate states were generated by diagonalizing the helium Hamiltonian \cite{yan96-1}. All relevant
matrix elements of the multipole transition operators were thus calculated, including the finite nuclear mass corrections \cite{zhang06}. With these,
we calculate
the first-order dipolar and second-order long-range dispersion coefficients for
the He${(n_0\,^{\lambda}S)}$-He${(n_0\,^{\lambda}S)}$-He${(n_0^{\prime}\,^{\lambda}L)}$ system.

According to the Eqs. (\ref{eee2})--(\ref{QQ4}), we can see that the dependence of these interaction coefficients on the geometrical configurations of the three atoms is due to two reasons. Firstly, all these coefficients contain the expansion coefficients $a$, $b$, $c$. The cause of this dependence is the existence of the three degenerate states that are shown in Eqs. (\ref{e6a_a})--(\ref{e6c_c}). Secondly, these nonadditive interaction coefficients contain the interior angles of the triangle $\alpha$, $\beta$, $\gamma$ formed by these three helium atoms. So in this paper, the expansion coefficients $a$, $b$, $c$ and the interior angles $\alpha$, $\beta$, $\gamma$ are called geometric parameters.

Now, let us reconsider the formulas of additive dispersion interaction coefficients that are shown in Eqs. (\ref{C6a1}), (\ref{C8a1}), (\ref{C3a}), (\ref{C6a}) and (\ref{C8a}). Since these coefficients do not contain the interior angles $\alpha$, $\beta$, $\gamma$, we can easily separate them into geometric parameters $a$, $b$, $c$, and the interaction $constants$ $\mathbb{T}_1$, $\mathbb{T}_2$, $\mathbb{T}_3$, $\mathbb{R}_1$, $\mathbb{R}_2$, $\mathbb{R}_3$, $\mathbb{D}_0(M=0)$, $\mathbb{D}_0(M=\pm 1)$, $\mathbb{D}_1(M=0)$, $\mathbb{D}_1(M=\pm1)$, $\mathbb{D}_2$, $\mathbb{Q}_1(M=0)$, $\mathbb{Q}_1(M=\pm1)$, $\mathbb{Q}_2$, $\mathbb{Q}_3(M=0)$, $\mathbb{Q}_3(M=\pm 1)$ that are independent of geometrical configuration of the three atoms.
% and only depend on the invovlving atoms.
For the He($1\,^{1}S$)-He($1\,^{1}S$)-He($2\,^{1}S$) system, $\mathbb{T}_1$, $\mathbb{T}_2$, $\mathbb{T}_3$ are connected to the second-order additive dispersion coefficient $C_6^{(IJ)}$. $\mathbb{R}_1$, $\mathbb{R}_2$, $\mathbb{R}_3$, are connected to the second-order additive dispersion coefficient $C_8^{(IJ)}$. For He($n_0\,^{\lambda}S$)-He($n_0\,^{\lambda}S$)-He($n_0^{\prime}\,^{\lambda}P$) system, $\mathbb{D}_0(M=0)$ and $\mathbb{D}_0(M=\pm 1)$ are connected to the first-order additive coefficient $C_3^{(IJ)}(L,M)$. $\mathbb{D}_1(M=0)$, $\mathbb{D}_1(M=\pm1)$, and $\mathbb{D}_2$ are connected to the second-order additive dispersion coefficient $C_6^{(IJ)}(L,M)$, and $\mathbb{Q}_1(M=0)$, $\mathbb{Q}_1(M=\pm1)$, $\mathbb{Q}_2$, $\mathbb{Q}_3(M=0)$, and $\mathbb{Q}_3(M=\pm1)$ are connected to the second-order additive dispersion coefficient $C_8^{(IJ)}(L,M)$.  The values of these interaction constants for two helium isotopes are listed in Table \ref{TabIS} and Table \ref{TabI}, respectively.
While, as shown in Eqs. (\ref{C6d}) and (\ref{C8d}), the nonadditive interaction coefficients are inseparable, because they contain both the expansion coefficients $a$, $b$, $c$ and the interior angles $\alpha$, $\beta$, $\gamma$.

Actually, with these interaction constants, we can easily find the connection between this work and previous studies of long-range interactions for two-body system. For example, if we set $a=\frac{1}{\sqrt{2}}$, $b=\pm\frac{1}{\sqrt{2}}$, $c=0$, our expressions describe the long-range interactions for the two-body $n_0S-n_0'L$ system.
For example, for He($1\,^{1}S$)-He($2\,^{1}S$) system, $\mathbb{T}_1$, the second-order dispersion coefficients $C_6=\mathbb{T}_1+\mathbb{T}_3$, $C_8=\mathbb{R}_1+\mathbb{R}_3$. For He($n_0\,^{\lambda}S$)-He($n_0\,^{\lambda}S$)-He($n_0^{\prime}\,^{\lambda}P$) system, the first-order dispersion interaction coefficient $C_3(M)=\mathbb{D}_0(M)$, the second-order dispersion coefficients $C_6(M)=\mathbb{D}_1(M)$, $C_8(M)=\mathbb{Q}_1(M)+\mathbb{Q}_3(M)$. Similarly, If we set $a=b=0$, $c=1$, our  expressions describe the long-range interactions for the two-body $n_0S-n_0'S$ system, where $C_6=\mathbb{D}_2$, $C_8=\mathbb{Q}_2$.

\subsection{Dipolar and dispersion coefficients for He${(1\,^{1}S)}$-He${(1\,^{1}S)}$-He${(2\,^{1}S)}$}

\subsubsection{zeroth-order wave function of He${(1\,^{1}S)}$-He${(1\,^{1}S)}$-He${(2\,^{1}S)}$}

In this work, we are only concerned with the equilateral triangle with $R_{12}$= $R_{23}$ =  $R_{31}$ =$R$. The perturbation matrix with respect to \{$\phi_1$, $\phi_2$, $\phi_3$\} thus becomes
\begin{equation}\label{e110}
{H}^{\prime}=
\left(
  \begin{array}{ccc}
    A & D & D\\
    D & A & D\\
    D & D & A\\
  \end{array}
\right)\,,
\end{equation}

where

\begin{eqnarray}
A=\Delta_{11}=\Delta_{22}=\Delta_{33}\,.
\end{eqnarray}

\begin{eqnarray}
D=\Delta_{12}=\Delta_{13}=\Delta_{21}=\Delta_{23}=\Delta_{31}=\Delta_{32}\,.
\end{eqnarray}

Solving the eigenvalue problem of the above matrix Eq. (\ref{e110}), one obtains the eigenvalues: $A+2D$, $A-D$, $A-D$, and the corresponding zeroth-order wave functions:
\begin{eqnarray}
\Psi_{1,\Delta}^{(0)}&=&\frac{1}{\sqrt{3}}|\phi_1\rangle+\frac{1}{\sqrt{3}}|\phi_2\rangle+\frac{1}{\sqrt{3}}|\phi_3\rangle \,, \label{delta11} \\
\Psi_{2,\Delta}^{(0)}&=&\frac{1}{\sqrt{2}}|\phi_1\rangle-\frac{1}{\sqrt{2}}|\phi_3\rangle \,, \label{delta22} \\
\Psi_{3,\Delta}^{(0)}&=&\frac{1}{\sqrt{6}}|\phi_1\rangle-\sqrt{\frac{2}{3}}|\phi_2\rangle+\frac{1}{\sqrt{6}}|\phi_3\rangle \,, \label{delta33}
\end{eqnarray}
where the symbol $\Delta$ denotes the equilateral triangle.

\subsubsection{Dipolar and dispersion coefficients for the equilateral triangle}

With these zeroth-order wave functions Eqs. (\ref{delta11})-(\ref{delta33}) the corresponding long-range interaction coefficients $C_6^{(IJ)}$ and $C_8^{(IJ)}$ for the isotopic He${(1\,^{1}S)}$-He${(1\,^{1}S)}$-He${(2\,^{1}S)}$ systems are listed in Table~\ref{TabIIS}. We note that these coefficients $C_6^{(IJ)}$ and $C_8^{(IJ)}$ are all same and positive for $\Psi_{1,\Delta}^{(0)}$ with the geometric parameter $a=b=c=1/\sqrt{3}$. For $\Psi_{2,\Delta}^{(0)}$ with the geometric parameter $b=0$, $a=-c=1/\sqrt{2}$, $C_6^{(12)}=C_6^{(23)}$ and $C_8^{(12)}=C_8^{(23)}$. For $\Psi_{3,\Delta}^{(0)}$ with the geometric parameter $b=-\sqrt{\frac{2}{3}}$, $a=c=1/\sqrt{6}$, we also have $C_6^{(12)}$= $C_6^{(23)}$ and $C_8^{(12)}$= $C_8^{(23)}$. These relationships between these coefficients are due to the different values of geometric parameters.

\begin{figure}
\begin{center}
\includegraphics[width=14cm,height=8cm]{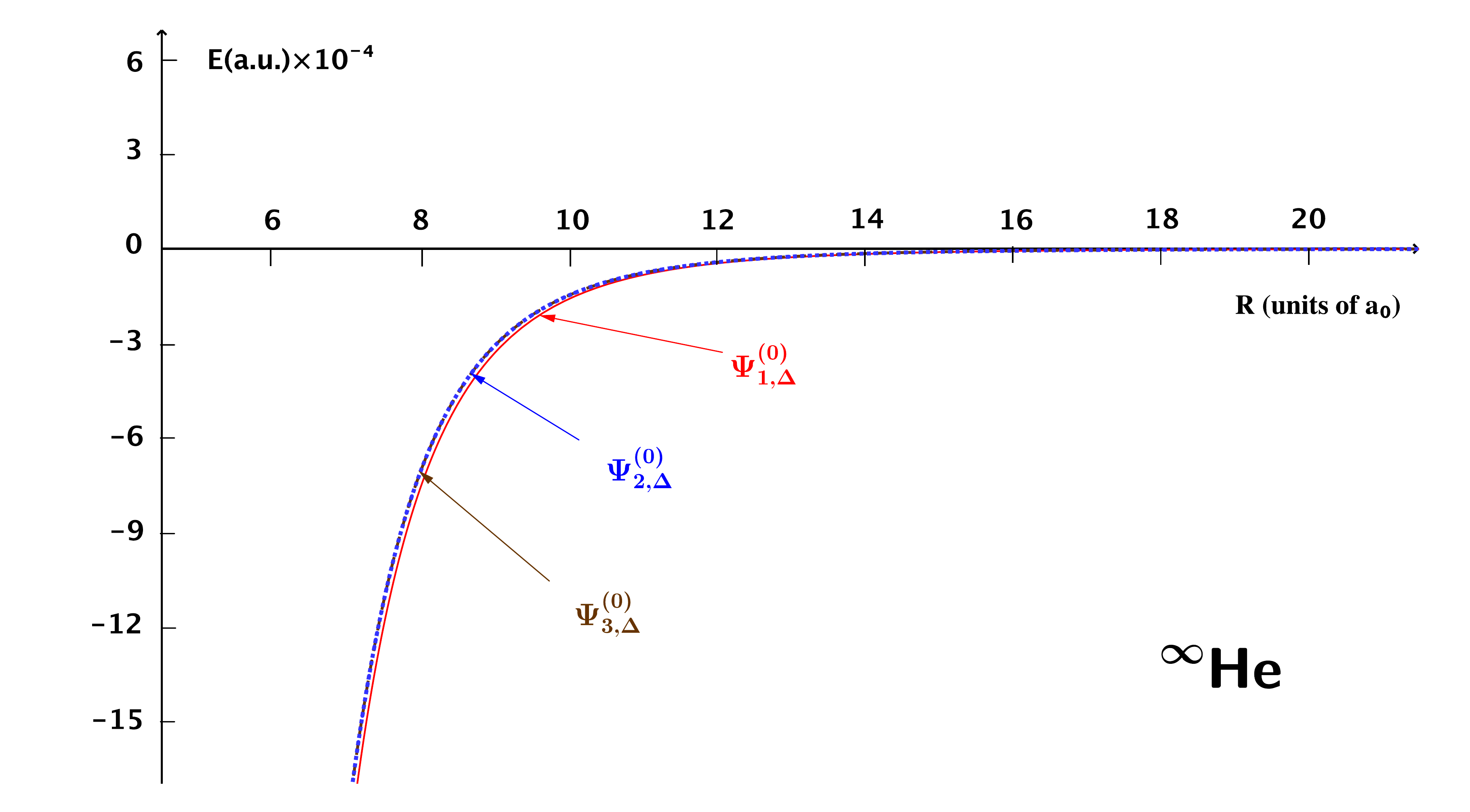}
\end{center}
\caption {\label{f1S} Long-range potentials for the He($1\,^1S$)-He($1\,^1S$)-He($2\,^{1}S$) system for three different types of the zeroth-order wave functions, where the three atoms form an equilateral triangle, in atomic units.
For each curve labeled by a wave function, the plotted curve is the sum of $\Delta E^{(1)}$ and $\Delta E^{(2)}$, where $\Delta E^{(1)}$=0.
}
\end{figure}

The curves of potential energy $(E)$, for the He${(1\,^{1}S)}$-He${(1\,^{1}S)}$-He${(2\,^{1}S)}$ system,  corresponding to the different zeroth-order wave functions are plotted in Fig.~\ref{f1S}.

\subsection{Dipolar and dispersion coefficients for He${(n_0\,^{\lambda}S)}$-He${(n_0\,^{\lambda}S)}$-He${(n_0^{\prime}\,^{\lambda}P)}$}

\subsubsection{zeroth-order wave function of He${(n_0\,^{\lambda}S)}$-He${(n_0\,^{\lambda}S)}$-He${(n_0^{\prime}\,^{\lambda}P)}$}

In this work, we need to obtain the values of these geometric parameters for specific configurations by using degenerate perturbation theory. Here, we are only concerned with the equilateral triangle with $R_{12}$=  $R_{23}$ =  $R_{31}$ =$R$. The perturbation matrix with respect to \{$\phi_1$, $\phi_2$, $\phi_3$\} thus becomes
\begin{equation}\label{e100}
{H}^{\prime}=H_{12}^{\prime}
\left(
  \begin{array}{ccc}
    0 & 1 & 1\\
    1 & 0 & 1\\
    1 & 1 & 0\\
  \end{array}
\right)\,,
\end{equation}
where
\begin{eqnarray}
H^{\prime}_{12}
  &=&\frac{4\pi}{R^{2L+1}}\frac{(-1)^{M}[(2L-1)!!]^2}{(2L+1)^2(L-M)!(L+M)!} |\langle\varphi_{n_0}(0;\boldsymbol{\sigma})\|T_L(\boldsymbol{\sigma})
\|\varphi_{n_0'}(L;\boldsymbol{\sigma})\rangle|^2\,.
\end{eqnarray}

Solving the eigenvalue problem of the above matrix (Eq. (\ref{e100})) for the equilateral triangle, one obtains the eigenvalues: $2H_{12}^{\prime}$, $-H_{12}^{\prime}$, $-H_{12}^{\prime}$, and the corresponding orthonormalized zeroth-order wave functions:
\begin{eqnarray}
\Psi_{1,\Delta}^{(0)}&=&\frac{1}{\sqrt{3}}|\phi_1\rangle+\frac{1}{\sqrt{3}}|\phi_2\rangle+\frac{1}{\sqrt{3}}|\phi_3\rangle \,, \label{delta1} \\
\Psi_{2,\Delta}^{(0)}&=&\frac{1}{\sqrt{2}}|\phi_1\rangle-\frac{1}{\sqrt{2}}|\phi_3\rangle \,, \label{delta2} \\
\Psi_{3,\Delta}^{(0)}&=&\frac{1}{\sqrt{6}}|\phi_1\rangle-\sqrt{\frac{2}{3}}|\phi_2\rangle+\frac{1}{\sqrt{6}}|\phi_3\rangle \,, \label{delta3}
\end{eqnarray}
where the symbol $\Delta$ denotes the equilateral triangle.
We note that Eq.~(\ref{delta1}) corresponds to the ``Dicke state''~\cite{FenLiZhu13}.

\subsubsection{Dipolar and dispersion coefficients for the equilateral triangle}

 With these zeroth-order wave functions Eqs. (\ref{delta1})-(\ref{delta3}), for the isotopic He${(1\,^{1}S)}$-He${(1\,^{1}S)}$-He${(2\,^{1}P)}$, He${(2\,^{1}S)}$-He${(2\,^{1}S)}$-He${(2\,^{1}P)}$, He${(2\,^{3}S)}$-He${(2\,^{3}S)}$-He${(2\,^{3}P)}$ systems, the corresponding long-range interaction coefficients are listed in Tables~\ref{TabII}-\ref{TabVI}.

Table~\ref{TabII} lists the first-order dipolar coefficients $C_3^{(IJ)}$(1,M) for the equilateral triangle. We note that these coefficients all satisfy $C_3^{(IJ)}(1,M=0)=$$-$$2C_3^{(IJ)}(1,M=\pm 1)$. For $\Psi_{1,\Delta}^{(0)}$ with the geometric parameter $a=b=c=1/\sqrt{3}$, $C_3^{(IJ)}(1,M=0)$ are all same and negative; $C_3^{(IJ)}(1,M=\pm1)$ are all same and positive. For $\Psi_{2,\Delta}^{(0)}$ with the geometric parameter $b=0$, $a=-c=1/\sqrt{2}$, $C_3^{(12)}(1,M)=C_3^{(23)(1,M)}=0$; $C_3^{(31)}(1,M=0)$ appears positive, $C_3^{(31)}(1,M=\pm1)$ negative. For $\Psi_{3,\Delta}^{(0)}$ with the geometric parameter $b=-\sqrt{\frac{2}{3}}$, $a=c=1/\sqrt{6}$, $C_3^{(12)}(1,M=0)$ and $C_3^{(23)}(1,M=0)$ are same and positive; $C_3^{(12)}(1,M=\pm1)$ and $C_3^{(23)}(1,M=\pm1)$ are same and negative. While $C_3^{(31)}(1,M=0)$ appears negative, $C_3^{(31)}(1,M=\pm1)$ positive. All these relationships between these coefficients are due to the different values of geometric parameters. These positive or negative signs mean that different states of a fixed system may lead to different types of interaction: attraction or repulsion.

Table~\ref{TabIII} lists the leading terms of the second-order long-range interaction $C_6^{(IJ)}(1,M=0)$ and $C^{(IJ,JK)}_6(1,M=0)$ for the equilateral triangle. Table~\ref{TabIV} lists $C_6^{(IJ)}(1,M=\pm1)$ and $C^{(IJ,JK)}_6(1,M=\pm1)$. We note that the absolute values of $C_6^{(IJ)}(1,M=0)$ and $C^{(IJ,JK)}_6(1,M=0)$ are always a little bit larger than those of $C_6^{(IJ)}(1,M=\pm1)$ and $C^{(IJ,JK)}_6(1,M=\pm1)$, respectively. The additive interactions coefficients $C_6^{(IJ)}(1,M)$ are always positive, while the nonadditive interactions coefficients $C_6^{(IJ,JK)}(1,M)$ can be postive, negative or zero. This is due to the different signs of $\mathbb{Q}_4(a,b,1,M,\gamma)$ and geometric parameters. For example, for $\Psi_{2,\Delta}^{(0)}$, the nonadditive coefficients $C_6^{(12,23)}(1,M=0)$ is negative because $\mathbb{Q}_4(a,b,1,M,\gamma)>0$ and $a=-c=1/\sqrt{2}$; the nonadditive coefficients $C_6^{(23,31)}(1,M)$ and $C_6^{(31,12)}(1,M)$ are zero because  $\mathbb{Q}_4(a,b,1,M,\gamma)=0$ and $b=0$. The second-order long-range interaction $C_8^{(IJ)}(1,M)$ and $C^{(IJ,JK)}_8(1,M)$have a very similar characteristic
as $C_6^{(IJ)}(1,M)$ and $C_6^{(IJ,JK)}(1,M)$, respectively. The numerical values are listed in Tables~\ref{TabV}-\ref{TabVI}.

\begin{figure}
\begin{center}
\includegraphics[width=14cm,height=8cm]{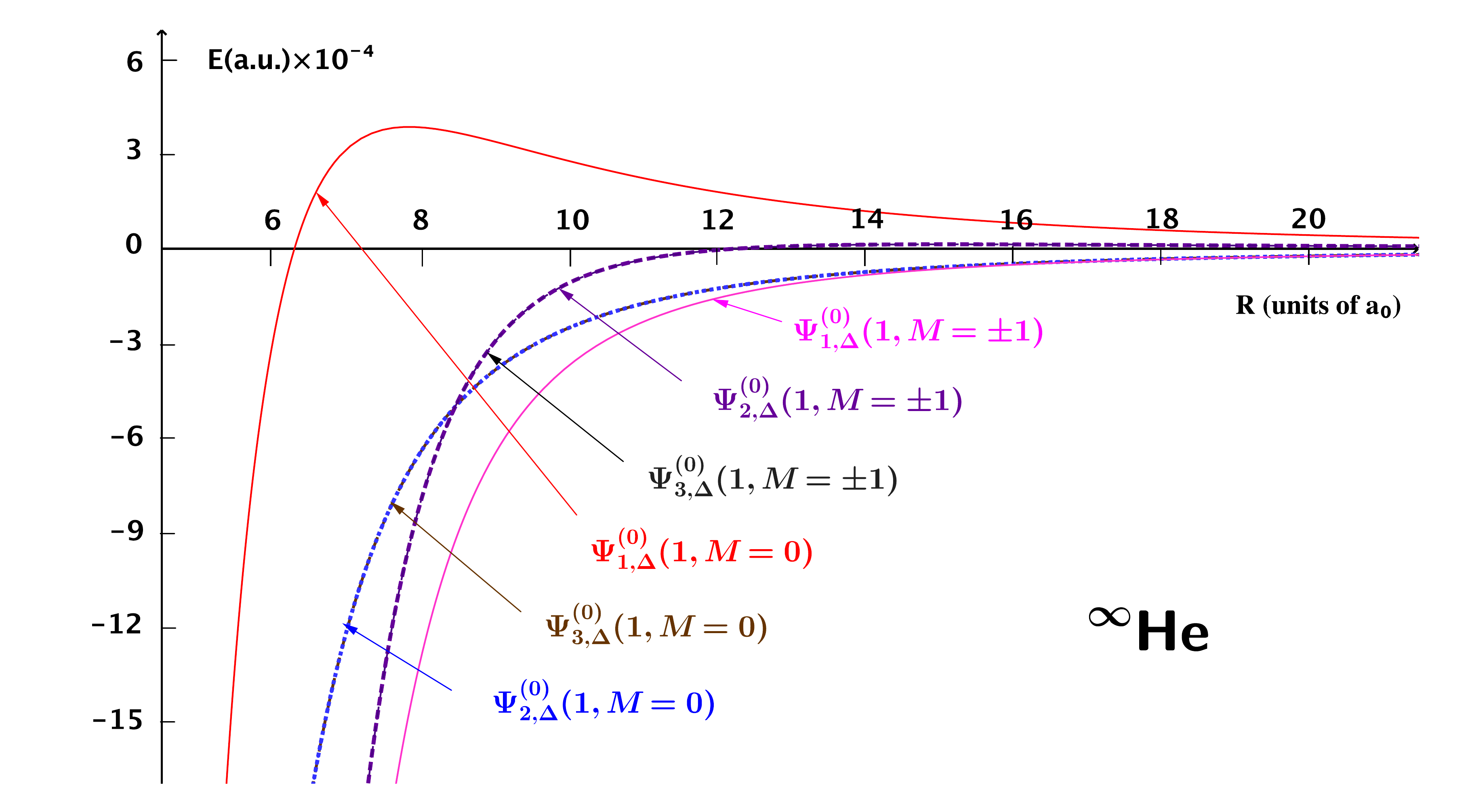}
\end{center}
\caption {\label{f2} Long-range potentials for the He($1\,^1S$)-He($1\,^1S$)-He($2\,^{1}P$) system for three different types of the zeroth-order wave functions, where the three atoms form an equilateral triangle, in atomic units.
For each curve labeled by a wave function, the plotted curve is the sum of $\Delta E^{(1)}$ and $\Delta E^{(2)}$.
}
\end{figure}

\begin{figure}
\begin{center}
\includegraphics[width=14cm,height=8cm]{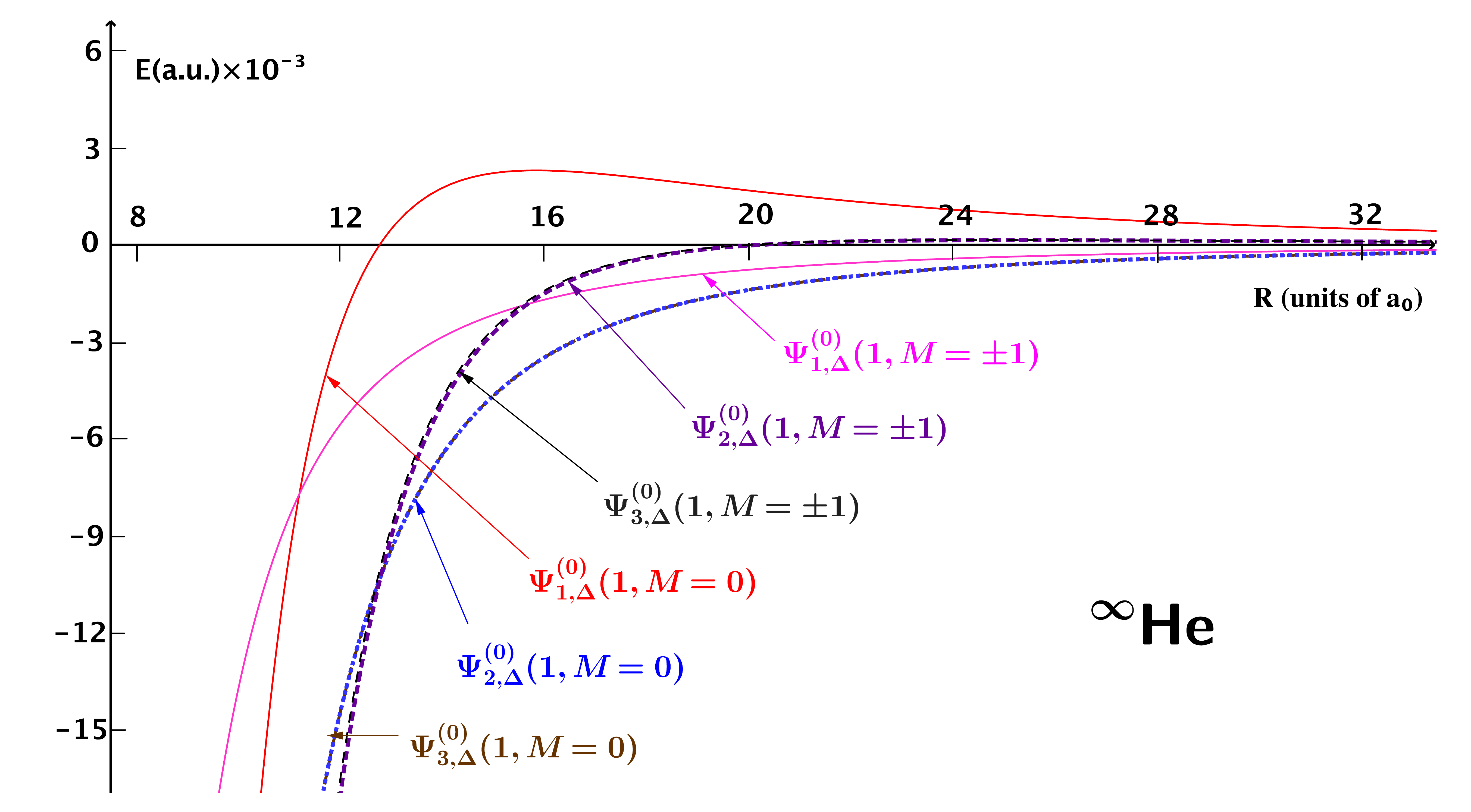}
\end{center}
\caption {\label{f3} Long-range potentials for the He($2\,^1S$)-He($2\,^1S$)-He($2\,^{1}P$) system for three different types of the zeroth-order wave functions, where the three atoms form an equilateral triangle, in atomic units.
For each curve labeled by a wave function, the plotted curve is the sum of $\Delta E^{(1)}$ and $\Delta E^{(2)}$.
}
\end{figure}

\begin{figure}
\begin{center}
\includegraphics[width=14cm,height=8cm]{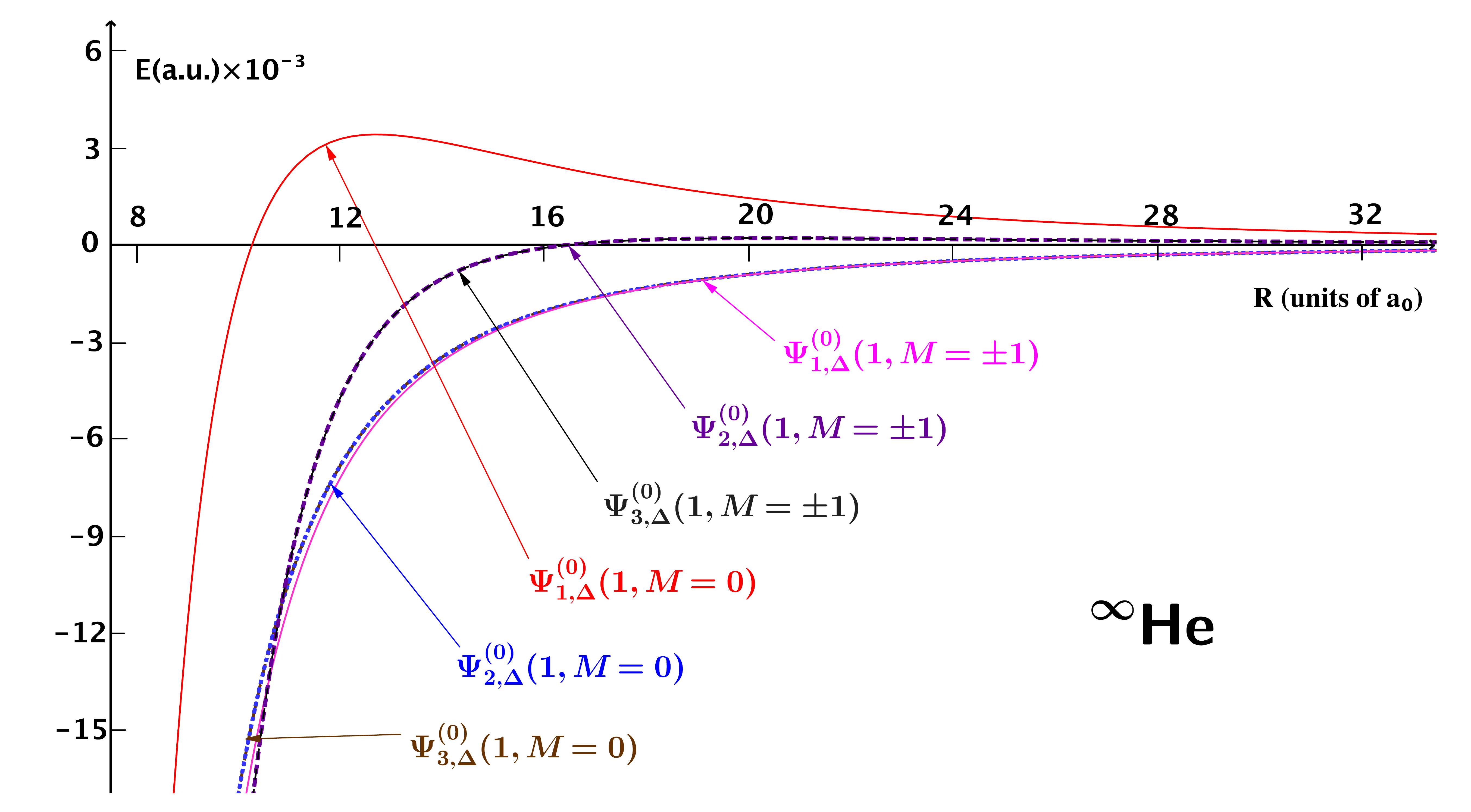}
\end{center}
\caption {\label{f4} Long-range potentials for the He($2\,^3S$)-He($2\,^3S$)-He($2\,^{3}P$) system for three different types of the zeroth-order wave functions, where the three atoms form an equilateral triangle, in atomic units.
For each curve labeled by a wave function, the plotted curve is the sum of $\Delta E^{(1)}$ and $\Delta E^{(2)}$.
}
\end{figure}

From Tables~\ref{TabIII}-\ref{TabVI}, we can see that the dispersion coefficients for the additive terms are always positive,
but the dispersion coefficients for the nonadditive terms can be positive or negative or zero. Furthermore, the absolute values of the non-zero nonadditive dispersion coefficients are less than the additive dispersion coefficients by one to two orders of  magnitude. However, the nonadditive terms may not be neglected in constructing an accurate potential surface. For example, the ratios of $(\frac{C_8^{(12,23)}(1,M=\pm 1)}{R_{12}^4R_{23}^4})/(\frac{C_8^{(12)}(1,M=\pm 1)}{R_{12}^8})$ for the $\Psi_{1,\Delta}^{(0)}$ of He${(1\,^{1}S)}$-He${(1\,^{1}S)}$-He${(2\,^{1}P)}$, He${(2\,^{1}S)}$-He${(2\,^{1}S)}$-He${(2\,^{1}P)}$ and He${(2\,^{3}S)}$-He${(2\,^{3}S)}$-He${(2\,^{3}P)}$ systems,
are 0.08\%, 26\%, 24\%, respectively. The curves of potential energy $(E)$, for the He${(1\,^{1}S)}$-He${(1\,^{1}S)}$-He${(2\,^{1}P)}$, He${(2\,^{1}S)}$-He${(2\,^{1}S)}$-He${(2\,^{1}P)}$ and He${(2\,^{3}S)}$-He${(2\,^{3}S)}$-He${(2\,^{3}P)}$ systems,  corresponding to the different zeroth-order wave functions are plotted in Fig.~\ref{f2} - Fig.~\ref{f4}, respectively.

%%%%%%%%%%%%%%%%%%%%%%%%%%%%%%%%%%%%%%%%%%%%%%%%%%%%%%%%%%%%%%%%%%%%%%%%%%%%%%%%%%%%%%%%%%%%%%%%%%%%%%%%%%%%%%%%%%%%%%%%%%%%%%%%%%%%%%%%%%

\subsection{Dipolar and dispersion coefficients for the isosceles right triangle}

For $\lambda_1=\frac{1}{\sqrt{2}}$ and $\lambda_2=1$, solving the eigenvalue problem of the above matrix (Eq.(\ref{e10})) for the isosceles right triangle, one obtains the eigenvalues: $\frac{H_{12}^{\prime}}{8} \left(\sqrt{2}+\sqrt{130}\right), \frac{H_{12}^{\prime}}{8} \left(\sqrt{2}-\sqrt{130}\right), -\frac{H_{12}^{\prime}}{2 \sqrt{2}}$, and the corresponding orthonormalized zeroth-order wave functions:
\begin{eqnarray}
\Psi_{1,\bot}^{(0)}&=&\frac{\sqrt{130}-\sqrt{2}}{2 \sqrt{65-\sqrt{65}}}|\phi_1\rangle+\frac{4}{\sqrt{65-\sqrt{65}}}|\phi_2\rangle
+\frac{4}{\sqrt{65-\sqrt{65}}}|\phi_3\rangle  \,,  \label{bot1} \\
\Psi_{2,\bot}^{(0)}&=&\frac{-(\sqrt{130}+\sqrt{2})}{2 \sqrt{65+\sqrt{65}}}|\phi_1\rangle+\frac{4}{\sqrt{65+\sqrt{65}}}|\phi_2\rangle
+ \frac{4}{\sqrt{65+\sqrt{65}}}|\phi_3\rangle \,,  \label{bot2} \\
\Psi_{3,\bot}^{(0)}&=&-\frac{1}{\sqrt{2}}|\phi_2\rangle+\frac{1}{\sqrt{2}}|\phi_3\rangle \,,  \label{bot3}
\end{eqnarray}
where the symbol $\bot$ denotes the isosceles right triangle. With these zero-order wave functions Eqs. (\ref{bot1})-(\ref{bot3}), for the isotopic He${(1\,^{1}S)}$-He${(1\,^{1}S)}$-He${(2\,^{1}P)}$, He${(2\,^{1}S)}$-He${(2\,^{1}S)}$-He${(2\,^{1}P)}$, He${(2\,^{3}S)}$-He${(2\,^{3}S)}$-He${(2\,^{3}P)}$ systems, the corresponding long-range interaction coefficients are listed in Tables~\ref{TabVII}-\ref{TabXI}.

Table~\ref{TabVII} lists the first-order dipolar coefficients $C_3^{(31)}(1,M)$ for the the isosceles right triangle. We note that $C_3^{(12)}(1,M)=C_3^{(31)}(1,M)$ for all three zero-order wave functions $\Psi_{1,\bot}^{(0)}$, $\Psi_{2,\bot}^{(0)}$ with the geometric parameters $b=c$ and $\Psi_{3,\bot}^{(0)}$ with the geometric parameters $a=0$. This is also why $C_3^{(12)}(1,M)=C_3^{(31)}(1,M)=0$ for $\Psi_{3,\bot}^{(0)}$. Similar with the coefficients shown in Table~\ref{TabII}, these coefficients also satisfy $C_3^{(IJ)}(1,M=0)=$$-$$2C_3^{(IJ)}(1,M=\pm 1)$. But the number values of these coefficients are totally different. This is because a change in geometric configuration would lead to the change in quantum state of the three-body system, then lead to the change in these number values of these long-range interaction coefficients.

\begin{figure}
\begin{center}
\includegraphics[width=14cm,height=8cm]{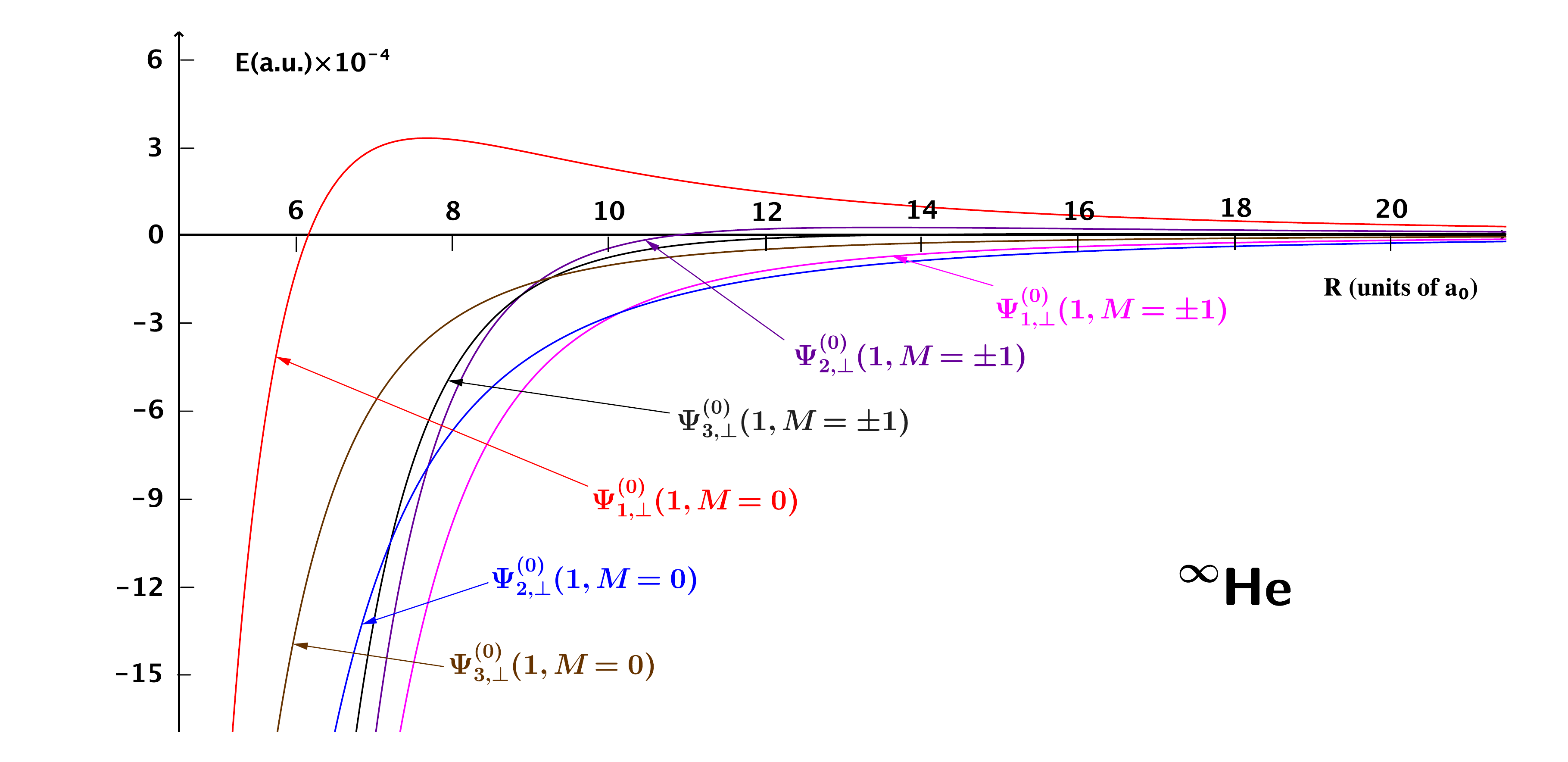}
\end{center}
\caption {\label{f5} Long-range potentials for the He($1\,^1S$)-He($1\,^1S$)-He($2\,^{1}P$) system for three different types of the zeroth-order wave functions, where the three atoms form an isosceles right triangle, in atomic units.
For each curve labeled by a wave function, the plotted curve is the sum of $\Delta E^{(1)}$ and $\Delta E^{(2)}$.
}
\end{figure}

\begin{figure}
\begin{center}
\includegraphics[width=14cm,height=8cm]{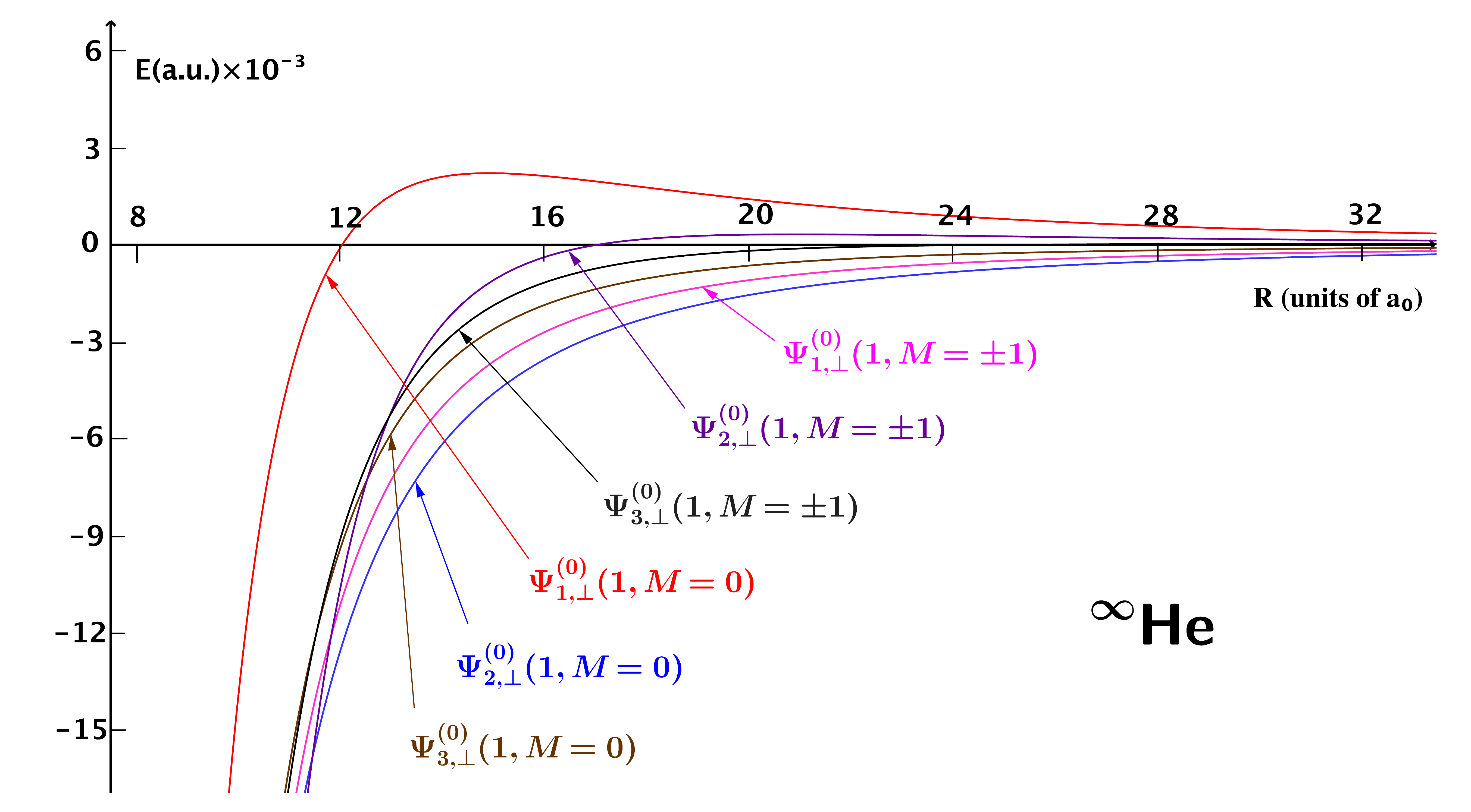}
\end{center}
\caption {\label{f6} Long-range potentials for the He($2\,^1S$)-He($2\,^1S$)-He($2\,^{1}P$) system for three different types of the zeroth-order wave functions, where the three atoms form an isosceles right triangle, in atomic units.
For each curve labeled by a wave function, the plotted curve is the sum of $\Delta E^{(1)}$ and $\Delta E^{(2)}$.
}
\end{figure}

\begin{figure}
\begin{center}
\includegraphics[width=14cm,height=8cm]{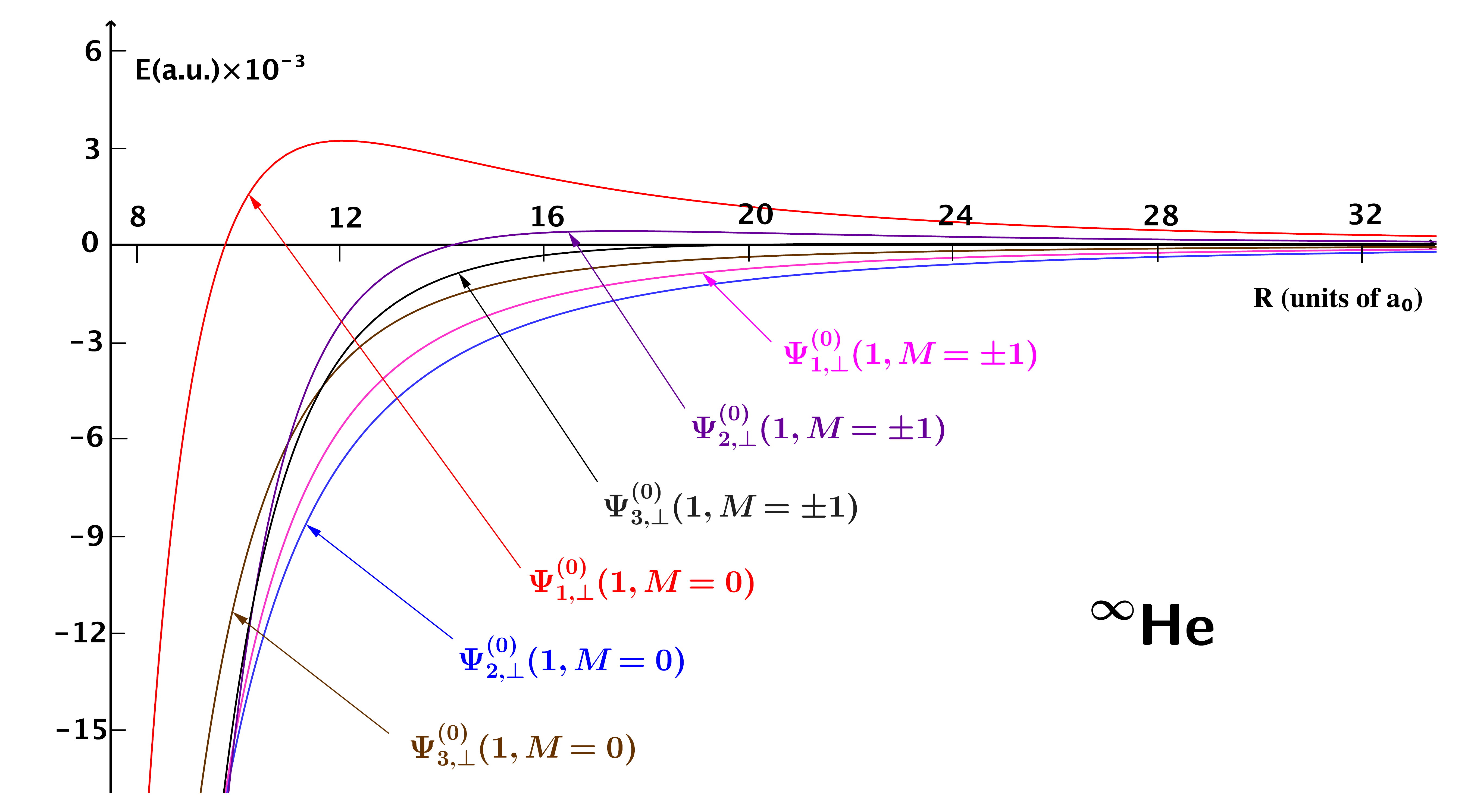}
\end{center}
\caption {\label{f7} Long-range potentials for the He($2\,^3S$)-He($2\,^3S$)-He($2\,^{3}P$) system for three different types of the zeroth-order wave functions, where the three atoms form an isosceles right triangle, in atomic units.
For each curve labeled by a wave function, the plotted curve is the sum of $\Delta E^{(1)}$ and $\Delta E^{(2)}$.
}
\end{figure}

Tables~\ref{TabVIII}-\ref{TabXI} list the second-order dispersion coefficients $C_6^{(IJ)}(1,M)$, $C_6^{(IJ,JK)}(1,M)$, $C_8^{(IJ)}(1,M)$ and $C_8^{(IJ,JK)}(1,M)$ for the the isosceles right triangle. We note that $C_6^{(12)}(1,M)=C_6^{(31)}(1,M)$, $C_6^{(12,23)}(1,M)=C_6^{(23,31)}(1,M)$, $C_8^{(12)}(1,M)=C_8^{(31)}(1,M)$ and $C_8^{(12,23)}(1,M)=C_8^{(23,31)}(1,M)$ because $b=\pm c$, $a=0$, and $\beta=\gamma$. Also $C_8^{(31,12)}(1,M)=0$ and $C_6^{(31,12)}(1,M)\neq 0$ because $\alpha=\pi/2$ and  $M_t-M$
can be even or odd in Eq.~(\ref{QQ4}).
We find
that allowing for finite nuclear mass increases the additive dispersion coefficients,
as shown in Tables~\ref{TabVIII}-\ref{TabXI}.
The nonadditive terms may also not be neglected in constructing a three-body potential surface for He($n_0\,^{\lambda}S$)-He($n_0\,^{\lambda}S$)-Li($n_0^{\prime}\,^{\lambda}P$). The curves of potential energy $(E)$ of the He${(1\,^{1}S)}$-He${(1\,^{1}S)}$-He${(2\,^{1}P)}$, He${(2\,^{1}S)}$-He${(2\,^{1}S)}$-He${(2\,^{1}P)}$ and He${(2\,^{3}S)}$-He${(2\,^{3}S)}$-He${(2\,^{3}P)}$ systems, resulting
from $\Delta E^{(1)}$ and $\Delta E^{(2)}$ for this geometrical structure are shown in Figs.~\ref{f5} - Fig.~\ref{f7}, respectively.

\subsection{Dipolar and dispersion coefficients for a straight line}

For $\lambda_1=\frac{1}{2}$ and $\lambda_2=1$, solving the eigenvalue problem of the above matrix (Eq.(\ref{e10})) for the isosceles right triangle, one obtains the eigenvalues: $\frac{H_{12}^{\prime}}{16} \left(1+3 \sqrt{57}\right), \frac{H_{12}^{\prime}}{16} \left(1-3 \sqrt{57}\right), -\frac{H_{12}^{\prime}}{8}$, and the corresponding orthonormalized zeroth-order wave functions:

\begin{eqnarray}
\Psi_{1,\text{---}}^{(0)}&=&\frac{3\sqrt{57}-1}{\sqrt{1026-6\sqrt{57}}}|\phi_1\rangle+\frac{16}{\sqrt{1026-6\sqrt{57}}}|\phi_2\rangle
+\frac{16}{\sqrt{1026-6\sqrt{57}}}|\phi_3\rangle  \,,  \label{line1}\\
\Psi_{2,\text{---}}^{(0)}&=&\frac{-(3\sqrt{57}+1)}{\sqrt{1026+6\sqrt{57}}}|\phi_1\rangle+\frac{16}{\sqrt{1026+6\sqrt{57}}}|\phi_2\rangle
+\frac{16}{\sqrt{1026+6\sqrt{57}}}|\phi_3\rangle  \,,  \label{line2} \\
\Psi_{3,\text{---}}^{(0)}&=&-\frac{1}{\sqrt{2}}|\phi_2\rangle + \frac{1}{\sqrt{2}}|\phi_3\rangle \,,  \label{line3}
\end{eqnarray}
where the symbol $\text{--}$ denotes the geometrical configuration of a straight line. With these zero-order wave functions Eqs. (\ref{line1})-(\ref{line3}), for the isotopic He${(1\,^{1}S)}$-He${(1\,^{1}S)}$-He${(2\,^{1}P)}$, He${(2\,^{1}S)}$-He${(2\,^{1}S)}$-He${(2\,^{1}P)}$, He${(2\,^{3}S)}$-He${(2\,^{3}S)}$-He${(2\,^{3}P)}$ systems, the corresponding long-range interaction coefficients are listed in Table~\ref{TabXII}-\ref{TabXVI}.

\begin{figure}
\begin{center}
\includegraphics[width=14cm,height=8cm]{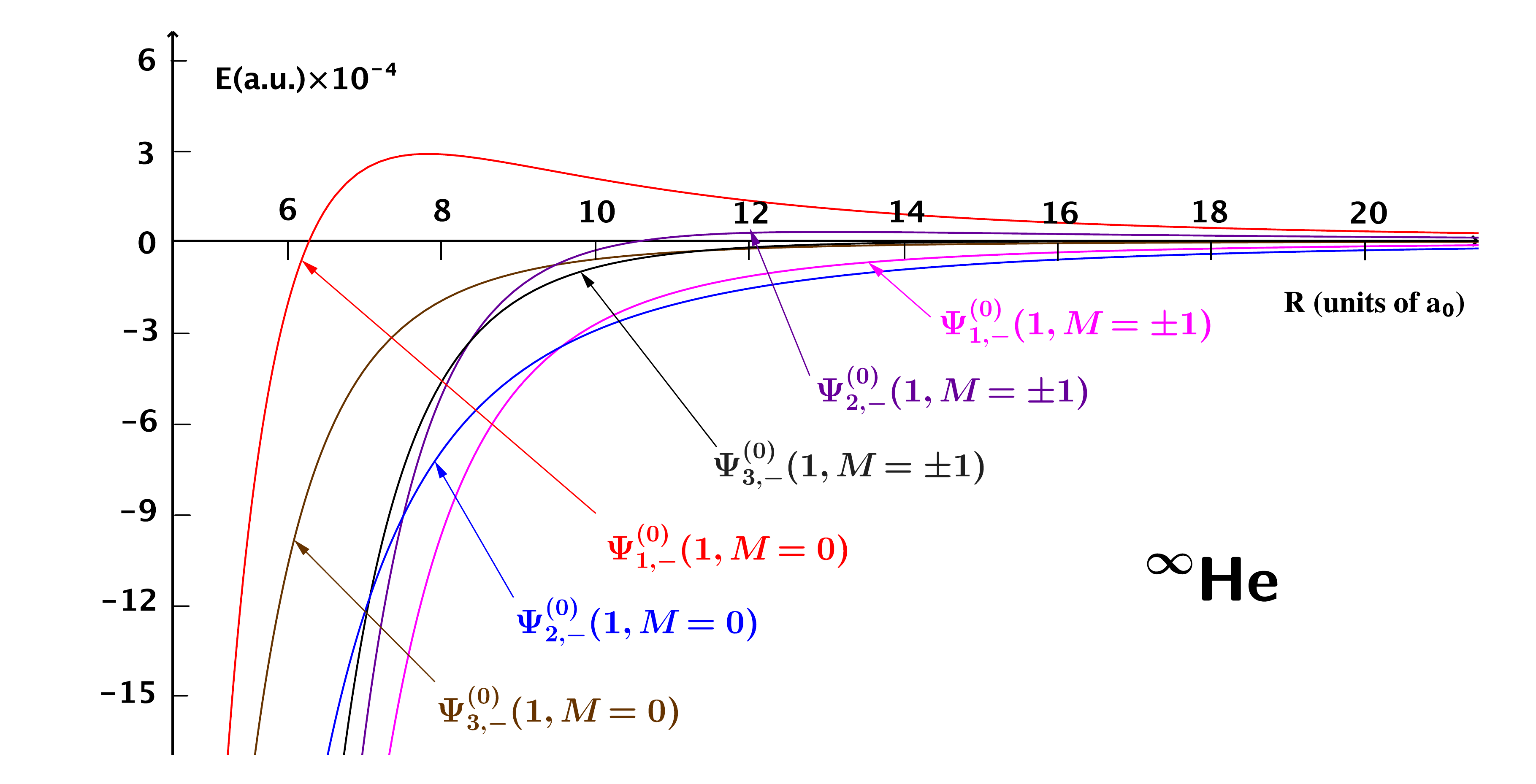}
\end{center}
\caption {\label{f8} Long-range potentials for the He($1\,^1S$)-He($1\,^1S$)-He($2\,^{1}P$) system for three different types of the zeroth-order wave functions, where the three atoms form a straight line, in atomic units.
For each curve labeled by a wave function, the plotted curve is the sum of $\Delta E^{(1)}$ and $\Delta E^{(2)}$.
}
\end{figure}

\begin{figure}
\begin{center}
\includegraphics[width=14cm,height=8cm]{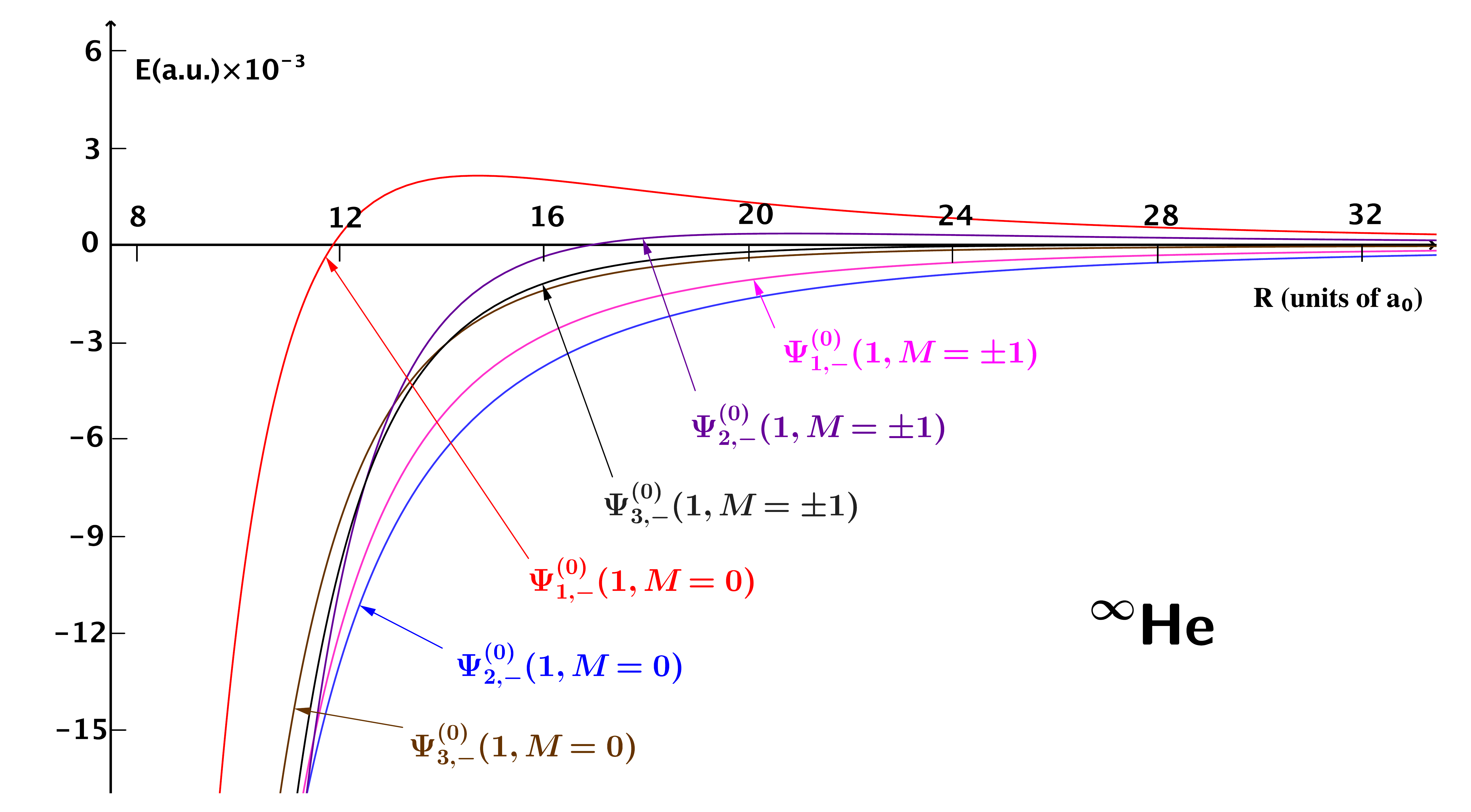}
\end{center}
\caption {\label{f9} Long-range potentials for the He($2\,^1S$)-He($2\,^1S$)-He($2\,^{1}P$) system for three different types of the zeroth-order wave functions, where the three atoms form a straight line, in atomic units.
For each curve labeled by a wave function, the plotted curve is the sum of $\Delta E^{(1)}$ and $\Delta E^{(2)}$.
}
\end{figure}

\begin{figure}
\begin{center}
\includegraphics[width=14cm,height=8cm]{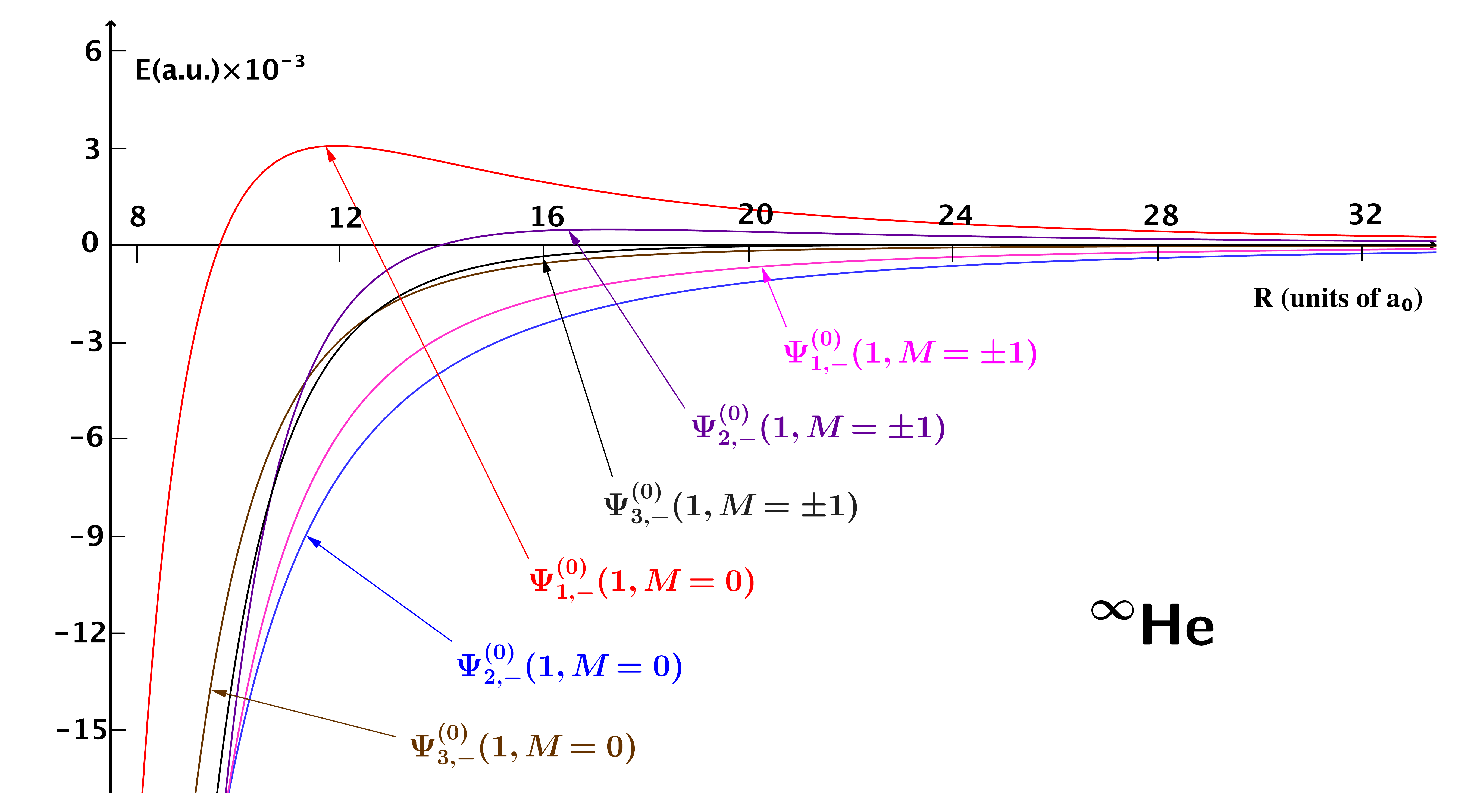}
\end{center}
\caption {\label{f10} Long-range potentials for the He($2\,^3S$)-He($2\,^3S$)-He($2\,^{3}P$) system for three different types of the zeroth-order wave functions, where the three atoms form a straight line, in atomic units.
For each curve labeled by a wave function, the plotted curve is the sum of $\Delta E^{(1)}$ and $\Delta E^{(2)}$.
}
\end{figure}

Since the zeroth-order wave function coefficients have $b=c$ in Eqs.~(\ref{line1}) and (\ref{line2}), and $a=0$ and $b=-c$ in Eq.~(\ref{line3}) the dispersion coefficients have similar characteristics as the case of the isosceles right triangle.
The only differences are the values of three interior angles: $\beta=\gamma=0$ and $\alpha=2\pi$, which leads to the relatively larger nonzero dispersion
coefficients  $C_8^{(31,12)}(1,M=0)$ and $C_8^{(31,12)}(1,M=\pm1)$.
The corresponding curves of  potential energy $(E)$ of the He${(1\,^{1}S)}$-He${(1\,^{1}S)}$-He${(2\,^{1}P)}$, He${(2\,^{1}S)}$-He${(2\,^{1}S)}$-He${(2\,^{1}P)}$ and He${(2\,^{3}S)}$-He${(2\,^{3}S)}$-He${(2\,^{3}P)}$ systems resulting from  $\Delta E^{(1)}$ and $\Delta E^{(2)}$  are shown in Fig.~\ref{f8} - Fig.~\ref{f10}, respectively.

%%%%%%%%%%%%%%%%%%%%%%%%%%%%%%%%%%%%%%%%%%%%%%%%%%%%%%%%%%%%%%%%%%%%%%%%%%%%%%%%%%%%%%%%%%%%%%%%%%%%%%%%%%%%%%%%%%%%%%%%%%%%%%%%%%%%%%%%%%

\section{Conclusion}\label{Con}

The long-range additive dipolar and additive dispersion interactions and nonadditive dispersion interactions $C_3$, $C_6$, $C_8$ for the He($n_0\,^{\lambda}S$)-He($n_0\,^{\lambda}S$)-He($n_0^{\prime}\,^{\lambda}L$) systems
He($1\,^{1}S$)-He($1\,^{1}S$)-He($2\,^{1}S$),
He${(1\,^{1}S)}$-He${(1\,^{1}S)}$-He${(2\,^{1}P)}$,
He${(2\,^{1}S)}$-He${(2\,^{1}S)}$-He${(2\,^{1}P)}$, and
He${(2\,^{3}S)}$-He${(2\,^{3}S)}$-He${(2\,^{3}P)}$
were obtained by perturbation theory.
We considered configurations of three-like nuclei for the hypothetical infinite mass He nucleus
and for the ${}^3\textrm{He}$ and the ${}^4\textrm{He}$ nucleus.

The coefficients are dependent on the geometrical configuration of the atoms.
We note that the geometry dependent nonadditive dispersion interactions for the degenerate He($n_0\,^{\lambda}S$)-He($n_0\,^{\lambda}S$)-He($n_0^{\prime}\,^{\lambda}P$) system start to appear at the \textit{second order} in the perturbative
treatment, in contrast to
the case of three $S$ atoms where nonadditive dispersion
interactions appear at \textit{third order}. The formulas obtained apply to all geometrical configurations and
we demonstrated the methodology for three basic types of geometrical configurations
(nuclei forming an equilateral triangle or an isosceles triangle, or nuclei equally-spaced and collinear)
by calculating coefficients to high precision
using variational wave functions in Hylleraas coordinates.
Our numerical results  may be useful in constructing accurate three-body potential curves
and for applications to scattering and recombination studies.
%----------

The present numerical approach differs from widely-used ``single-electron'' approaches
(e.g. ``model potentials'')
used for the calculation of trimer interactions
involving three highly-excited alkali-metal atoms
each in a Rydberg state, see for example Ref.~\cite{CanFor12}, in that
we include fully the correlation of the electrons in the He atoms.
On the other hand, the electronic structure of metastable triplet He atoms
is similar to alkali-metal atoms in the sense of having a single active electron,
so a comparison with the theory of Ref.~\cite{CanFor12} might be possible.

\begin{acknowledgments}
This work was supported by NSERC of Canada, the CAS/SAFEA International Partnership Program for Creative Research Teams,
and by the National Basic Research Program of China under Grant No. 2012CB821305 and by NNSF of China under Grant Nos. 11474319, 11104323.
JFB was supported in part by the U. S. NSF through a grant for the Institute of Theoretical Atomic, Molecular, and Optical Physics at
Harvard University and Smithsonian Astrophysical Observatory.
\end{acknowledgments}

%\bibliography{positron}

\newpage

\clearpage

%============================================================================================
%%%%%%%%%%%%%%%%%%%%%%%%%%%%%%%%%%%%%%%%%%%%%%%%%%%%

\section{appendix}\label{App}

In the present work, we take the mutual electrostatic interactions $V_{IJ}$ between pairs of atoms for the He($n_0\,^{\lambda}S$)-He($n_0\,^{\lambda}S$)-He($n_0^{\prime}\,^{\lambda}S$) system as a perturbation. According to degenerate perturbation theory, the zero-order wave function of the unperturbed system can be written as,
\begin{equation}\label{e7}
\ket{\Psi^{(0)}}=a\ket{\phi_1}+b\ket{\phi_2}+c\ket{\phi_3} \,.
\end{equation}
where $\phi_1$, $\phi_2$, $\phi_3$ are three orthogonal eigenvectors corresponding to the same energy eigenvalue $E_{n_0n_0n_0'}^{(0)}=2E_{n_0S}^{(0)}+E_{n_0'S}^{(0)}$. With the zeroth-order wavefunction of Eq. (\ref{e7}), we can derive general formulas of the dispersion coefficients for the He($1\,^{1}S$)-He($1\,^{1}S$)-He($2\,^{1}S$) system. The values of coefficients $a$, $b$ and $c$ can be obtained by the degenerate perturbation theory.

\subsection{The zeroth-order wave function}\label{AppA1}

According to degenerate perturbation theory, the zeroth-order energy correction is obtained by the perturbation matrix with respect to $\{\phi_1, \phi_2, \phi_3\}$,

\begin{equation}\label{e10}
{H}^{\prime}=
\left(
  \begin{array}{ccc}
    \Delta_{11} & \Delta_{12} & \Delta_{13}\\
    \Delta_{12}^* & \Delta_{22} & \Delta_{23}\\
    \Delta_{13}^* & \Delta_{23}^* & \Delta_{33}\\
  \end{array}
\right)\,,
\end{equation}

where

\begin{eqnarray}\label{Delta_11}
\Delta_{11}&=&-\sum_{n_sn_tn_u}\sum_{L_sL_tL_u}\sum_{M_sM_tM_u}
\frac{|\langle\phi_1|V_{123}|\chi_{n_s}(L_sM_s;\boldsymbol{\sigma})\chi_{n_t}(L_tM_t;\boldsymbol{\rho})
\chi_{n_u}(L_uM_u;\boldsymbol{\varsigma})\rangle |^{2}}{E_{n_sL_s;n_tL_t;n_uL_u}-E_{n_0S;n_0S;n_0'S}^{(0)}} \nonumber \\
&=&
-\sum_{n_sn_t}\sum_{L_sL_t}\sum_{M_sM_t}\frac{16\pi^2}{R_{12}^{2L_{s}+2L_{t}+2}}
\frac{[P_{L_{s}+L_{t}}^{M_{s}+M_{t}}(0)(L_{s}+L_{t}-M_{s}-M_{t})!]^2(L_{s},L_{t})^{-2}}{(L_{s}+M_{s})!(L_{s}-M_{s})!(L_{t}+M_{t})!(L_{t}-M_{t})!
} \nonumber \\
&& \times
\frac{|\langle\varphi_{n'_0}(0;\boldsymbol{\sigma})\|T_{L_{s}}(\boldsymbol{\sigma})\|\chi_{n_s}(L_s;\boldsymbol{\sigma})\rangle|^{2}
|\langle\varphi_{n_0}(0;\boldsymbol{\rho})\|T_{L_t}(\boldsymbol{\rho})\|\chi_{n_t}(L_t;\boldsymbol{\rho})\rangle|^{2}}
{E_{n_sL_s}+E_{n_tL_t}-E_{n_0S}^{(0)}-E_{n'_0S}^{(0)}} \nonumber \\
&-&\sum_{n_tn_u}\sum_{L_tL_u}\sum_{M_tM_u}\frac{16\pi^2}{R_{23}^{2L_{t}+2L_{u}+2}}
\frac{[P_{L_{t}+L_{u}}^{M_{t}+M_{u}}(0)(L_{t}+L_{u}-M_{t}-M_{u})!]^2(L_{t},L_{u})^{-2}}{(L_{t}+M_{t})!(L_{t}-M_{t})!(L_{u}+M_{u})!
(L_{u}-M_{u})!} \nonumber \\
&& \times
\frac{|\langle\varphi_{n_0}(0;\boldsymbol{\rho})\|T_{L_{t}}(\boldsymbol{\rho})\|\chi_{n_t}(L_t;\boldsymbol{\rho})\rangle|^{2}
|\langle\varphi_{n_0}(0;\boldsymbol{\varsigma})\|T_{L_u}(\boldsymbol{\varsigma})\|\chi_{n_u}(L_u;\boldsymbol{\varsigma})\rangle|^{2}}
{E_{n_tL_t}+E_{n_uL_u}-2E_{n_0S}^{(0)}} \nonumber \\
&-&\sum_{n_sn_u}\sum_{L_sL_u}\sum_{M_sM_u} \frac{16\pi^2}{R_{31}^{2L_{s}+2L_{u}+2}} \frac{[P_{L_{u}+L_{s}}^{M_{u}+M_{s}}(0)(L_{u}+L_{s}-M_{u}-M_{s})!]^2(L_{u},L_{s})^{-2}}{(L_{u}+M_{u})!(L_{u}-M_{u})!
(L_{s}+M_{s})!(L_{s}-M_{s})!} \nonumber \\
&& \times
\frac{|\langle\varphi_{n'_0}(0;\boldsymbol{\sigma})\|T_{L_{s}}(\boldsymbol{\sigma})\|\chi_{n_s}(L_s;\boldsymbol{\sigma})\rangle|^{2}
|\langle\varphi_{n_0}(0;\boldsymbol{\varsigma})\|T_{L_{u}}(\boldsymbol{\varsigma})\|\chi_{n_u}(L_u;\boldsymbol{\varsigma})\rangle|^{2}}
{E_{n_sL_s}+E_{n_uL_u}-E_{n_0S}-E_{n'_0S}}  \nonumber \,, \\
\end{eqnarray}

\begin{eqnarray}\label{Delta_22}
\Delta_{22}&=&-\sum_{n_sn_tn_u}\sum_{L_sL_tL_u}\sum_{M_sM_tM_u}
\frac{|\langle\phi_2|V_{123}|\chi_{n_s}(L_sM_s;\boldsymbol{\sigma})\chi_{n_t}(L_tM_t;\boldsymbol{\rho})
\chi_{n_u}(L_uM_u;\boldsymbol{\varsigma})\rangle |^{2}}{E_{n_sL_s;n_tL_t;n_uL_u}-E_{n_0S;n_0S;n_0'S}^{(0)}} \nonumber \\
&=&
-\sum_{n_sn_t}\sum_{L_sL_t}\sum_{M_sM_t}\frac{16\pi^2}{R_{12}^{2L_{s}+2L_{t}+2}}
\frac{[P_{L_{s}+L_{t}}^{M_{s}+M_{t}}(0)(L_{s}+L_{t}-M_{s}-M_{t})!]^2(L_{s},L_{t})^{-2}}{(L_{s}+M_{s})!(L_{s}-M_{s})!(L_{t}+M_{t})!(L_{t}-M_{t})!
} \nonumber \\
&& \times
\frac{|\langle\varphi_{n_0}(0;\boldsymbol{\sigma})\|T_{L_{s}}(\boldsymbol{\sigma})\|\chi_{n_s}(L_s;\boldsymbol{\sigma})\rangle|^{2}
|\langle\varphi_{n'_0}(0;\boldsymbol{\rho})\|T_{L_t}(\boldsymbol{\rho})\|\chi_{n_t}(L_t;\boldsymbol{\rho})\rangle|^{2}}
{E_{n_sL_s}+E_{n_tL_t}-E_{n_0S}^{(0)}-E_{n'_0S}^{(0)}} \nonumber \\
&-&\sum_{n_tn_u}\sum_{L_tL_u}\sum_{M_tM_u}\frac{16\pi^2}{R_{23}^{2L_{t}+2L_{u}+2}}
\frac{[P_{L_{t}+L_{u}}^{M_{t}+M_{u}}(0)(L_{t}+L_{u}-M_{t}-M_{u})!]^2(L_{t},L_{u})^{-2}}{(L_{t}+M_{t})!(L_{t}-M_{t})!(L_{u}+M_{u})!
(L_{u}-M_{u})!} \nonumber \\
&&  \times
\frac{|\langle\varphi_{n'_0}(0;\boldsymbol{\rho})\|T_{L_{t}}(\boldsymbol{\rho})\|\chi_{n_t}(L_t;\boldsymbol{\rho})\rangle|^{2}
|\langle\varphi_{n_0}(0;\boldsymbol{\varsigma})\|T_{L_u}(\boldsymbol{\varsigma})\|\chi_{n_u}(L_u;\boldsymbol{\varsigma})\rangle|^{2}}
{E_{n_tL_t}+E_{n_uL_u}-E_{n_0S}^{(0)}-E_{n'_0S}^{(0)}} \nonumber \\
&-&\sum_{n_sn_u}\sum_{L_sL_u}\sum_{M_sM_u} \frac{16\pi^2}{R_{31}^{2L_{s}+2L_{u}+2}} \frac{[P_{L_{u}+L_{s}}^{M_{u}+M_{s}}(0)(L_{u}+L_{s}-M_{u}-M_{s})!]^2(L_{u},L_{s})^{-2}}{(L_{u}+M_{u})!(L_{u}-M_{u})!
(L_{s}+M_{s})!(L_{s}-M_{s})!} \nonumber \\
&&  \times
\frac{|\langle\varphi_{n_0}(0;\boldsymbol{\sigma})\|T_{L_{s}}(\boldsymbol{\sigma})\|\chi_{n_s}(L_s;\boldsymbol{\sigma})\rangle|^{2}
|\langle\varphi_{n_0}(0;\boldsymbol{\varsigma})\|T_{L_{u}}(\boldsymbol{\varsigma})\|\chi_{n_u}(L_u;\boldsymbol{\varsigma})\rangle|^{2}}
{E_{n_sL_s}+E_{n_uL_u}-2E_{n_0S}}  \nonumber \,,\\
\end{eqnarray}

\begin{eqnarray}\label{Delta_33}
\Delta_{33}&=&-\sum_{n_sn_tn_u}\sum_{L_sL_tL_u}\sum_{M_sM_tM_u}
\frac{|\langle\phi_3|V_{123}|\chi_{n_s}(L_sM_s;\boldsymbol{\sigma})\chi_{n_t}(L_tM_t;\boldsymbol{\rho})
\chi_{n_u}(L_uM_u;\boldsymbol{\varsigma})\rangle |^{2}}{E_{n_sL_s;n_tL_t;n_uL_u}-E_{n_0S;n_0S;n_0'S}^{(0)}} \nonumber \\
&=&
-\sum_{n_sn_t}\sum_{L_sL_t}\sum_{M_sM_t}\frac{16\pi^2}{R_{12}^{2L_{s}+2L_{t}+2}}
\frac{[P_{L_{s}+L_{t}}^{M_{s}+M_{t}}(0)(L_{s}+L_{t}-M_{s}-M_{t})!]^2(L_{s},L_{t})^{-2}}{(L_{s}+M_{s})!(L_{s}-M_{s})!(L_{t}+M_{t})!(L_{t}-M_{t})!
} \nonumber \\
&& \times
\frac{|\langle\varphi_{n_0}(0;\boldsymbol{\sigma})\|T_{L_{s}}(\boldsymbol{\sigma})\|\chi_{n_s}(L_s;\boldsymbol{\sigma})\rangle|^{2}
|\langle\varphi_{n_0}(0;\boldsymbol{\rho})\|T_{L_t}(\boldsymbol{\rho})\|\chi_{n_t}(L_t;\boldsymbol{\rho})\rangle|^{2}}
{E_{n_sL_s}+E_{n_tL_t}-2E_{n_0S}^{(0)}} \nonumber \\
&-&\sum_{n_tn_u}\sum_{L_tL_u}\sum_{M_tM_u}\frac{16\pi^2}{R_{23}^{2L_{t}+2L_{u}+2}}
\frac{[P_{L_{t}+L_{u}}^{M_{t}+M_{u}}(0)(L_{t}+L_{u}-M_{t}-M_{u})!]^2(L_{t},L_{u})^{-2}}{(L_{t}+M_{t})!(L_{t}-M_{t})!(L_{u}+M_{u})!
(L_{u}-M_{u})!} \nonumber \\
&& \times
\frac{|\langle\varphi_{n_0}(0;\boldsymbol{\rho})\|T_{L_{t}}(\boldsymbol{\rho})\|\chi_{n_t}(L_t;\boldsymbol{\rho})\rangle|^{2}
|\langle\varphi_{n'_0}(0;\boldsymbol{\varsigma})\|T_{L_u}(\boldsymbol{\varsigma})\|\chi_{n_u}(L_u;\boldsymbol{\varsigma})\rangle|^{2}}
{E_{n_tL_t}+E_{n_uL_u}-E_{n_0S}^{(0)}-E_{n'_0S}^{(0)}} \nonumber \\
&-&\sum_{n_sn_u}\sum_{L_sL_u}\sum_{M_sM_u} \frac{16\pi^2}{R_{31}^{2L_{s}+2L_{u}+2}} \frac{[P_{L_{u}+L_{s}}^{M_{u}+M_{s}}(0)(L_{u}+L_{s}-M_{u}-M_{s})!]^2(L_{u},L_{s})^{-2}}{(L_{u}+M_{u})!(L_{u}-M_{u})!
(L_{s}+M_{s})!(L_{s}-M_{s})!} \nonumber \\
&& \times
\frac{|\langle\varphi_{n_0}(0;\boldsymbol{\sigma})\|T_{L_{s}}(\boldsymbol{\sigma})\|\chi_{n_s}(L_s;\boldsymbol{\sigma})\rangle|^{2}
|\langle\varphi_{n_0}(0;\boldsymbol{\varsigma})\|T_{L_{u}}(\boldsymbol{\varsigma})\|\chi_{n_u}(L_u;\boldsymbol{\varsigma})\rangle|^{2}}
{E_{n_sL_s}+E_{n_uL_u}-E_{n_0S}-E_{n'_0S}}  \nonumber \,,\\
\end{eqnarray}

\begin{eqnarray}\label{Delta_12}
\Delta_{12}&=&-\sum_{n_sn_tn_u}\sum_{L_sL_tL_u}\sum_{M_sM_tM_u}\langle\phi_1|V_{123}|\chi_{n_s}(L_sM_s;\boldsymbol{\sigma})\chi_{n_t}(L_tM_t;\boldsymbol{\rho})
\chi_{n_u}(L_uM_u;\boldsymbol{\varsigma})\rangle^*
\nonumber \\ && \times
\frac{\langle\phi_2|V_{123}|\chi_{n_s}(L_sM_s;\boldsymbol{\sigma})\chi_{n_t}(L_tM_t;\boldsymbol{\rho})
\chi_{n_u}(L_uM_u;\boldsymbol{\varsigma})\rangle}{E_{n_sL_s;n_tL_t;n_uL_u}-E_{n_0S;n_0S;n_0'S}^{(0)}} \nonumber \\&=&
-\sum_{n_sn_t}\sum_{L_sL_t }\sum_{M_sM_t}
\frac{16\pi^2}{R_{12}^{2L_{s}+2L_{t}+2}}
\frac{[P_{L_{s}+L_{t}}^{M_{s}+M_{t}}(0)(L_{s}+L_{t}-M_{s}-M_{t})!]^2(L_{s},L_{t})^{-2}}{(L_{s}+M_{s})!(L_{s}-M_{s})!(L_{t}+M_{t})!(L_{t}-M_{t})!
} \nonumber \\
&& \times
\langle\varphi_{n_0'}(0;\boldsymbol{\sigma})\|T_{L_s}(\boldsymbol{\sigma})\|\chi_{n_s}(L_s;\boldsymbol{\sigma})\rangle^*
\langle\varphi_{n_0}(0;\boldsymbol{\rho})\|T_{L_t}(\boldsymbol{\rho})\|\chi_{n_{t}}(L_t;\boldsymbol{\rho})\rangle^*
\nonumber \\
&& \times
\frac{\langle\varphi_{n_0}(0;\boldsymbol{\sigma})\|T_{L_s}(\boldsymbol{\sigma})\|\chi_{n_s}(L_s;\boldsymbol{\sigma})\rangle
\langle\varphi_{n_0'}(0;\boldsymbol{\rho})\|T_{L_t}(\boldsymbol{\rho})\|\chi_{n_t}(L_t;\boldsymbol{\rho})\rangle}
{E_{n_sL_s}+E_{n_tL_t}-E_{n_0S}^{(0)}-E_{n_0'S}^{(0)}}
 \nonumber \,,\\
\end{eqnarray}

\begin{eqnarray}\label{Delta_13}
\Delta_{13}&=&-\sum_{n_sn_tn_u}\sum_{L_sL_tL_u}\sum_{M_sM_tM_u}\langle\phi_1|V_{123}|\chi_{n_s}(L_sM_s;\boldsymbol{\sigma})\chi_{n_t}(L_tM_t;\boldsymbol{\rho})
\chi_{n_u}(L_uM_u;\boldsymbol{\varsigma})\rangle^*
\nonumber \\ && \times
\frac{\langle\phi_3|V_{123}|\chi_{n_s}(L_sM_s;\boldsymbol{\sigma})\chi_{n_t}(L_tM_t;\boldsymbol{\rho})
\chi_{n_u}(L_uM_u;\boldsymbol{\varsigma})\rangle}{E_{n_sL_s;n_tL_t;n_uL_u}-E_{n_0S;n_0S;n_0'S}^{(0)}} \nonumber \\ &=&
-\sum_{n_sn_u}\sum_{L_sL_u}\sum_{M_sM_u} \frac{16\pi^2}{R_{31}^{2L_{s}+2L_{u}+2}} \frac{[P_{L_{u}+L_{s}}^{M_{u}+M_{s}}(0)(L_{u}+L_{s}-M_{u}-M_{s})!]^2(L_{u},L_{s})^{-2}}{(L_{u}+M_{u})!(L_{u}-M_{u})!
(L_{s}+M_{s})!(L_{s}-M_{s})!} \nonumber \\
&& \times \langle\varphi_{n_0'}(0;\boldsymbol{\sigma})\|T_{L_s}(\boldsymbol{\sigma})\|\chi_{n_s}(L_s;\boldsymbol{\sigma})\rangle^*
\langle\varphi_{n_0}(0;\boldsymbol{\varsigma})\|T_{L_{u}}(\boldsymbol{\varsigma})\|\chi_{n_u}(L_u;\boldsymbol{\varsigma})\rangle^*
\nonumber \\
&& \times
\frac{\langle\varphi_{n_0}(0;\boldsymbol{\sigma})\|T_{L_{s}}(\boldsymbol{\sigma})\|\chi_{n_s}(L_s;\boldsymbol{\sigma})\rangle
\langle\varphi_{n_0'}(0;\boldsymbol{\varsigma})\|T_{L_u}(\boldsymbol{\varsigma})\|\chi_{n_u}(L_u;\boldsymbol{\varsigma})\rangle}
{E_{n_sL_s}+E_{n_uL_u}-E_{n_0S}-E_{n_0'S}} \nonumber \,,\\
\end{eqnarray}

\begin{eqnarray}\label{Delta_23}
\Delta_{23}&=&-\sum_{n_sn_tn_u}\sum_{L_sL_tL_u}\sum_{M_sM_tM_u}\langle\phi_2|V_{123}|\chi_{n_s}(L_sM_s;\boldsymbol{\sigma})\chi_{n_t}(L_tM_t;\boldsymbol{\rho})
\chi_{n_u}(L_uM_u;\boldsymbol{\varsigma})\rangle^*
\nonumber \\ && \times
\frac{\langle\phi_3|V_{123}|\chi_{n_s}(L_sM_s;\boldsymbol{\sigma})\chi_{n_t}(L_tM_t;\boldsymbol{\rho})
\chi_{n_u}(L_uM_u;\boldsymbol{\varsigma})\rangle}{E_{n_sL_s;n_tL_t;n_uL_u}-E_{n_0S;n_0S;n_0'S}^{(0)}} \nonumber \\
&=&
-\sum_{n_tn_u}\sum_{L_tL_u}\sum_{M_tM_u}\frac{16\pi^2}{R_{23}^{2L_{t}+2L_{u}+2}}
\frac{[P_{L_{t}+L_{u}}^{M_{t}+M_{u}}(0)(L_{t}+L_{u}-M_{t}-M_{u})!]^2(L_{t},L_{u})^{-2}}{(L_{t}+M_{t})!(L_{t}-M_{t})!(L_{u}+M_{u})!
(L_{u}-M_{u})!} \nonumber \\
&&\times
\langle\varphi_{n_0'}(0;\boldsymbol{\rho})\|T_{L_t}(\boldsymbol{\rho})\|\chi_{n_t}(L_t;\boldsymbol{\rho})\rangle^*
\langle\varphi_{n_0}(0;\boldsymbol{\varsigma})\|T_{L_{u}}(\boldsymbol{\varsigma})\|\chi_{n_u}(L_u;\boldsymbol{\varsigma})\rangle^*
\nonumber \\
&&\times
\frac{\langle\varphi_{n_0}(0;\boldsymbol{\rho})\|T_{L_{t}}(\boldsymbol{\rho})\|\chi_{n_t}(L_t;\boldsymbol{\rho})\rangle
\langle\varphi_{n_0'}(0;\boldsymbol{\varsigma})\|T_{L_u}(\boldsymbol{\varsigma})\|\chi_{n_u}(L_u;\boldsymbol{\varsigma})\rangle}
{E_{n_tL_t}+E_{n_uL_u}-E_{n_0S}^{(0)}-E_{n_0'S}^{(0)}} \nonumber \,,\\
\end{eqnarray}

Then solve this eigenvalue problem to get the eigenvalues and corresponding zeroth-order wave functions.

\subsection{The second-order energy correction}\label{AppA2}

The second-order energy correction is given by
\begin{eqnarray}\label{E2}
\Delta E^{(2)}&=&-\sum_{n_sn_tn_u}\sum_{L_sL_tL_u}\sum_{M_sM_tM_u}
\frac{|\langle\Psi^{(0)}|V_{123}|\chi_{n_s}(L_sM_s;\boldsymbol{\sigma})\chi_{n_t}(L_tM_t;\boldsymbol{\rho})
\chi_{n_u}(L_uM_u;\boldsymbol{\varsigma})\rangle |^{2}}{E_{n_sL_s;n_tL_t;n_uL_u}-E_{n_0S;n_0S;n_0'S}^{(0)}} \nonumber \\
&=&V_{12}^{(2)}+V_{23}^{(2)}+V_{31}^{(2)}\,,
\end{eqnarray}
where $\chi_{n_s}(L_sM_s;\boldsymbol{\sigma})\chi_{n_t}(L_tM_t;\boldsymbol{\rho})
\chi_{n_u}(L_uM_u;\boldsymbol{\varsigma})$ is an intermediate state of the system with the energy eigenvalue $E_{n_sL_s;n_tL_t;n_uL_u}=E_{n_sL_s}+E_{n_tL_t}+E_{n_uL_u}$. It is noted that the above summations should exclude terms
with $E_{n_sL_s;n_tL_t;n_uL_u}=E_{n_0S;n_0S;n_0'L}^{(0)}$. Then the three additive terms in the second-order energy correction, denoted by $V_{12}^{(2)}$, $V_{23}^{(2)}$, and $V_{31}^{(2)}$, become respectively

\begin{eqnarray}\label{V12-12b}
V_{12}^{(2)}&=&-|a|^2\sum_{n_sn_t}\sum_{L_sL_t}\sum_{M_sM_t}\frac{16\pi^2}{R_{12}^{2L_{s}+2L_{t}+2}}
\frac{[P_{L_{s}+L_{t}}^{M_{s}+M_{t}}(0)(L_{s}+L_{t}-M_{s}-M_{t})!]^2(L_{s},L_{t})^{-2}}{(L_{s}+M_{s})!(L_{s}-M_{s})!(L_{t}+M_{t})!(L_{t}-M_{t})!
} \nonumber \\
&& \times
\frac{|\langle\varphi_{n'_0}(0;\boldsymbol{\sigma})\|T_{L_{s}}(\boldsymbol{\sigma})\|\chi_{n_s}(L_s;\boldsymbol{\sigma})\rangle|^{2}
|\langle\varphi_{n_0}(0;\boldsymbol{\rho})\|T_{L_t}(\boldsymbol{\rho})\|\chi_{n_t}(L_t;\boldsymbol{\rho})\rangle|^{2}}
{E_{n_sL_s}+E_{n_tL_t}-E_{n_0S}^{(0)}-E_{n'_0S}^{(0)}} \nonumber \\
&-&|b|^2\sum_{n_sn_t}\sum_{L_sL_t}\sum_{M_sM_t}\frac{16\pi^2}{R_{12}^{2L_{s}+2L_{t}+2}}
\frac{[P_{L_{s}+L_{t}}^{M_{s}+M_{t}}(0)(L_{s}+L_{t}-M_{s}-M_{t})!]^2(L_{s},L_{t})^{-2}}{(L_{s}+M_{s})!(L_{s}-M_{s})!(L_{t}+M_{t})!(L_{t}-M_{t})!
} \nonumber \\
&&\times
\frac{|\langle\varphi_{n_0}(0;\boldsymbol{\sigma})\|T_{L_{s}}(\boldsymbol{\sigma})\|\chi_{n_s}(L_s;\boldsymbol{\sigma})\rangle|^{2}
|\langle\varphi_{n'_0}(0;\boldsymbol{\rho})\|T_{L_t}(\boldsymbol{\rho})\|\chi_{n_t}(L_t;\boldsymbol{\rho})\rangle|^{2}}
{E_{n_sL_s}+E_{n_tL_t}-E_{n_0S}^{(0)}-E_{n'_0S}^{(0)}} \nonumber \\
&-&|c|^2\sum_{n_sn_t}\sum_{L_sL_t}\sum_{M_sM_t}\frac{16\pi^2}{R_{12}^{2L_{s}+2L_{t}+2}}
\frac{[P_{L_{s}+L_{t}}^{M_{s}+M_{t}}(0)(L_{s}+L_{t}-M_{s}-M_{t})!]^2(L_{s},L_{t})^{-2}}{(L_{s}+M_{s})!(L_{s}-M_{s})!(L_{t}+M_{t})!(L_{t}-M_{t})!
} \nonumber \\
&&\times
\frac{|\langle\varphi_{n_0}(0;\boldsymbol{\sigma})\|T_{L_{s}}(\boldsymbol{\sigma})\|\chi_{n_s}(L_s;\boldsymbol{\sigma})\rangle|^{2}
|\langle\varphi_{n_0}(0;\boldsymbol{\rho})\|T_{L_t}(\boldsymbol{\rho})\|\chi_{n_t}(L_t;\boldsymbol{\rho})\rangle|^{2}}
{E_{n_sL_s}+E_{n_tL_t}-2E_{n_0S}^{(0)}} \nonumber \\
&-&a^{*}b\sum_{n_sn_t}\sum_{L_sL_t }\sum_{M_sM_t}
\frac{16\pi^2}{R_{12}^{2L_{s}+2L_{t}+2}}
\frac{[P_{L_{s}+L_{t}}^{M_{s}+M_{t}}(0)(L_{s}+L_{t}-M_{s}-M_{t})!]^2(L_{s},L_{t})^{-2}}{(L_{s}+M_{s})!(L_{s}-M_{s})!(L_{t}+M_{t})!(L_{t}-M_{t})!
} \nonumber \\
&&\times
\langle\varphi_{n_0'}(0;\boldsymbol{\sigma})\|T_{L_s}(\boldsymbol{\sigma})\|\chi_{n_s}(L_s;\boldsymbol{\sigma})\rangle^*
\langle\varphi_{n_0}(0;\boldsymbol{\rho})\|T_{L_t}(\boldsymbol{\rho})\|\chi_{n_{t}}(L_t;\boldsymbol{\rho})\rangle^*
\nonumber \\
&&\times
\frac{\langle\varphi_{n_0}(0;\boldsymbol{\sigma})\|T_{L_s}(\boldsymbol{\sigma})\|\chi_{n_s}(L_s;\boldsymbol{\sigma})\rangle
\langle\varphi_{n_0'}(0;\boldsymbol{\rho})\|T_{L_t}(\boldsymbol{\rho})\|\chi_{n_t}(L_t;\boldsymbol{\rho})\rangle}
{E_{n_sL_s}+E_{n_tL_t}-E_{n_0S}^{(0)}-E_{n_0'S}^{(0)}}
 \nonumber \\
&-&b^{*}a\sum_{n_sn_t}\sum_{L_sL_t }\sum_{M_sM_t}
\frac{16\pi^2}{R_{12}^{2L_{s}+2L_{t}+2}}
\frac{[P_{L_{s}+L_{t}}^{M_{s}+M_{t}}(0)(L_{s}+L_{t}-M_{s}-M_{t})!]^2(L_{s},L_{t})^{-2}}{(L_{s}+M_{s})!(L_{s}-M_{s})!(L_{t}+M_{t})!(L_{t}-M_{t})!
} \nonumber \\
&&\times
\langle\varphi_{n_0}(0;\boldsymbol{\sigma})\|T_{L_{s}}(\boldsymbol{\sigma})\|\chi_{n_s}(L_s;\boldsymbol{\sigma})\rangle^*
\langle\varphi_{n_0'}(0;\boldsymbol{\rho})\|T_{L_t}(\boldsymbol{\rho})\|\chi_{n_t}(L_t;\boldsymbol{\rho})\rangle^*
\nonumber \\
&&\times
\frac{\langle\varphi_{n_0'}(0;\boldsymbol{\sigma})\|T_{L_s}(\boldsymbol{\sigma})\|\chi_{n_s}(L_s;\boldsymbol{\sigma})\rangle\langle\varphi_{n_0}(0;\boldsymbol{\rho})\|T_{L_t}(\boldsymbol{\rho})\|\chi_{n_t}(L_t;\boldsymbol{\rho})\rangle
}
{E_{n_sL_s}+E_{n_tL_t}-E_{n_0S}^{(0)}-E_{n_0'S}^{(0)}} \nonumber \\
&=&-\bigg\{|a|^2\sum_{n_sn_t}\sum_{L_sL_t} \frac{K_1(n_s,n_t,L_s,L_t)}{R_{12}^{2L_{s}+2L_{t}+2}}
+|b|^2\sum_{n_sn_t}\sum_{L_sL_t}
\frac{K_1(n_t,n_s,L_t,L_s)}
{R_{12}^{2L_{s}+2L_{t}+2}} \nonumber  \\
&+&|c|^2\sum_{n_sn_t}\sum_{L_sL_t}
\frac{K_2(n_s,n_t,L_s,L_t)}{R_{12}^{2L_{s}+2L_{t}+2}}
+ a^{*}b\sum_{n_sn_t}\sum_{L_sL_t}
\frac{K_3(n_s,n_t,L_s,L_t)}{R_{12}^{2L_{s}+2L_{t}+2}} \nonumber \\
&+& b^{*}a\sum_{n_sn_t}\sum_{L_sL_t}
\frac{ K_3^*(n_s,n_t,L_s,L_t)}{R_{12}^{2L_{s}+2L_{t}+2}}\bigg\} \,,
\end{eqnarray}

\begin{eqnarray}\label{V23-23b}
V_{23}^{(2)}&=&-|a|^2\sum_{n_tn_u}\sum_{L_tL_u}\sum_{M_tM_u}\frac{16\pi^2}{R_{23}^{2L_{t}+2L_{u}+2}}
\frac{[P_{L_{t}+L_{u}}^{M_{t}+M_{u}}(0)(L_{t}+L_{u}-M_{t}-M_{u})!]^2(L_{t},L_{u})^{-2}}{(L_{t}+M_{t})!(L_{t}-M_{t})!(L_{u}+M_{u})!
(L_{u}-M_{u})!} \nonumber \\
&& \times
\frac{|\langle\varphi_{n_0}(0;\boldsymbol{\rho})\|T_{L_{t}}(\boldsymbol{\rho})\|\chi_{n_t}(L_t;\boldsymbol{\rho})\rangle|^{2}
|\langle\varphi_{n_0}(0;\boldsymbol{\varsigma})\|T_{L_u}(\boldsymbol{\varsigma})\|\chi_{n_u}(L_u;\boldsymbol{\varsigma})\rangle|^{2}}
{E_{n_tL_t}+E_{n_uL_u}-2E_{n_0S}^{(0)}} \nonumber \\
&-&|b|^2\sum_{n_tn_u}\sum_{L_tL_u}\sum_{M_tM_u}\frac{16\pi^2}{R_{23}^{2L_{t}+2L_{u}+2}}
\frac{[P_{L_{t}+L_{u}}^{M_{t}+M_{u}}(0)(L_{t}+L_{u}-M_{t}-M_{u})!]^2(L_{t},L_{u})^{-2}}{(L_{t}+M_{t})!(L_{t}-M_{t})!(L_{u}+M_{u})!
(L_{u}-M_{u})!} \nonumber \\
&&\times
\frac{|\langle\varphi_{n'_0}(0;\boldsymbol{\rho})\|T_{L_{t}}(\boldsymbol{\rho})\|\chi_{n_t}(L_t;\boldsymbol{\rho})\rangle|^{2}
|\langle\varphi_{n_0}(0;\boldsymbol{\varsigma})\|T_{L_u}(\boldsymbol{\varsigma})\|\chi_{n_u}(L_u;\boldsymbol{\varsigma})\rangle|^{2}}
{E_{n_tL_t}+E_{n_uL_u}-E_{n_0S}^{(0)}-E_{n'_0S}^{(0)}} \nonumber \\
&-&|c|^2\sum_{n_tn_u}\sum_{L_tL_u}\sum_{M_tM_u}\frac{16\pi^2}{R_{23}^{2L_{t}+2L_{u}+2}}
\frac{[P_{L_{t}+L_{u}}^{M_{t}+M_{u}}(0)(L_{t}+L_{u}-M_{t}-M_{u})!]^2(L_{t},L_{u})^{-2}}{(L_{t}+M_{t})!(L_{t}-M_{t})!(L_{u}+M_{u})!
(L_{u}-M_{u})!} \nonumber \\
&&\times
\frac{|\langle\varphi_{n_0}(0;\boldsymbol{\rho})\|T_{L_{t}}(\boldsymbol{\rho})\|\chi_{n_t}(L_t;\boldsymbol{\rho})\rangle|^{2}
|\langle\varphi_{n'_0}(0;\boldsymbol{\varsigma})\|T_{L_u}(\boldsymbol{\varsigma})\|\chi_{n_u}(L_u;\boldsymbol{\varsigma})\rangle|^{2}}
{E_{n_tL_t}+E_{n_uL_u}-E_{n_0S}^{(0)}-E_{n'_0S}^{(0)}} \nonumber \\
&-&b^{*}c\sum_{n_tn_u}\sum_{L_tL_u}\sum_{M_tM_u}\frac{16\pi^2}{R_{23}^{2L_{t}+2L_{u}+2}}
\frac{[P_{L_{t}+L_{u}}^{M_{t}+M_{u}}(0)(L_{t}+L_{u}-M_{t}-M_{u})!]^2(L_{t},L_{u})^{-2}}{(L_{t}+M_{t})!(L_{t}-M_{t})!(L_{u}+M_{u})!
(L_{u}-M_{u})!} \nonumber \\
&&\times
\langle\varphi_{n_0'}(0;\boldsymbol{\rho})\|T_{L_t}(\boldsymbol{\rho})\|\chi_{n_t}(L_t;\boldsymbol{\rho})\rangle^*
\langle\varphi_{n_0}(0;\boldsymbol{\varsigma})\|T_{L_{u}}(\boldsymbol{\varsigma})\|\chi_{n_u}(L_u;\boldsymbol{\varsigma})\rangle^*
\nonumber \\
&&\times
\frac{\langle\varphi_{n_0}(0;\boldsymbol{\rho})\|T_{L_{t}}(\boldsymbol{\rho})\|\chi_{n_t}(L_t;\boldsymbol{\rho})\rangle
\langle\varphi_{n_0'}(0;\boldsymbol{\varsigma})\|T_{L_u}(\boldsymbol{\varsigma})\|\chi_{n_u}(L_u;\boldsymbol{\varsigma})\rangle}
{E_{n_tL_t}+E_{n_uL_u}-E_{n_0S}^{(0)}-E_{n_0'S}^{(0)}} \nonumber \\
&-&c^{*}b\sum_{n_tn_u}\sum_{L_tL_u}\sum_{M_tM_u}\frac{16\pi^2}{R_{23}^{2L_{t}+2L_{u}+2}}
\frac{[P_{L_{t}+L_{u}}^{M_{t}+M_{u}}(0)(L_{t}+L_{u}-M_{t}-M_{u})!]^2(L_{t},L_{u})^{-2}}{(L_{t}+M_{t})!(L_{t}-M_{t})!(L_{u}+M_{u})!
(L_{u}-M_{u})!} \nonumber \\
&&\times
\langle\varphi_{n_0}(0;\boldsymbol{\rho})\|T_{L_{t}}(\boldsymbol{\rho})\|\chi_{n_t}(L_t;\boldsymbol{\rho})\rangle^*
\langle\varphi_{n_0'}(0;\boldsymbol{\varsigma})\|T_{L_u}(\boldsymbol{\varsigma})\|\chi_{n_u}(L_u;\boldsymbol{\varsigma})\rangle^*
\nonumber \\
&&\times
\frac{\langle\varphi_{n_0'}(0;\boldsymbol{\rho})\|T_{L_t}(\boldsymbol{\rho})\|\chi_{n_t}(L_t;\boldsymbol{\rho})\rangle
\langle\varphi_{n_0}(0;\boldsymbol{\varsigma})\|T_{L_{u}}(\boldsymbol{\varsigma})\|\chi_{n_u}(L_u;\boldsymbol{\varsigma})\rangle}
{E_{n_tL_t}+E_{n_uL_u}-E_{n_0S}^{(0)}-E_{n_0'S}^{(0)}} \nonumber \\
&=&-\bigg\{|a|^2\sum_{n_tn_u}\sum_{L_tL_u}
\frac{K_2(n_t,n_u,L_t,L_u)}{R_{23}^{2L_{t}+2L_u+2}}
+|b|^2\sum_{n_tn_u}\sum_{L_tL_u}
\frac{K_1(n_t,n_u,L_t,L_u)}{R_{23}^{2L_{t}+2L_{u}+2}} \nonumber \\
&+&|c|^2\sum_{n_tn_u}\sum_{L_tL_u}
\frac{K_1(n_u,n_t,L_u,L_t)}{R_{23}^{2L_{t}+2L_{u}+2}} +b^{*}c\sum_{n_tn_u}\sum_{L_tL_u}
\frac{K_3(n_t,n_u,L_t,L_u)}{R_{23}^{2L_{u}+2L_{t}+2}}\nonumber \\
&+& c^{*}b\sum_{n_tn_u}\sum_{L_tL_u}
\frac{K_3^{*}(n_t,n_u,L_t,L_u)}{R_{23}^{2L_{t}+2L_{u}+2}}
\bigg\} \,,
\end{eqnarray}

\begin{eqnarray}\label{V31-31b}
V_{31}^{(2)}&=&-|a|^2\sum_{n_sn_u}\sum_{L_sL_u}\sum_{M_sM_u} \frac{16\pi^2}{R_{31}^{2L_{s}+2L_{u}+2}} \frac{[P_{L_{u}+L_{s}}^{M_{u}+M_{s}}(0)(L_{u}+L_{s}-M_{u}-M_{s})!]^2(L_{u},L_{s})^{-2}}{(L_{u}+M_{u})!(L_{u}-M_{u})!
(L_{s}+M_{s})!(L_{s}-M_{s})!} \nonumber \\
&&\times
\frac{|\langle\varphi_{n'_0}(0;\boldsymbol{\sigma})\|T_{L_{s}}(\boldsymbol{\sigma})\|\chi_{n_s}(L_s;\boldsymbol{\sigma})\rangle|^{2}
|\langle\varphi_{n_0}(0;\boldsymbol{\varsigma})\|T_{L_{u}}(\boldsymbol{\varsigma})\|\chi_{n_u}(L_u;\boldsymbol{\varsigma})\rangle|^{2}}
{E_{n_sL_s}+E_{n_uL_u}-E_{n_0S}-E_{n'_0S}}  \nonumber \\
&-&|b|^2\sum_{n_sn_u}\sum_{L_sL_u}\sum_{M_sM_u} \frac{16\pi^2}{R_{31}^{2L_{s}+2L_{u}+2}} \frac{[P_{L_{u}+L_{s}}^{M_{u}+M_{s}}(0)(L_{u}+L_{s}-M_{u}-M_{s})!]^2(L_{u},L_{s})^{-2}}{(L_{u}+M_{u})!(L_{u}-M_{u})!
(L_{s}+M_{s})!(L_{s}-M_{s})!} \nonumber \\
&&\times
\frac{|\langle\varphi_{n_0}(0;\boldsymbol{\sigma})\|T_{L_{s}}(\boldsymbol{\sigma})\|\chi_{n_s}(L_s;\boldsymbol{\sigma})\rangle|^{2}
|\langle\varphi_{n_0}(0;\boldsymbol{\varsigma})\|T_{L_{u}}(\boldsymbol{\varsigma})\|\chi_{n_u}(L_u;\boldsymbol{\varsigma})\rangle|^{2}}
{E_{n_sL_s}+E_{n_uL_u}-2E_{n_0S}}  \nonumber \\
&-&|c|^2\sum_{n_sn_u}\sum_{L_sL_u}\sum_{M_sM_u} \frac{16\pi^2}{R_{31}^{2L_{s}+2L_{u}+2}} \frac{[P_{L_{u}+L_{s}}^{M_{u}+M_{s}}(0)(L_{u}+L_{s}-M_{u}-M_{s})!]^2(L_{u},L_{s})^{-2}}{(L_{u}+M_{u})!(L_{u}-M_{u})!
(L_{s}+M_{s})!(L_{s}-M_{s})!} \nonumber \\
&&\times
\frac{|\langle\varphi_{n_0}(0;\boldsymbol{\sigma})\|T_{L_{s}}(\boldsymbol{\sigma})\|\chi_{n_s}(L_s;\boldsymbol{\sigma})\rangle|^{2}
|\langle\varphi_{n_0}(0;\boldsymbol{\varsigma})\|T_{L_{u}}(\boldsymbol{\varsigma})\|\chi_{n_u}(L_u;\boldsymbol{\varsigma})\rangle|^{2}}
{E_{n_sL_s}+E_{n_uL_u}-E_{n_0S}-E_{n'_0S}}  \nonumber \\
&-&c^{*}a\sum_{n_sn_u}\sum_{L_sL_u}\sum_{M_sM_u} \frac{16\pi^2}{R_{31}^{2L_{s}+2L_{u}+2}} \frac{[P_{L_{u}+L_{s}}^{M_{u}+M_{s}}(0)(L_{u}+L_{s}-M_{u}-M_{s})!]^2(L_{u},L_{s})^{-2}}{(L_{u}+M_{u})!(L_{u}-M_{u})!
(L_{s}+M_{s})!(L_{s}-M_{s})!} \nonumber \\
&&\times
\langle\varphi_{n_0'}(0;\boldsymbol{\varsigma})\|T_{L_u}(\boldsymbol{\varsigma})\|\chi_{n_u}(L_u;\boldsymbol{\varsigma})\rangle^*
\langle\varphi_{n_0}(0;\boldsymbol{\sigma})\|T_{L_{s}}(\boldsymbol{\sigma})\|\chi_{n_s}(L_s;\boldsymbol{\sigma})\rangle^*
\nonumber \\
&&\times
\frac{\langle\varphi_{n_0}(0;\boldsymbol{\varsigma})\|T_{L_u}(\boldsymbol{\varsigma})\|\chi_{n_u}(L_u;\boldsymbol{\varsigma})\rangle
\langle\varphi_{n_0'}(0;\boldsymbol{\sigma})\|T_{L_s}(\boldsymbol{\sigma})\|\chi_{n_s}(L_s;\boldsymbol{\sigma})\rangle
}
{E_{n_sL_s}+E_{n_uL_u}-E_{n_0S}-E_{n_0'S}} \nonumber \\
&-&a^{*}c\sum_{n_sn_u}\sum_{L_sL_u}\sum_{M_sM_u} \frac{16\pi^2}{R_{31}^{2L_{s}+2L_{u}+2}} \frac{[P_{L_{u}+L_{s}}^{M_{u}+M_{s}}(0)(L_{u}+L_{s}-M_{u}-M_{s})!]^2(L_{u},L_{s})^{-2}}{(L_{u}+M_{u})!(L_{u}-M_{u})!
(L_{s}+M_{s})!(L_{s}-M_{s})!} \nonumber \\
&& \times
\langle\varphi_{n_0'}(0;\boldsymbol{\sigma})\|T_{L_s}(\boldsymbol{\sigma})\|\chi_{n_s}(L_s;\boldsymbol{\sigma})\rangle^*
\langle\varphi_{n_0}(0;\boldsymbol{\varsigma})\|T_{L_{u}}(\boldsymbol{\varsigma})\|\chi_{n_u}(L_u;\boldsymbol{\varsigma})\rangle^*
\nonumber \\
&&\times
\frac{\langle\varphi_{n_0}(0;\boldsymbol{\sigma})\|T_{L_{s}}(\boldsymbol{\sigma})\|\chi_{n_s}(L_s;\boldsymbol{\sigma})\rangle
\langle\varphi_{n_0'}(0;\boldsymbol{\varsigma})\|T_{L_u}(\boldsymbol{\varsigma})\|\chi_{n_u}(L_u;\boldsymbol{\varsigma})\rangle}
{E_{n_sL_s}+E_{n_uL_u}-E_{n_0S}-E_{n_0'S}} \nonumber \\
&=&-\bigg\{
|a|^2 \sum_{n_sn_u}\sum_{L_sL_u}
\frac{K_1(n_s,n_u,L_s,L_u)}{R_{31}^{2L_{u}+2L_{s}+2}}
+ |b|^2\sum_{n_sn_u}\sum_{L_sL_u} \frac{K_2(n_s,n_u,L_s,L_u)}{R_{31}^{2L_{s}+2L_{u}+2}}\nonumber \\
&+& |c|^2 \sum_{n_sn_u}\sum_{L_sL_u }
\frac{K_1(n_u,n_s,L_u,L_s)}{R_{31}^{2L_{s}+2L_{u}+2}}
+ a^{*}c \sum_{n_sn_u}\sum_{L_sL_u}
\frac{K_3(n_u,n_s,L_u,L_s)}{R_{31}^{2L_{s}+2L_{u}+2}}\nonumber \\
&+& c^{*}a\sum_{n_sn_u}\sum_{L_sL_u}
\frac{K_3^*(n_u,n_s,L_u,L_s)}{R_{31}^{2L_{s}+2L_{u}+2}}\bigg\}  \,,
\end{eqnarray}

In the above, the $K_i$ functions are defined by
\begin{eqnarray}\label{AK1}
K_1(n_s,n_t,L_s,L_t) &=&G_1(L_s,L_t)
\frac{|\langle\varphi_{n'_0}(0;\boldsymbol{\sigma})\|T_{L_{s}}(\boldsymbol{\sigma})\|\chi_{n_s}(L_s;\boldsymbol{\sigma})\rangle|^{2}
|\langle\varphi_{n_0}(0;\boldsymbol{\rho})\|T_{L_t}(\boldsymbol{\rho})\|\chi_{n_t}(L_t;\boldsymbol{\rho})\rangle|^{2}}
{E_{n_sL_s}+E_{n_tL_t}-E_{n_0S}^{(0)}-E_{n'_0S}^{(0)}} \,, \nonumber \\
\end{eqnarray}
\begin{eqnarray}\label{AK2}
K_2(n_s,n_t,L_s,L_t) &=&G_1(L_s,L_t)
\frac{|\langle\varphi_{n_0}(0;\boldsymbol{\sigma})\|T_{L_{s}}(\boldsymbol{\sigma})\|\chi_{n_s}(L_s;\boldsymbol{\sigma})\rangle|^{2}
|\langle\varphi_{n_0}(0;\boldsymbol{\rho})\|T_{L_t}(\boldsymbol{\rho})\|\chi_{n_t}(L_t;\boldsymbol{\rho})\rangle|^{2}}
{E_{n_sL_s}+E_{n_tL_t}-2E_{n_0S}^{(0)}} \,, \nonumber \\
\end{eqnarray}
\begin{eqnarray}\label{AK3}
K_3(n_s,n_t,L_s,L_t) &=&G_1(L_s,L_t)
\langle\varphi_{n_0}(0;\boldsymbol{\sigma})\|T_{L_s}(\boldsymbol{\sigma})\|\chi_{n_s}(L_s;\boldsymbol{\sigma})\rangle^*
\langle\varphi_{n_0'}(0;\boldsymbol{\rho})\|T_{l_{2}^{\prime}}(\boldsymbol{\rho})\|\chi_{n_t}(L_t;\boldsymbol{\rho})\rangle^*
 \nonumber \\
&\times&
\frac{\langle\varphi_{n_0'}(0;\boldsymbol{\sigma})\|T_{l_1}(\boldsymbol{\sigma})\|\chi_{n_s}(L_s;\boldsymbol{\sigma})\rangle
\langle\varphi_{n_0}(0;\boldsymbol{\rho})\|T_{L_t}(\boldsymbol{\rho})\|\chi_{n_{t}}(L_t;\boldsymbol{\rho})\rangle
}
{E_{n_sL_s}+E_{n_tL_t}-E_{n_0S}^{(0)}-E_{n_0'S}^{(0)}} \,.
 \nonumber \\
\end{eqnarray}
where  $G_1(L_i,L_j)$ is further defined by:
\begin{eqnarray}\label{G1}
G_1(L_i,L_j)&=& 16\pi^2(L_i,L_j)^{-2}\sum_{M_iM_j}
\frac{[P_{L_i+L_j}^{M_i+M_j}(0)(L_i+L_j-M_i-M_j)!]^2}{(L_i+M_i)!
(L_i-M_i)!(L_j+M_j)!(L_j-M_j)!} \, .
\end{eqnarray}

Then the second-order energy correction is simplified as,
\begin{eqnarray}
\Delta E^{(2)}
&=&-\sum_{n\geq 3}\bigg(\frac{C_{2n}^{(12)}}{R_{12}^{2n}}+\frac{C_{2n}^{(23)}}{R_{23}^{2n}}+\frac{C_{2n}^{(31)}}{R_{31}^{2n}}
\bigg) \, .
\end{eqnarray}
where $C_{2n}^{(IJ)}$ are the additive dispersion coefficients. These coefficients can be expressed as
\begin{eqnarray}
C_{2n}^{(12)}(L,M)&=&|a|^2\sum_{n_sn_t}\sum_{L_sL_t}
K_1(n_s,n_t,L_s,L_t)
+|b|^2\sum_{n_sn_t}\sum_{L_sL_t}
K_1(n_t,n_s,L_t,L_s) \nonumber  \\
&+&|c|^2\sum_{n_sn_t} \sum_{L_sL_t}
K_2(n_s,n_t,L_s,L_t)
+a^{*}b\sum_{n_sn_t}\sum_{L_sL_t}
K_3(n_s,n_t,L_s,L_t) \nonumber \\
&+& b^{*}a\sum_{n_sn_t}\sum_{L_sL_t }K_3^*(n_s,n_t,L_s,L_t) \,,
\end{eqnarray}
\begin{eqnarray}
C_{2n}^{(23)}(L,M)&=&|a|^2\sum_{n_tn_u}\sum_{L_tL_u} K_2(n_t,n_u,L_t,L_u)
+|b|^2\sum_{n_tn_u}\sum_{L_tL_u} K_1(n_t,n_u,L_t,L_u)
 \nonumber \\
&+&|c|^2\sum_{n_tn_u}\sum_{L_tL_u} K_1(n_u,n_t,L_u,L_t) +b^{*}c\sum_{n_tn_u}\sum_{L_tL_u}
K_3(n_t,n_u,L_t,L_u)\nonumber \\
&+& c^{*}b\sum_{n_tn_u}\sum_{L_tL_u}
K_3^*(n_t,n_u,L_t,L_u) \,,
\end{eqnarray}
\begin{eqnarray}
C_{2n}^{(31)}(L,M)&=&|a|^2 \sum_{n_sn_u}\sum_{L_sL_u}
K_1(n_s,n_u,L_s,L_u)
+ |b|^2\sum_{n_sn_u}\sum_{L_sL_u } K_2(n_s,n_u,L_s,L_u) \nonumber \\
&+& |c|^2 \sum_{n_sn_u}\sum_{L_sL_u }
K_1(n_u,n_s,L_u,L_s)
+ a^{*}c\sum_{n_sn_u}\sum_{L L_sL_u}
K_3^*(n_u,n_s,L_u,L_s) \nonumber \\
&+& c^{*}a \sum_{n_sn_u}\sum_{L_sL_u }
K_3(n_u,n_s,L_u,L_s) \,,
\end{eqnarray}

\begingroup
\squeezetable
\begin{table*}%%%% I		
\caption{Values of $\mathbb{T}_1$, $\mathbb{T}_2$, $\mathbb{T}_3$, $\mathbb{R}_1$, $\mathbb{R}_2$, $\mathbb{R}_3$ for the He($1\,^1S$)-He($1\,^1S$)-He($2\,^{1}S$) system, in atomic units. All these quantities are independent of the geometrical configuration formed by the three atoms. The numbers in parentheses represent the computational uncertainties.}\label{TabIS}
\begin{ruledtabular}
\begin{tabular}{lccccccc}
\multicolumn{1}{c}{Atom}  & \multicolumn{1}{c}{$\mathbb{T}_1$}  & \multicolumn{1}{c}{$\mathbb{T}_2$}  & \multicolumn{1}{c}{$\mathbb{T}_3$} & \multicolumn{1}{c}{$\mathbb{R}_1$}  & \multicolumn{1}{c}{$\mathbb{R}_2$}  & \multicolumn{1}{c}{$\mathbb{R}_3$}\\
\hline
\\
\multicolumn{1}{c}{$^{\infty}$He}    &41.8413161(1) &1.4609778377(1)   &2.90911603(1)  &3310.4418(1)    &14.11785735(1)    &96.090681(2)\\
\\
\multicolumn{1}{c}{$^{4}$He}   &41.8723173(1)  &1.4621228531(1) &2.91128666(1)  &3312.0535(1)    &14.12578804(1)    &96.141508(2) \\
\\
\multicolumn{1}{c}{$^{3}$He}    &41.8824647(1)  &1.4624976699(1) &2.91199719(1)  &3312.5809(1)    & 14.12838381(1)   &96.158143(2) \\
\\
\end{tabular}
\end{ruledtabular}
\end{table*}
\endgroup

%====================== All the tables for He(1s)-He(1s)-He(2p)-(Singlet)  =======================================================
\begingroup
\squeezetable
\begin{table*}%%%% I
\caption{Values of $\mathbb{D}_0(M=0)$, $\mathbb{D}_0(M=\pm 1)$, $\mathbb{D}_1(M=0)$, $\mathbb{D}_1(M=\pm1)$, $\mathbb{D}_2$, $\mathbb{Q}_1(M=0)$, $\mathbb{Q}_1(M=\pm1)$, $\mathbb{Q}_2$, $\mathbb{Q}_3(M=0)$, and $\mathbb{Q}_3(M=\pm1)$ for the He($n_0\,^{\lambda}S$)-He($n_0\,^{\lambda}S$)-He($n_0^{\prime}\,^{\lambda}P$) system, in atomic units. All these quantities are independent of the geometrical configuration formed by the three atoms. The numbers in parentheses represent the computational uncertainties.}\label{TabI}
\begin{ruledtabular}
\begin{tabular}{lcccccc}
\multicolumn{1}{c}{Atom} & \multicolumn{1}{c}{$\mathbb{D}_0(M=0)$} & \multicolumn{1}{c}{$\mathbb{D}_0(M=\pm 1)$} & \multicolumn{1}{c}{$\mathbb{D}_1(M=0)$}  & \multicolumn{1}{c}{$\mathbb{D}_1(M=\pm1)$} & \multicolumn{1}{c}{$\mathbb{D}_2$} \\
\multicolumn{1}{c}{} & \multicolumn{1}{c}{$\mathbb{Q}_1(M=0)$} & \multicolumn{1}{c}{$\mathbb{Q}_1(M=\pm 1)$} & \multicolumn{1}{c}{$\mathbb{Q}_2$}  & \multicolumn{1}{c}{$\mathbb{Q}_3(M=0)$} & \multicolumn{1}{c}{$\mathbb{Q}_3(M=\pm 1)$} \\
\hline
\\
& & \multicolumn{3}{c}{He($1\,^1S$)-He($1\,^1S$)-He($2\,^{1}P$)}\\
\cline{3-5}
\\
\multicolumn{1}{c}{$^{\infty}$He} & $-$0.1770556027(1) &0.0885278013(1) &32.6430764067(3) &45.758183356(1) &1.4609778377(1) \\
\multicolumn{1}{c}{} & 377.0136(1) &5123.8556(1) & 14.1178573524(3) &19.5357464(2) &$-$78.9244427(5)\\
\\
\multicolumn{1}{c}{$^{4}$He}      & $-$0.1770765305(1) &0.0885382652(1) &32.6709827359(1) &45.797120320(1) &1.4621228531(1) \\
\multicolumn{1}{c}{}      & 376.9508(1) &5127.4835(1) &14.1257880415(1) &19.5451867(2) &$-$78.9606567(6)\\
\\
\multicolumn{1}{c}{$^{3}$He}      & $-$0.1770833800(1) &0.0885416900(1) &32.6801176662(2) &45.809866024(1) &1.4624976699(1) \\
\multicolumn{1}{c}{}      & 376.9301(1) &5128.6709(1) &14.1283838158(2) &19.5482763(2) &$-$78.9725084(6)\\
\\
& & \multicolumn{3}{c}{He($2\,^1S$)-He($2\,^1S$)-He($2\,^{1}P$)}\\
\cline{3-5}
\\
\multicolumn{1}{c}{$^{\infty}$He} & $-$8.5048343026(1) &4.2524171513(1) &4068.2(1) &4849.0(2) &11241.0468(1) \\
\multicolumn{1}{c}{} & 456379(2) &1547973(4) &817250.251(2) &36573.7501(1) &$-$962005.7512(1)\\
\\
\multicolumn{1}{c}{$^{4}$He}      & $-$8.5078707885(1) &4.2539353942(1) &4071.8(1) &4853.3(2) &11247.7393(1) \\
\multicolumn{1}{c}{}      & 456573(2) &1548946(5) &817626.242(2) &36641.8289(1) &$-$962545.9503(1)\\
\\
\multicolumn{1}{c}{$^{3}$He}      & $-$8.5088643582(1) &4.2544321791(1) &4073.0(1) &4854.7(2) &11249.9297(1) \\
\multicolumn{1}{c}{}      & 456635(3) &1549265(5) &817749.274(2) &36664.1093(1) &$-$962722.7314(1)\\
\\
& & \multicolumn{3}{c}{He($2\,^3S$)-He($2\,^3S$)-He($2\,^{3}P$)}\\
\cline{3-5}
\\
\multicolumn{1}{c}{$^{\infty}$He} & $-$6.4077465538(1) &3.2038732769(1) &1862.5727(2) &2251.4034(2) &3276.6801(3) \\
\multicolumn{1}{c}{} & 135961.610(2) &531453.96(2) &210566.55(2) &32944.5218(1) &$-$291050.4284(1)\\
\\
\multicolumn{1}{c}{$^{4}$He}      & $-$6.4090875602(1) &3.2045437801(1) &1863.4728(2) &2252.4905(1) &3279.4590(3) \\
\multicolumn{1}{c}{}      & 135980.281(2) &531566.81(2) &210667.81(2) &32941.0762(1) &$-$291128.5937(1)\\
\\
\multicolumn{1}{c}{$^{3}$He}      & $-$6.4095263009(2) &3.2047631504(1) &1863.7672(1) &2252.8465(2) &3280.3686(2) \\
\multicolumn{1}{c}{}      & 135986.388(2) &531603.71(1) &210700.95(2) &32939.9454(1) &$-$291154.1703(1)\\
\end{tabular}
\end{ruledtabular}
\end{table*}
\endgroup

%====================== Tables for equilateral triangle =======================================================

\begingroup
\squeezetable
\begin{table*}%%%% IIS		
\caption{The additive dispersion coefficients $C_6^{(IJ)}$ and $C_8^{(IJ)}$
of the He($1\,^1S$)-He($1\,^1S$)-He($2\,^{1}S$) system for three different types of the zeroth-order wave functions, where the three atoms form an equilateral triangle, in atomic units.
The numbers in parentheses represent the computational uncertainties.}\label{TabIIS}
\begin{ruledtabular}
\begin{tabular}{lccccccc}
\multicolumn{1}{c}{Atom}  &\multicolumn{1}{c}{State}  & \multicolumn{1}{c}{$C_6^{(12)}$}  & \multicolumn{1}{c}{$C_6^{(23)}$}  & \multicolumn{1}{c}{$C_6^{(31)}$} & \multicolumn{1}{c}{$C_8^{(12)}$}  & \multicolumn{1}{c}{$C_8^{(23)}$}  & \multicolumn{1}{c}{$C_8^{(31)}$}\\
\hline
\\
\multicolumn{1}{c}{$^{\infty}$He}  &$\Psi_{1,\Delta}^{(0)}$  &30.320614(1)  &30.320614(1) &30.320614(1)  &2275.727(1)    &2275.727(1)    &2275.727(1)\\
\multicolumn{1}{c}{}  &$\Psi_{2,\Delta}^{(0)}$  &21.651146(1)  &21.651146(1) &38.932200(1)  &1662.279(1) &1662.279(1)  &3214.351(1)\\
\multicolumn{1}{c}{}  &$\Psi_{3,\Delta}^{(0)}$  &33.171849(1)  &33.171849(1) &15.890795(1)    &2696.994(1)    &2696.994(1) &1144.922(1)\\
\\
\multicolumn{1}{c}{$^{4}$He}  &$\Psi_{1,\Delta}^{(0)}$  &30.343110(1)  &30.343110(1) &30.343110(1)  &2276.838(1)    &2276.838(1)    &2276.838(1)\\
\multicolumn{1}{c}{}  &$\Psi_{2,\Delta}^{(0)}$  &21.667220(1)  &21.667220(1) &38.961030(1)  &1663.089(1) &1663.089(1) &3215.912(1)\\
\multicolumn{1}{c}{}  &$\Psi_{3,\Delta}^{(0)}$  &33.196427(1)  &33.196427(1) &15.902616(1)    &2698.304(1)   &2698.304(1) &1145.482(1)\\
\\
\multicolumn{1}{c}{$^{3}$He}  &$\Psi_{1,\Delta}^{(0)}$  &30.350473(1)  &30.350473(1) &30.350473(1)  &2277.202(1)    &2277.202(1)    &2277.202(1)\\
\multicolumn{1}{c}{}  &$\Psi_{2,\Delta}^{(0)}$  &21.672481(1)  &21.672481(1) &38.970467(1)  &1663.354(1) &1663.354(1) &3216.422(1)\\
\multicolumn{1}{c}{}  &$\Psi_{3,\Delta}^{(0)}$  &33.204472(1)  &33.204472(1) &15.906485(1)    &2698.733(1)    &2698.733(1) &1145.665(1)\\
\end{tabular}
\end{ruledtabular}
\end{table*}
\endgroup

\begingroup
\squeezetable
%\begin{table*}%%%% II
\begin{sidewaystable}[h]
\caption{The additive long-range coefficients $C_3^{(IJ)}(1,M)$ of the He($n_0\,^{\lambda}S$)-He($n_0\,^{\lambda}S$)-He($n_0^{\prime}\,^{\lambda}P$) system
for three different types of the zeroth-order wave functions, where the three atoms form an equilateral triangle,
in atomic units. The numbers in parentheses represent the computational uncertainties. }\label{TabII}
\centering
\begin{ruledtabular}
\begin{tabular}{lccccccc}
\multicolumn{1}{c}{Atom}  &\multicolumn{1}{c}{State}  & \multicolumn{1}{c}{$C_3^{(12)}(1,M=0)$}  & \multicolumn{1}{c}{$C_3^{(23)}(1,M=0)$} & \multicolumn{1}{c}{$C_3^{(31)}(1,M=0)$}
& \multicolumn{1}{c}{$C_3^{(12)}(1,M=\pm1)$} & \multicolumn{1}{c}{$C_3^{(23)}(1,M=\pm1)$}  & \multicolumn{1}{c}{$C_3^{(31)}(1,M=\pm1)$}\\
\hline
\\
& & & & \multicolumn{2}{c}{He($1\,^1S$)-He($1\,^1S$)-He($2\,^{1}P$)}\\
\cline{5-6}
\\
\multicolumn{1}{c}{$^{\infty}$He} &
$\Psi_{1,\Delta}^{(0)}$   &$-$0.1180370684(1)  &$-$0.1180370684(1)  &$-$0.1180370684(1) &0.0590185342(1)     &0.0590185342(1)    &0.0590185342(1)\\
&$\Psi_{2,\Delta}^{(0)}$   &0                &0                &0.1770556027(1)      &0                &0               &$-$0.0885278013(1)\\
&$\Psi_{3,\Delta}^{(0)}$   &0.1180370684(1)     &0.1180370684(1)     &$-$0.0590185342(1) &$-$0.0590185342(1)&$-$0.0590185342(1) &0.0295092671(1)\\
\\
\multicolumn{1}{c}{$^{4}$He} &
$\Psi_{1,\Delta}^{(0)}$   &$-$0.1180510203(1)   &$-$0.1180510203(1)   &$-$0.1180510203(1)   &0.0590255101(1)     &0.0590255101(1)     &0.0590255101(1)\\
&$\Psi_{2,\Delta}^{(0)}$   &0                &0                &0.1770765305(1)      &0                &0                &$-$0.0885382652(1)\\
&$\Psi_{3,\Delta}^{(0)}$   &0.1180510203(1)      &0.1180510203(1)      &$-$0.0590255101(1)  &$-$0.0590255101(1)  &$-$0.0590255101(1)  &0.0295127550(1)\\
\\
\multicolumn{1}{c}{$^{3}$He}
&$\Psi_{1,\Delta}^{(0)}$   &$-$0.1180555867(1)   &$-$0.1180555867(1)   &$-$0.1180555867(1)  &0.0590277933(1)      &0.0590277933(1)      &0.0590277933(1)\\
&$\Psi_{2,\Delta}^{(0)}$   &0                &0                &0.1770833800(1)      &0                &0                &$-$0.0885416900(1)\\
&$\Psi_{3,\Delta}^{(0)}$   &0.1180555867(1)      &0.1180555867(1)      &$-$0.0590277933(1)   &$-$0.0590277933(1)   &$-$0.0590277933(1)   &0.0295138966(1)\\
\\
& & & & \multicolumn{2}{c}{He($2\,^1S$)-He($2\,^1S$)-He($2\,^{1}P$)}\\
\cline{5-6}
\\
\multicolumn{1}{c}{$^{\infty}$He}   & $\Psi_{1,\Delta}^{(0)}$   &$-$5.6698895350(1)  &$-$5.6698895350(1)  &$-$5.6698895350(1) &2.8349447675(1)     &2.8349447675(1)    &2.8349447675(1)\\
&$\Psi_{2,\Delta}^{(0)}$   &0                &0                &8.5048343026(1)      &0                &0               &$-$4.2524171513(1)\\
&$\Psi_{3,\Delta}^{(0)}$   &5.6698895350(1)     &5.6698895350(1)     &$-$2.8349447675(1) &$-$2.8349447675(1) &$-$2.8349447675(1) &1.4174723837(1)\\
\\
\multicolumn{1}{c}{$^{4}$He}  &$\Psi_{1,\Delta}^{(0)}$   &$-$5.6719138590(1)   &$-$5.6719138590(1)   &$-$5.6719138590(1)   &2.8359569295(1)     &2.8359569295(1)     &2.8359569295(1)\\
&$\Psi_{2,\Delta}^{(0)}$   &0                &0                &8.5078707885(1)      &0                &0                &$-$4.2539353942(1)\\
&$\Psi_{3,\Delta}^{(0)}$   &5.6719138590(1)      &5.6719138590(1)      &$-$2.8359569295(1)  &$-$2.8359569295(1)  &$-$2.8359569295(1)  &1.4179784647(1)\\
\\
\multicolumn{1}{c}{$^{3}$He}  &$\Psi_{1,\Delta}^{(0)}$   &$-$5.6725762388(1)   &$-$5.6725762388(1)   &$-$5.6725762388(1)  &2.8362881194(1)      &2.8362881194(1)      &2.8362881194(1)\\
&$\Psi_{2,\Delta}^{(0)}$   &0                &0                &8.5088643582(1)      &0                &0                &$-$4.2544321791(1)\\
&$\Psi_{3,\Delta}^{(0)}$   &5.6725762388(1)      &5.6725762388(1)      &$-$2.8362881194(1)   &$-$2.8362881194(1)   &$-$2.8362881194(1)   &1.4181440597(1)\\
\\
& & & & \multicolumn{2}{c}{He($2\,^3S$)-He($2\,^3S$)-He($2\,^{3}P$)}\\
\cline{5-6}
\\
\multicolumn{1}{c}{$^{\infty}$He}   &$\Psi_{1,\Delta}^{(0)}$   &$-$4.2718310359(1)  &$-$4.2718310359(1)  &$-$4.2718310359(1) &2.1359155179(1)     &2.1359155179(1)    &2.1359155179(1)\\
&$\Psi_{2,\Delta}^{(0)}$   &0                &0                &6.4077465538(1)      &0                &0               &$-$3.2038732769(1)\\
&$\Psi_{3,\Delta}^{(0)}$   &4.2718310359(1)     &4.2718310359(1)     &$-$2.1359155179(1) &$-$2.1359155179(1)&$-$2.1359155179(1) &1.0679577589(1)\\
\\
\multicolumn{1}{c}{$^{4}$He}  &$\Psi_{1,\Delta}^{(0)}$   &$-$4.2727250401(1)   &$-$4.2727250401(1)   &$-$4.2727250401(1)   &2.1363625200(1)     &2.1363625200(1)     &2.1363625200(1)\\
&$\Psi_{2,\Delta}^{(0)}$   &0                &0                &6.4090875602(1)      &0                &0                &$-$3.2045437801(1)\\
&$\Psi_{3,\Delta}^{(0)}$   &4.2727250401(1)      &4.2727250401(1)      &$-$2.1363625200(1)  &$-$2.1363625200(1)  &$-$2.1363625200(1)  &1.0681812600(1)\\
\\
\multicolumn{1}{c}{$^{3}$He}  &$\Psi_{1,\Delta}^{(0)}$   &$-$4.2730175339(1)   &$-$4.2730175339(1)   &$-$4.2730175339(1)  &2.1365087669(1)      &2.1365087669(1)      &2.1365087669(1)\\
&$\Psi_{2,\Delta}^{(0)}$   &0                &0                &6.4095263009(1)      &0                &0                &$-$3.2047631504(1)\\
 &$\Psi_{3,\Delta}^{(0)}$   &4.2730175339(1)      &4.2730175339(1)      &$-$2.1365087669(1)   &$-$2.1365087669(1)   &$-$2.1365087669(1)   &1.0682543834(1)\\
 \\
\end{tabular}
\end{ruledtabular}
\label{tab:LPer}
\end{sidewaystable}
%\end{table*}
\endgroup

\begingroup
\squeezetable
\begin{sidewaystable}[h]
%\begin{table*}%%%% III	
\caption{The additive and nonadditive dispersion coefficients $C_6^{(IJ)}(1,M=0)$ and $C_6^{(IJ,JK)}(1,M=0)$ of the He($n_0\,^{\lambda}S$)-He($n_0\,^{\lambda}S$)-He($n_0^{\prime}\,^{\lambda}P$) system for three different types of the zeroth-order wave functions, where the three atoms form an equilateral triangle, in atomic units. The numbers in parentheses represent the computational uncertainties.}\label{TabIII}
\centering
\begin{ruledtabular}
\begin{tabular}{lccccccc}
\multicolumn{1}{c}{Atom}  &\multicolumn{1}{c}{State}  & \multicolumn{1}{c}{$C_6^{(12)}(1,M=0)$}  & \multicolumn{1}{c}{$C_6^{(23)}(1,M=0)$}  & \multicolumn{1}{c}{$C_6^{(31)}(1,M=0)$} & \multicolumn{1}{c}{$C_6^{(12,23)}(1,M=0)$}  & \multicolumn{1}{c}{$C_6^{(23,31)}(1,M=0)$}  & \multicolumn{1}{c}{$C_6^{(31,12)}(1,M=0)$}\\
\hline
\\
& & & & \multicolumn{2}{c}{He($1\,^1S$)-He($1\,^1S$)-He($2\,^{1}P$)}\\
\cline{5-6}
\\
\multicolumn{1}{c}{$^{\infty}$He}  &$\Psi_{1,\Delta}^{(0)}$  &22.2490435504(2) &22.2490435504(2) &22.2490435504(2)  &0.3183512641(1)     &0.3183512641(1)    &0.3183512641(1)\\
\multicolumn{1}{c}{}  &$\Psi_{2,\Delta}^{(0)}$  &17.0520271221(1) &17.0520271221(1) &32.6430764067(3)  &$-$0.4775268962(1)  &0             &0\\
\multicolumn{1}{c}{}  &$\Psi_{3,\Delta}^{(0)}$  &27.4460599786(3) &27.4460599786(3) &11.8550106940(1)  &0.1591756320(1)      &$-$0.3183512641(1) &$-$0.3183512641(1)\\
\\
\multicolumn{1}{c}{$^{4}$He}  &$\Psi_{1,\Delta}^{(0)}$  &22.2680294416(1)  &22.2680294416(1)  &22.2680294416(1)  &0.3185403283(1)     &0.3185403283(1)    &0.3185403283(1)\\
\multicolumn{1}{c}{}  &$\Psi_{2,\Delta}^{(0)}$  &17.0665527945(1)  &17.0665527945(1)  &32.6709827359(1)  &$-$0.4778104925(1)  &0             &0\\
\multicolumn{1}{c}{}  &$\Psi_{3,\Delta}^{(0)}$  &27.4695060888(1)  &27.4695060888(1)  &11.8650761474(1)  &0.1592701641(1)      &$-$0.3185403283(1) &$-$0.3185403283(1)\\
\\
\multicolumn{1}{c}{$^{3}$He}  &$\Psi_{1,\Delta}^{(0)}$  &22.2742443341(1)  &22.2742443341(1)  &22.2742443341(1)  &0.3186022100(1)     &0.3186022100(1)    &0.3186022100(1)\\
\multicolumn{1}{c}{}  &$\Psi_{2,\Delta}^{(0)}$  &17.0713076681(1)  &17.0713076681(1)  &32.6801176662(2)  &$-$0.4779033149(2)  &0             &0\\
\multicolumn{1}{c}{}  &$\Psi_{3,\Delta}^{(0)}$  &27.4771810002(2)  &27.4771810002(2)  &11.8683710021(1)  &0.1593011050(1)      &$-$0.3186022100(1) &$-$0.3186022100(1)\\
\\
& & & & \multicolumn{2}{c}{He($2\,^1S$)-He($2\,^1S$)-He($2\,^{1}P$)}\\
\cline{5-6}
\\
\multicolumn{1}{c}{$^{\infty}$He}  &$\Psi_{1,\Delta}^{(0)}$  &6459.14(5) &6459.14(5) &6459.14(5)  &1277.7398(1)     &1277.7398(1)    &1277.7398(1)\\
\multicolumn{1}{c}{}  &$\Psi_{2,\Delta}^{(0)}$  &7654.62(4) &7654.62(4) &4068.2(1)  &$-$1916.6097(1)  &0             &0\\
\multicolumn{1}{c}{}  &$\Psi_{3,\Delta}^{(0)}$  &5263.7(1) &5263.7(1) &8850.10(3)  &638.8699(1)      &$-$1277.7398(1) &$-$1277.7398(1)\\
\\
\multicolumn{1}{c}{$^{4}$He}  &$\Psi_{1,\Delta}^{(0)}$  &6463.79(6)  &6463.79(6)  &6463.79(6)  &1278.5076(1)     &1278.5076(1)    &1278.5076(1)\\
\multicolumn{1}{c}{}  &$\Psi_{2,\Delta}^{(0)}$  &7659.77(4)  &7659.77(4)  &4071.8(1)  &$-$1917.7614(1)  &0             &0\\
\multicolumn{1}{c}{}  &$\Psi_{3,\Delta}^{(0)}$  &5267.8(1)  &5267.8(1)  &8855.76(3)  &639.2538(1)      &$-$1278.5076(1) &$-$1278.5076(1)\\
\\
\multicolumn{1}{c}{$^{3}$He}  &$\Psi_{1,\Delta}^{(0)}$  &6465.31(6)  &6465.31(6)  &6465.31(6)  &1278.7588(1)     &1278.7588(1)    &1278.7588(1)\\
\multicolumn{1}{c}{}  &$\Psi_{2,\Delta}^{(0)}$  &7661.47(5)  &7661.47(5)  &4073.0(1)  &$-$1918.1383(1)  &0             &0\\
\multicolumn{1}{c}{}  &$\Psi_{3,\Delta}^{(0)}$  &5269.1(1)  &5269.1(1)  &8857.62(3)  &639.3794(1)      &$-$1278.7588(1) &$-$1278.7588(1)\\
\\
& & & & \multicolumn{2}{c}{He($2\,^3S$)-He($2\,^3S$)-He($2\,^{3}P$)}\\
\cline{5-6}
\\
\multicolumn{1}{c}{$^{\infty}$He}  &$\Psi_{1,\Delta}^{(0)}$  &2333.9418(2) &2333.9418(2) &2333.9418(2)  &376.6260(1)     &376.6260(1)    &376.6260(1)\\
\multicolumn{1}{c}{}  &$\Psi_{2,\Delta}^{(0)}$  &2569.6263(2) &2569.6263(2) &1862.5727(2)  &$-$564.9390(1)  &0             &0\\
\multicolumn{1}{c}{}  &$\Psi_{3,\Delta}^{(0)}$  &2098.2571(1) &2098.2571(1) &2805.3109(2)  &188.3130(1)      &$-$376.6260(1) &$-$376.6260(1)\\
\\
\multicolumn{1}{c}{$^{4}$He}  &$\Psi_{1,\Delta}^{(0)}$  &2335.4680(1)  &2335.4680(1)  &2335.4680(1)  &376.9477(1)     &376.9477(1)    &376.9477(1)\\
\multicolumn{1}{c}{}  &$\Psi_{2,\Delta}^{(0)}$  &2571.4658(2)  &2571.4658(2)  &1863.4728(2)  &$-$565.4215(1)  &0             &0\\
\multicolumn{1}{c}{}  &$\Psi_{3,\Delta}^{(0)}$  &2099.4703(1)  &2099.4703(1)  &2807.4635(2)  &188.4738(1)     &$-$376.9477(1) &$-$376.9477(1)\\
\\
\multicolumn{1}{c}{$^{3}$He}  &$\Psi_{1,\Delta}^{(0)}$  &2335.9678(2)  &2335.9678(2)  &2335.9678(2)  &377.0530(1)     &377.0530(1)    &377.0530(1)\\
\multicolumn{1}{c}{}  &$\Psi_{2,\Delta}^{(0)}$  &2572.0680(2)  &2572.0680(2)  &1863.7672(1)  &$-$565.5795(1)  &0             &0\\
\multicolumn{1}{c}{}  &$\Psi_{3,\Delta}^{(0)}$  &2099.8674(1)  &2099.8674(1)  &2808.1682(2)  &188.5265(1)      &$-$377.0530(1) &$-$377.0530(1)\\
\\
\end{tabular}
\end{ruledtabular}
%\end{table*}
\label{tab:LPer}
\end{sidewaystable}
\endgroup

\begingroup
\squeezetable
\begin{sidewaystable}
%\begin{table*}%%%% IV	!!!!!!	
\caption{The additive and nonadditive dispersion coefficients $C_6^{(IJ)}(1,M=\pm 1)$ and $C_6^{(IJ,JK)}(1,M=\pm 1)$
of the He($n_0\,^{\lambda}S$)-He($n_0\,^{\lambda}S$)-He($n_0^{\prime}\,^{\lambda}P$) system
for three different types of the zeroth-order wave functions, where the three atoms form an equilateral triangle, in atomic units.
The numbers in parentheses represent the computational uncertainties.}\label{TabIV}
\centering
\begin{ruledtabular}
\begin{tabular}{lccccccc}
\multicolumn{1}{c}{Atom}  & \multicolumn{1}{c}{State}  & \multicolumn{1}{c}{$C_6^{(12)}(1,M=\pm 1)$}  & \multicolumn{1}{c}{$C_6^{(23)}(1,M=\pm 1)$}  & \multicolumn{1}{c}{$C_6^{(31)}(1,M=\pm1)$} & \multicolumn{1}{c}{$C_6^{(12,23)}(1,M=\pm1)$}  & \multicolumn{1}{c}{$C_6^{(23,31)}(1,M=\pm1)$}  & \multicolumn{1}{c}{$C_6^{(31,12)}(1,M=\pm1)$}\\
\hline
\\
& & & & \multicolumn{2}{c}{He($1\,^1S$)-He($1\,^1S$)-He($2\,^{1}P$)}\\
\cline{5-6}
\\
\multicolumn{1}{c}{$^{\infty}$He}  & $\Psi_{1,\Delta}^{(0)}$  &30.9924481829(1) &30.9924481829(1)  &30.9924481829(1)  &$-$0.2785573560(1) &$-$0.2785573560(1)  &$-$0.2785573560(1)\\
\multicolumn{1}{c}{}  & $\Psi_{2,\Delta}^{(0)}$  &23.6095805966(1) &23.6095805966(1)  &45.758183355(1)  &0.4178360340(2)    &0              &0\\
\multicolumn{1}{c}{}  & $\Psi_{3,\Delta}^{(0)}$  &38.3753157692(1) &38.3753157692(1)  &16.2267130103(1)  &$-$0.13927867804(5)  &0.2785573560(1)     &0.2785573560(1)\\
\\
\multicolumn{1}{c}{$^{4}$He}  & $\Psi_{1,\Delta}^{(0)}$  &31.0187878317(2) &31.0187878317(2)  &31.0187878317(2)  &$-$0.2787227872(1) &$-$0.2787227872(1)  &$-$0.2787227872(1)\\
\multicolumn{1}{c}{}  & $\Psi_{2,\Delta}^{(0)}$  &23.6296215870(2) &23.6296215870(2)  &45.797120320(1)  &0.4180841809(1)    &0              &0\\
\multicolumn{1}{c}{}  & $\Psi_{3,\Delta}^{(0)}$  &38.4079540763(3) &38.4079540763(3)  &16.2404553424(1)  &$-$0.1393613936(1)  &0.2787227872(1)     &0.2787227872(1)\\
\\
\multicolumn{1}{c}{$^{3}$He}  & $\Psi_{1,\Delta}^{(0)}$  &31.0274099057(1)  &31.0274099057(1)  &31.0274099057(1)  &$-$0.2787769337(1) &$-$0.2787769337(1) &$-$0.2787769337(1)\\
\multicolumn{1}{c}{}  & $\Psi_{2,\Delta}^{(0)}$  &23.6361818467(2)  &23.6361818467(2)  &45.809866023(1)  &0.4181654005(2)    &0             &0\\
\multicolumn{1}{c}{}  & $\Psi_{3,\Delta}^{(0)}$  &38.4186379646(2)  &38.4186379646(2)  &16.2449537878(1)  &$-$0.1393884668(1)  &0.2787769337(1)    &0.2787769337(1)\\
\\
& & & & \multicolumn{2}{c}{He($2\,^1S$)-He($2\,^1S$)-He($2\,^{1}P$)}\\
\cline{5-6}
\\
\multicolumn{1}{c}{$^{\infty}$He}  & $\Psi_{1,\Delta}^{(0)}$  &6979.8(2) &6979.8(2)  &6979.8(2)  &$-$1118.0223(1) &$-$1118.0223(1)  &$-$1118.0223(1)\\
\multicolumn{1}{c}{}  & $\Psi_{2,\Delta}^{(0)}$  &8045.0(1) &8045.0(1)  &4849.0(2)  &1677.0335(1)    &0              &0\\
\multicolumn{1}{c}{}  & $\Psi_{3,\Delta}^{(0)}$  &5914.4(2) &5914.4(2)  &9110.4(1)  &$-$559.0111(1)  &1118.0223(1)     &1118.0223(1)\\
\\
\multicolumn{1}{c}{$^{4}$He}  & $\Psi_{1,\Delta}^{(0)}$  &6984.7(1) &6984.7(1)  &6984.7(1)  &$-$1118.6941(1) &$-$1118.6941(1)  &$-$1118.6941(1)\\
\multicolumn{1}{c}{}  & $\Psi_{2,\Delta}^{(0)}$  &8050.5(1) &8050.5(1)  &4853.3(2)  &1678.0412(1)    &0              &0\\
\multicolumn{1}{c}{}  & $\Psi_{3,\Delta}^{(0)}$  &5919.0(2) &5919.0(2)  &9116.2(1)  &$-$559.3470(1)  &1118.6941(1)     &1118.6941(1)\\
\\
\multicolumn{1}{c}{$^{3}$He}  & $\Psi_{1,\Delta}^{(0)}$  &6986.5(2)  &6986.5(2)  &6986.5(2)  &$-$1118.9140(1) &$-$1118.9140(1) &$-$1118.9140(1)\\
\multicolumn{1}{c}{}  & $\Psi_{2,\Delta}^{(0)}$  &8052.3(1)  &8052.3(1)  &4854.7(2)  &1678.3710(1)    &0             &0\\
\multicolumn{1}{c}{}  & $\Psi_{3,\Delta}^{(0)}$  &5920.6(2)  &5920.6(2)  &9118.1(1)  &$-$559.4570(1)  &1118.9140(1)    &1118.9140(1)\\
\\
& & & & \multicolumn{2}{c}{He($2\,^3S$)-He($2\,^3S$)-He($2\,^{3}P$)}\\
\cline{5-6}
\\
\multicolumn{1}{c}{$^{\infty}$He}  & $\Psi_{1,\Delta}^{(0)}$  &2593.1622(2) &2593.1622(2)  &2593.1622(2)  &$-$329.5477(1) &$-$329.5477(1)  &$-$329.5477(1)\\
\multicolumn{1}{c}{}  & $\Psi_{2,\Delta}^{(0)}$  &2764.0417(2) &2764.0417(2)  &2251.4034(2)  &494.3216(1)    &0              &0\\
\multicolumn{1}{c}{}  & $\Psi_{3,\Delta}^{(0)}$  &2422.2828(2) &2422.2828(2)  &2934.9211(2)  &$-$164.7738(1)  &329.5477(1)     &329.5477(1)\\
\\
\multicolumn{1}{c}{$^{4}$He}  & $\Psi_{1,\Delta}^{(0)}$  &2594.8134(2) &2594.8134(2)  &2594.8134(2)  &$-$329.8292(1) &$-$329.8292(1)  &$-$329.8292(1)\\
\multicolumn{1}{c}{}  & $\Psi_{2,\Delta}^{(0)}$  &2765.9749(3) &2765.9749(3)  &2252.4905(1)  &494.7438(1)    &0              &0\\
\multicolumn{1}{c}{}  & $\Psi_{3,\Delta}^{(0)}$  &2423.6520(2) &2423.6520(2)  &2937.1361(2)  &$-$164.9146(1)  &329.8292(1)     &329.8292(1)\\
\\
\multicolumn{1}{c}{$^{3}$He}  & $\Psi_{1,\Delta}^{(0)}$  &2595.3540(3)  &2595.3540(3) &2595.3540(3)  &$-$329.9214(1) &$-$329.9214(1) &$-$329.9214(1)\\
\multicolumn{1}{c}{}  & $\Psi_{2,\Delta}^{(0)}$  &2766.6075(2)  &2766.6075(2)  &2252.8465(2)  &494.8821(1)    &0             &0\\
\multicolumn{1}{c}{}  & $\Psi_{3,\Delta}^{(0)}$  &2424.1002(2)  &2424.1002(2)  &2937.8612(2)  &$-$164.9607(1)  &329.9214(1)   &329.9214(1)\\
\\
\end{tabular}
\end{ruledtabular}
%\end{table*}
\label{tab:LPer}
\end{sidewaystable}
\endgroup

\begingroup
\squeezetable
\begin{sidewaystable}
%\begin{table*}%%%% V		
\caption{The additive and nonadditive dispersion coefficients $C_8^{(IJ)}(1,M=0)$ and $C_8^{(IJ,JK)}(1,M=0)$
of the He($n_0\,^{\lambda}S$)-He($n_0\,^{\lambda}S$)-He($n_0^{\prime}\,^{\lambda}P$) system for three different types of the zeroth-order wave functions, where the three atoms form an equilateral triangle, in atomic units.
The numbers in parentheses represent the computational uncertainties.}\label{TabV}
\centering
\begin{ruledtabular}
\begin{tabular}{lccccccc}
\multicolumn{1}{c}{Atom}  &\multicolumn{1}{c}{State}  & \multicolumn{1}{c}{$C_8^{(12)}(1,M=0)$}  & \multicolumn{1}{c}{$C_8^{(23)}(1,M=0)$}  & \multicolumn{1}{c}{$C_8^{(31)}(1,M=0)$} & \multicolumn{1}{c}{$C_8^{(12,23)}(1,M=0)$}  & \multicolumn{1}{c}{$C_8^{(23,31)}(1,M=0)$}  & \multicolumn{1}{c}{$C_8^{(31,12)}(1,M=0)$}\\
\hline
\\
& & & & \multicolumn{2}{c}{He($1\,^1S$)-He($1\,^1S$)-He($2\,^{1}P$)}\\
\cline{5-6}
\\
\multicolumn{1}{c}{$^{\infty}$He}  &$\Psi_{1,\Delta}^{(0)}$  &269.0722(1)  &269.0722(1) &269.0722(1)  &1.007423570(1)    &1.007423570(1)    &1.007423570(1)\\
\multicolumn{1}{c}{}  &$\Psi_{2,\Delta}^{(0)}$  &195.5657(1)  &195.5657(1) &357.4779(1)  &$-$1.511135354(1) &0              &0\\
\multicolumn{1}{c}{}  &$\Psi_{3,\Delta}^{(0)}$  &303.5072(1)  &303.5072(1) &141.5950(1)    &0.5037117851(5)    &$-$1.007423570(1) &$-$1.007423570(1)\\
\\
\multicolumn{1}{c}{$^{4}$He}  &$\Psi_{1,\Delta}^{(0)}$  &269.0392(1)  &269.0392(1) &269.0392(1)  &1.007830781(1)    &1.007830781(1)    &1.007830781(1)\\
\multicolumn{1}{c}{}  &$\Psi_{2,\Delta}^{(0)}$  &195.5383(1)  &195.5383(1) &357.4056(1)  &$-$1.511746172(2) &0              &0\\
\multicolumn{1}{c}{}  &$\Psi_{3,\Delta}^{(0)}$  &303.4498(1)  &303.4498(1) &141.5825(1)    &0.5039153905(5)    &$-$1.007830781(1) &$-$1.007830781(1)\\
\\
\multicolumn{1}{c}{$^{3}$He}  &$\Psi_{1,\Delta}^{(0)}$  &269.0284(1)  &269.0284(1) &269.0284(1)  &1.007964062(1)    &1.007964062(1)    &1.007964062(1)\\
\multicolumn{1}{c}{}  &$\Psi_{2,\Delta}^{(0)}$  &195.5292(1)  &195.5292(1) &357.3819(1)  &$-$1.511946094(2) &0              &0\\
\multicolumn{1}{c}{}  &$\Psi_{3,\Delta}^{(0)}$  &303.4310(1)  &303.4310(1) &141.5784(1)    &0.5039820312(5)    &$-$1.007964062(1) &$-$1.007964062(1)\\
\\
& & & & \multicolumn{2}{c}{He($2\,^1S$)-He($2\,^1S$)-He($2\,^{1}P$)}\\
\cline{5-6}
\\
\multicolumn{1}{c}{$^{\infty}$He}  &$\Psi_{1,\Delta}^{(0)}$  &601051(2)  &601051(2) &601051(2)  &64166.8492(1)    &64166.8492(1)    &64166.8492(1)\\
\multicolumn{1}{c}{}  &$\Psi_{2,\Delta}^{(0)}$  &636815(1)  &636815(1) &419806(2)  &$-$96250.2737(1) &0              &0\\
\multicolumn{1}{c}{}  &$\Psi_{3,\Delta}^{(0)}$  &492143(1)  &492143(1) &709151(1)    &32083.42460(5)    &$-$64166.8492(1) &$-$64166.8492(1)\\
\\
\multicolumn{1}{c}{$^{4}$He}  &$\Psi_{1,\Delta}^{(0)}$  &601351(6)  &601351(6) &601351(2)  &64203.6588(1)    &64203.6588(1)    &64203.6588(1)\\
\multicolumn{1}{c}{}  &$\Psi_{2,\Delta}^{(0)}$  &637098(2)  &637098(2) &419931(2)  &$-$96305.4883(2) &0              &0\\
\multicolumn{1}{c}{}  &$\Psi_{3,\Delta}^{(0)}$  &492320(2)  &492320(2) &709488(1)    &32101.82942(5)    &$-$64203.6588(1) &$-$64203.6588(1)\\
\\
\multicolumn{1}{c}{$^{3}$He}  &$\Psi_{1,\Delta}^{(0)}$  &601449(2)  &601449(2) &601449(2)  &64215.7044(1)    &64215.7044(1)    &64215.7044(1)\\
\multicolumn{1}{c}{}  &$\Psi_{2,\Delta}^{(0)}$  &637193(1)  &637193(1) &419971(3)  &$-$96323.5566(1) &0              &0\\
\multicolumn{1}{c}{}  &$\Psi_{3,\Delta}^{(0)}$  &492379(2)  &492379(2) &709599(1)    &32107.85225(6)    &$-$64215.7044(1) &$-$64215.7044(1)\\
\\
& & & & \multicolumn{2}{c}{He($2\,^3S$)-He($2\,^3S$)-He($2\,^{3}P$)}\\
\cline{5-6}
\\
\multicolumn{1}{c}{$^{\infty}$He}  &$\Psi_{1,\Delta}^{(0)}$  &182792.93(1)  &182792.93(1) &182792.93(1)  &19616.9709(1)    &19616.9709(1)    &19616.9709(1)\\
\multicolumn{1}{c}{}  &$\Psi_{2,\Delta}^{(0)}$  &173264.08(1)  &173264.08(1) &103017.09(1)  &$-$29425.4564(1) &0              &0\\
\multicolumn{1}{c}{}  &$\Psi_{3,\Delta}^{(0)}$  &126432.75(1)  &126432.75(1) &196679.74(1)    &9808.4854(1)    &$-$19616.9709(1) &$-$19616.9709(1)\\
\\
\multicolumn{1}{c}{$^{4}$He}  &$\Psi_{1,\Delta}^{(0)}$  &182836.84(1)  &182836.84(1) &182836.84(1)  &19623.2331(1)    &19623.2331(1)    &19623.2331(1)\\
\multicolumn{1}{c}{}  &$\Psi_{2,\Delta}^{(0)}$  &173324.04(1)  &173324.04(1) &103039.20(1)  &$-$29434.8497(1) &0              &0\\
\multicolumn{1}{c}{}  &$\Psi_{3,\Delta}^{(0)}$  &126467.48(1)  &126467.48(1) &196752.32(1)    &9811.6165(1)    &$-$19623.2331(1) &$-$19623.2331(1)\\
\\
\multicolumn{1}{c}{$^{3}$He}  &$\Psi_{1,\Delta}^{(0)}$  &182851.20(1)  &182851.20(1) &182851.20(1)  &19625.2823(1)    &19625.2823(1)    &19625.2823(1)\\
\multicolumn{1}{c}{}  &$\Psi_{2,\Delta}^{(0)}$  &173343.67(1)  &173343.67(1) &103046.44(1)  &$-$29437.9234(1) &0              &0\\
\multicolumn{1}{c}{}  &$\Psi_{3,\Delta}^{(0)}$  &126478.85(1)  &126478.85(1) &196776.07(1)    &9812.6411(1)    &$-$19625.2823(1) &$-$19625.2823(1)\\
\\
\end{tabular}
\end{ruledtabular}
%\end{table*}
\label{tab:LPer}
\end{sidewaystable}
\endgroup

\begingroup
\squeezetable
\begin{sidewaystable}
%\begin{table*}%%%% VI		
\caption{The additive and nonadditive dispersion coefficients $C_8^{(IJ)}(1,M=\pm 1)$ and $C_8^{(IJ,JK)}(1,M=\pm 1)$
of the He($n_0\,^{\lambda}S$)-He($n_0\,^{\lambda}S$)-He($n_0^{\prime}\,^{\lambda}P$) system
for three different types of the zeroth-order wave functions, where the three atoms form an equilateral triangle, in atomic units.
The numbers in parentheses represent the computational uncertainties.}\label{TabVI}
\centering
\begin{ruledtabular}
\begin{tabular}{lccccccc}
\multicolumn{1}{c}{Atom}  &\multicolumn{1}{c}{State}  & \multicolumn{1}{c}{$C_8^{(12)}(1,M=\pm1)$}  & \multicolumn{1}{c}{$C_8^{(23)}(1,M=\pm1)$}  & \multicolumn{1}{c}{$C_8^{(31)}(1,M=\pm1)$} & \multicolumn{1}{c}{$C_8^{(12,23)}(1,M=\pm1)$}  & \multicolumn{1}{c}{$C_8^{(23,31)}(1,M=\pm1)$}  & \multicolumn{1}{c}{$C_8^{(31,12)}(1,M=\pm1)$}\\
\hline
\\
& & & & \multicolumn{2}{c}{He($1\,^1S$)-He($1\,^1S$)-He($2\,^{1}P$)}\\
\cline{5-6}
\\
\multicolumn{1}{c}{$^{\infty}$He}  &$\Psi_{1,\Delta}^{(0)}$  &3367.9934(1)  &3367.9934(1) &3367.9934(1) &$-$2.707450845(3) &$-$2.707450845(3) &$-$2.707450845(3)\\
\multicolumn{1}{c}{}  &$\Psi_{2,\Delta}^{(0)}$  &2568.9867(1)  &2568.9867(1) &5202.7801(1) &4.061176267(4)    &0              &0\\
\multicolumn{1}{c}{}  &$\Psi_{3,\Delta}^{(0)}$  &4324.8490(1)  &4324.8490(1) &1691.0556(1) &$-$1.353725423(2) &2.707450845(3)    &2.707450845(3)\\
\\
\multicolumn{1}{c}{$^{4}$He}  &$\Psi_{1,\Delta}^{(0)}$  &3370.3905(1) &3370.3905(1) &3370.3905(1)  &$-$2.708545224(3) &$-$2.708545224(3) &$-$2.708545224(3)\\
\multicolumn{1}{c}{}  &$\Psi_{2,\Delta}^{(0)}$  &2570.8046(1) &2570.8046(1) &5206.4442(1)  &4.062817836(4)    &0              &0\\
\multicolumn{1}{c}{}  &$\Psi_{3,\Delta}^{(0)}$  &4327.8977(1) &4327.8977(1) &1692.2581(1)  &$-$1.354272611(1) &2.708545224(3)    &2.708545224(3)\\
\\
\multicolumn{1}{c}{$^{3}$He}  &$\Psi_{1,\Delta}^{(0)}$  &3371.1751(1) &3371.1751(1) &3371.1751(1) &$-$2.708903418(3) &$-$2.708903418(3)  &$-$2.708903418(3)\\
\multicolumn{1}{c}{}  &$\Psi_{2,\Delta}^{(0)}$  &2571.3996(1) &2571.3996(1) &5207.6434(1) &4.063355127(4)    &0               &0\\
\multicolumn{1}{c}{}  &$\Psi_{3,\Delta}^{(0)}$  &4328.8955(1) &4328.8955(1) &1692.6517(1) &$-$1.354451708(1) &2.708903418(3)     &2.708903418(3)\\
\\
& & & & \multicolumn{2}{c}{He($2\,^1S$)-He($2\,^1S$)-He($2\,^{1}P$)}\\
\cline{5-6}
\\
\multicolumn{1}{c}{$^{\infty}$He}  &$\Psi_{1,\Delta}^{(0)}$  &663061(3)  &663061(3) &663061(3) &$-$172448.4071(2) &$-$172448.4071(2) &$-$172448.4071(2)\\
\multicolumn{1}{c}{}  &$\Psi_{2,\Delta}^{(0)}$  &1182611(2)  &1182611(2) &2509977(5) &258672.6108(4)    &0              &0\\
\multicolumn{1}{c}{}  &$\Psi_{3,\Delta}^{(0)}$  &2067523(3)  &2067523(3) &740156(1) &$-$86224.2035(1) &172448.4071(2)    &172448.4071(2)\\
\\
\multicolumn{1}{c}{$^{4}$He}  &$\Psi_{1,\Delta}^{(0)}$  &663476(3) &663476(3) &663476(3)  &$-$172547.3332(3) &$-$172547.3332(3) &$-$172547.3332(3)\\
\multicolumn{1}{c}{}  &$\Psi_{2,\Delta}^{(0)}$  &1183287(2) &1183287(2) &2511492(5)  &258820.9997(4)    &0              &0\\
\multicolumn{1}{c}{}  &$\Psi_{3,\Delta}^{(0)}$  &2068757(4) &2068757(4) &740550(2)  &$-$86273.6665(1) &172547.3332(3)    &172547.3332(3)\\
\\
\multicolumn{1}{c}{$^{3}$He}  &$\Psi_{1,\Delta}^{(0)}$  &663612(3) &663612(3) &663612(3) &$-$172579.7058(3) &$-$172579.7058(3)  &$-$172579.7058(3)\\
\multicolumn{1}{c}{}  &$\Psi_{2,\Delta}^{(0)}$  &1183508(2) &1183508(2) &2511988(5) &258869.5588(5)    &0               &0\\
\multicolumn{1}{c}{}  &$\Psi_{3,\Delta}^{(0)}$  &2069161(4) &2069161(4) &740681(1) &$-$86289.8528(1) &172579.7058(3)     &172579.7058(3)\\
\\
& & & & \multicolumn{2}{c}{He($2\,^3S$)-He($2\,^3S$)-He($2\,^{3}P$)}\\
\cline{5-6}
\\
\multicolumn{1}{c}{$^{\infty}$He}  &$\Psi_{1,\Delta}^{(0)}$  &230457.87(2)  &230457.87(2) &230457.87(2) &$-$52720.6094(1) &$-$52720.6094(1) &$-$52720.6094(1)\\
\multicolumn{1}{c}{}  &$\Psi_{2,\Delta}^{(0)}$  &371010.24(1)  &371010.24(1) &822504.39(2) &79080.9141(1)    &0              &0\\
\multicolumn{1}{c}{}  &$\Psi_{3,\Delta}^{(0)}$  &672006.34(2)  &672006.34(2) &220512.21(2) &$-$26360.3047(1) &52720.6094(1)    &52720.6094(1)\\
\\
\multicolumn{1}{c}{$^{4}$He}  &$\Psi_{1,\Delta}^{(0)}$  &230514.73(1) &230514.73(1) &230514.73(1)  &$-$52737.4391(1) &$-$52737.4391(1) &$-$52737.4391(1)\\
\multicolumn{1}{c}{}  &$\Psi_{2,\Delta}^{(0)}$  &371117.31(2) &371117.31(2) &822695.40(2)  &79106.1586(1)    &0              &0\\
\multicolumn{1}{c}{}  &$\Psi_{3,\Delta}^{(0)}$  &672169.37(2) &672169.37(2) &220591.28(2)  &$-$26368.7195(1) &52737.4391(1)    &52737.4391(1)\\
\\
\multicolumn{1}{c}{$^{3}$He}  &$\Psi_{1,\Delta}^{(0)}$  &230533.34(1) &230533.34(1) &230533.34(1) &$-$52742.9462(1) &$-$52742.9462(1)  &$-$52742.9462(1)\\
\multicolumn{1}{c}{}  &$\Psi_{2,\Delta}^{(0)}$  &371152.34(2) &371152.34(2) &822757.90(2) &79114.4192(1)    &0               &0\\
\multicolumn{1}{c}{}  &$\Psi_{3,\Delta}^{(0)}$  &672222.71(2) &672222.71(2) &220617.15(2) &$-$26371.4731(1) &52742.9462(1)    &52742.9462(1)\\
\\
\end{tabular}
\end{ruledtabular}
%\end{table*}
\label{tab:LPer}
\end{sidewaystable}
\endgroup

%%%%%%%%%%%%%%%%%%%%%%%%%%%%%%%%%%%%%%%%%%%%%%%%%%%%%%%%%%%%%%%%%%%%%%%%%%%%%%%%%%%%%%%%%%%%%%%%%%%%%%%%%%%%%%%%%%%%%%%%%%%%%%%%%%%%%%%%%%%%%%%%%%%%%%%%%%%%%%%%%%%%%%%%%%%%%%%

\begingroup
\squeezetable
\begin{sidewaystable}
%\begin{table*}%%%% VII		
\caption{The additive long-range coefficients $C_3^{(IJ)}(1,M)$ of the He($n_0\,^{\lambda}S$)-He($n_0\,^{\lambda}S$)-He($n_0^{\prime}\,^{\lambda}P$) system
for three different types of the zeroth-order wave functions, where the three atoms form an isosceles right triangle,
in atomic units. The numbers in parentheses represent the computational uncertainties.}
\label{TabVII}
\centering
\begin{ruledtabular}
\begin{tabular}{lccccccc}
\multicolumn{1}{c}{Atom} & \multicolumn{1}{c}{State} & \multicolumn{1}{c}{$C_3^{(12)}(1,M=0)$} & \multicolumn{1}{c}{$C_3^{(23)}(1,M=0)$} & \multicolumn{1}{c}{$C_3^{(31)}(1,M=0)$} & \multicolumn{1}{c}{$C_3^{(12)}(1,M=\pm1)$}   & \multicolumn{1}{c}{$C_3^{(23)}(1,M=\pm1)$}    & \multicolumn{1}{c}{$C_3^{(31)}(1,M=\pm1)$}\\
\hline
\\
& & & & \multicolumn{2}{c}{He($1\,^1S$)-He($1\,^1S$)-He($2\,^{1}P$)}\\
\cline{5-6}
\\
\multicolumn{1}{c}{$^{\infty}$He} & $\Psi_{1,\bot}^{(0)}$  &$-$0.1242304290(1)  &$-$0.0995083237(1)  &$-$0.1242304290(1) &0.0621152144(1)      &0.0497541618(1)    &0.0621152144(1)\\
\multicolumn{1}{c}{} & $\Psi_{2,\bot}^{(0)}$  &0.1242304290(1)     &$-$0.0775472790(1)  &0.1242304290(1)    &$-$0.0621152144(1)   &0.0387736395(1)    &$-$0.0621152144(1)\\
\multicolumn{1}{c}{} & $\Psi_{3,\bot}^{(0)}$  &0                &0.1770556027(1)      &0               &0                 &$-$0.0885278013(1) &0\\
\\
\multicolumn{1}{c}{$^{4}$He} & $\Psi_{1,\bot}^{(0)}$  &$-$0.1242451129(1) &$-$0.0995200855(1) &$-$0.1242451129(1) &0.0621225564(1)      &0.0497600427(1)    &0.0621225564(1)\\
\multicolumn{1}{c}{} & $\Psi_{2,\bot}^{(0)}$  &0.1242451129(1)    &$-$0.0775564450(1) &0.1242451129(1)    &$-$0.0621225564(1)   &0.0387782225(1)    &$-$0.0621225564(1)\\
\multicolumn{1}{c}{} & $\Psi_{3,\bot}^{(0)}$  &0               &0.1770765305(1)     &0               &0                 &$-$0.0885382652(1) &0\\
\\
\multicolumn{1}{c}{$^{3}$He} & $\Psi_{1,\bot}^{(0)}$  &$-$0.1242499188(1) &$-$0.0995239350(1) &$-$0.1242499188(1) &0.0621249594(1)      &0.0497619675(1)     &0.0621249594(1)\\
\multicolumn{1}{c}{} & $\Psi_{2,\bot}^{(0)}$  &0.1242499188(1)    &$-$0.0775594450(1) &0.1242499188(1)    &$-$0.0621249594(1)   &0.0387797225(1)     &$-$0.0621249594(1)\\
\multicolumn{1}{c}{} & $\Psi_{3,\bot}^{(0)}$  &0               &0.1770833800(1)     &0               &0                 &$-$0.0885416900(1)  &0\\
\\
& & & & \multicolumn{2}{c}{He($2\,^1S$)-He($2\,^1S$)-He($2\,^{1}P$)}\\
\cline{5-6}
\\
\multicolumn{1}{c}{$^{\infty}$He} & $\Psi_{1,\bot}^{(0)}$  &$-$5.9673865023(1)  &$-$4.7798645840(1)  &$-$5.9673865023(1) &2.9836932511(1)      &2.3899322920(1)    &2.9836932511(1)\\
\multicolumn{1}{c}{} & $\Psi_{2,\bot}^{(0)}$  &5.9673865023(1)     &$-$3.7249697185(1)  &5.9673865023(1)    &$-$2.9836932511(1)   &1.8624848592(1)    &$-$2.9836932511(1)\\
\multicolumn{1}{c}{} & $\Psi_{3,\bot}^{(0)}$  &0                &8.5048343026(1)      &0               &0                 &$-$4.2524171513(1) &0\\
\\
\multicolumn{1}{c}{$^{4}$He} & $\Psi_{1,\bot}^{(0)}$  &$-$5.9695170418(1) &$-$4.7815711418(1) &$-$5.9695170418(1) &2.9847585209(1)      &2.3907855709(1)    &2.9847585209(1)\\
\multicolumn{1}{c}{} & $\Psi_{2,\bot}^{(0)}$  &5.9695170418(1)    &$-$3.7262996467(1) &5.9695170418(1)    &$-$2.9847585209(1)   &1.8631498233(1)    &$-$2.9847585209(1)\\
\multicolumn{1}{c}{} & $\Psi_{3,\bot}^{(0)}$  &0               &8.5078707885(1)     &0               &0                 &$-$4.2539353942(1) &0\\
\\
\multicolumn{1}{c}{$^{3}$He} & $\Psi_{1,\bot}^{(0)}$  &$-$5.9702141764(1) &$-$4.7821295452(1) &$-$5.9702141764(1) &2.9851070882(1)      &2.3910647726(1)     &2.9851070882(1)\\
\multicolumn{1}{c}{} & $\Psi_{2,\bot}^{(0)}$  &5.9702141764(1)    &$-$3.7267348129(1) &5.9702141764(1)    &$-$2.9851070882(1)   &1.8633674064(1)     &$-$2.9851070882(1)\\
\multicolumn{1}{c}{} & $\Psi_{3,\bot}^{(0)}$  &0               &8.5088643582(1)     &0               &0                 &$-$4.2544321791(1)  &0\\
\\
& & & & \multicolumn{2}{c}{He($2\,^3S$)-He($2\,^3S$)-He($2\,^{3}P$)}\\
\cline{5-6}
\\
\multicolumn{1}{c}{$^{\infty}$He} & $\Psi_{1,\bot}^{(0)}$  &$-$4.4959724006(1) &$-$3.6012648484(1) &$-$4.4959724006(1) &2.2479862003(1)      &1.8006324242(1)    &2.2479862003(1)\\
\multicolumn{1}{c}{} & $\Psi_{2,\bot}^{(0)}$  &4.4959724006(1)     &$-$2.8064817053(1)  &4.4959724006(1)    &$-$2.2479862003(1)   &1.4032408526(1)    &$-$2.2479862003(1)\\
\multicolumn{1}{c}{} & $\Psi_{3,\bot}^{(0)}$  &0                &6.4077465538(1)      &0               &0                 &$-$3.2038732769(1) &0\\
\\
\multicolumn{1}{c}{$^{4}$He} & $\Psi_{1,\bot}^{(0)}$  &$-$4.4969133128(1) &$-$3.6020185173(1) &$-$4.4969133128(1) &2.2484566564(1)     &1.8010092586(1)    &2.2484566564(1)\\
\multicolumn{1}{c}{} & $\Psi_{2,\bot}^{(0)}$  &4.4969133128(1)    &$-$2.8070690428(1) &4.4969133128(1)    &$-$2.2484566564(1)   &1.4035345214(1)    &$-$2.2484566564(1)\\
\multicolumn{1}{c}{} & $\Psi_{3,\bot}^{(0)}$  &0               &6.4090875602(1)     &0               &0                 &$-$3.2045437801(1) &0\\
\\
\multicolumn{1}{c}{$^{3}$He} & $\Psi_{1,\bot}^{(0)}$  &$-$4.4972211537(1) &$-$3.6022650972(1) &$-$4.4972211537(1) &2.2486105768(1)      &1.8011325486(1)     &2.2486105768(1)\\
\multicolumn{1}{c}{} & $\Psi_{2,\bot}^{(0)}$  &4.4972211537(1)    &$-$2.8072612036(1) &4.4972211537(1)    &$-$2.2486105768(1)   &1.4036306018(1)     &$-$2.2486105768(1)\\
\multicolumn{1}{c}{} & $\Psi_{3,\bot}^{(0)}$  &0               &6.4095263009(2)     &0               &0                 &$-$3.2047631504(1)  &0\\
\\
\end{tabular}
\end{ruledtabular}
%\end{table*}
\label{tab:LPer}
\end{sidewaystable}
\endgroup

\begingroup
\squeezetable
\begin{sidewaystable}
%\begin{table*}%%%% VIII		
\caption{The additive and nonadditive dispersion coefficients $C_6^{(IJ)}(1,M=0)$ and $C_6^{(IJ,JK)}(1,M=0)$ of the He($n_0\,^{\lambda}S$)-He($n_0\,^{\lambda}S$)-He($n_0^{\prime}\,^{\lambda}P$) system for three different types of the zeroth-order wave functions, where the three atoms form an isosceles right triangle, in atomic units. The numbers in parentheses represent the computational uncertainties.
}\label{TabVIII}
\centering
\begin{ruledtabular}
\begin{tabular}{lccccccc}
\multicolumn{1}{c}{Atom}  &\multicolumn{1}{c}{State}  & \multicolumn{1}{c}{$C_6^{(12)}(1,M=0)$}  & \multicolumn{1}{c}{$C_6^{(23)}(1,M=0)$}  & \multicolumn{1}{c}{$C_6^{(31)}(1,M=0)$} & \multicolumn{1}{c}{$C_6^{(12,23)}(1,M=0)$}  & \multicolumn{1}{c}{$C_6^{(23,31)}(1,M=0)$}  & \multicolumn{1}{c}{$C_6^{(31,12)}(1,M=0)$}\\
\hline
\\
& & & & \multicolumn{2}{c}{He($1\,^1S$)-He($1\,^1S$)-He($2\,^{1}P$)}\\
\cline{5-6}
\\
\multicolumn{1}{c}{$^{\infty}$He}  &$\Psi_{1,\bot}^{(0)}$  &23.8806359344(2) &18.9858587822(2) &23.8806359344(2)  &0.3350550348(2)     &0.3350550348(2)     &0.2683784091(1)\\
\multicolumn{1}{c}{}  &$\Psi_{2,\bot}^{(0)}$  &25.8144675944(2) &15.1181954622(1) &25.8144675944(2)  &$-$0.3350550348(2)  &$-$0.3350550348(2)  &0.2091484871(1)\\
\multicolumn{1}{c}{}  &$\Psi_{3,\bot}^{(0)}$  &17.0520271221(1) &32.6430764067(3) &17.0520271221(1)  &0              &0              &$-$0.4775268962(1)\\
\\
\multicolumn{1}{c}{$^{4}$He}  &$\Psi_{1,\bot}^{(0)}$  &23.9010221021(1) &19.0020441207(1) &23.9010221021(1)  &0.3352540193(1)     &0.3352540193(1)     &0.2685377950(1)\\
\multicolumn{1}{c}{}  &$\Psi_{2,\bot}^{(0)}$  &25.8365134283(1) &15.1310614683(1) &25.8365134283(1)  &$-$0.3352540193(1)  &$-$0.3352540193(1)  &0.2092726974(1)\\
\multicolumn{1}{c}{}  &$\Psi_{3,\bot}^{(0)}$  &17.0665527946(1) &32.6709827359(1) &17.0665527946(1)  &0              &0              &$-$0.4778104925(1)\\
\\
\multicolumn{1}{c}{$^{3}$He}  &$\Psi_{1,\bot}^{(0)}$  &23.9076953644(1) &19.0073422734(1) &23.9076953644(1)  &0.3353191479(1)     &0.3353191479(1)     &0.2685899629(1)\\
\multicolumn{1}{c}{}  &$\Psi_{2,\bot}^{(0)}$  &25.8437299699(2) &15.1352730627(1) &25.8437299699(2)  &$-$0.3353191479(1)  &$-$0.3353191479(1)  &0.2093133520(1)\\
\multicolumn{1}{c}{}  &$\Psi_{3,\bot}^{(0)}$  &17.0713076681(1) &32.6801176662(2) &17.0713076681(1)  &0              &0              &$-$0.4779033149(2)\\
\\
& & & & \multicolumn{2}{c}{He($2\,^1S$)-He($2\,^1S$)-He($2\,^{1}P$)}\\
\cline{5-6}
\\
\multicolumn{1}{c}{$^{\infty}$He}  &$\Psi_{1,\bot}^{(0)}$  &6083.83(6) &7209.77(4) &6083.83(6)  &1344.7823(1)     &1344.7823(1)     &1077.1679(1)\\
\multicolumn{1}{c}{}  &$\Psi_{2,\bot}^{(0)}$  &5639.00(7) &8099.47(4) &5639.00(7)  &$-$1344.7823(1)  &$-$1344.7823(1)  &839.4418(1)\\
\multicolumn{1}{c}{}  &$\Psi_{3,\bot}^{(0)}$  &7654.62(4) &4068.2(1) &7654.62(4)  &0              &0              &$-$1916.6097(1)\\
\\
\multicolumn{1}{c}{$^{4}$He}  &$\Psi_{1,\bot}^{(0)}$  &6088.31(6) &7214.73(4) &6088.31(6)  &1345.5904(1)     &1345.5904(1)     &1077.8152(1)\\
\multicolumn{1}{c}{}  &$\Psi_{2,\bot}^{(0)}$  &5643.28(7) &8104.81(4) &5643.28(7)  &$-$1345.5904(1)  &$-$1345.5904(1)  &839.9461(1)\\
\multicolumn{1}{c}{}  &$\Psi_{3,\bot}^{(0)}$  &7659.77(4) &4071.8(1) &7659.77(4)  &0              &0              &$-$1917.7614(1)\\
\\
\multicolumn{1}{c}{$^{3}$He}  &$\Psi_{1,\bot}^{(0)}$  &6089.77(6) &7216.37(5) &6089.77(6)  &1345.8548(1)     &1345.8548(1)     &1078.0270(1)\\
\multicolumn{1}{c}{}  &$\Psi_{2,\bot}^{(0)}$  &5644.67(6) &8106.56(4) &5644.67(6)  &$-$1345.8548(1)  &$-$1345.8548(1)  &840.1112(1)\\
\multicolumn{1}{c}{}  &$\Psi_{3,\bot}^{(0)}$  &7661.47(5) &4073.0(1) &7661.47(5)  &0              &0              &$-$1918.1383(1)\\
\\
& & & & \multicolumn{2}{c}{He($2\,^3S$)-He($2\,^3S$)-He($2\,^{3}P$)}\\
\cline{5-6}
\\
\multicolumn{1}{c}{$^{\infty}$He}  &$\Psi_{1,\bot}^{(0)}$  &2259.9491(2) &2481.9271(2) &2259.9491(2)  &396.38743(3)     &396.38743(3)     &317.50553(2)\\
\multicolumn{1}{c}{}  &$\Psi_{2,\bot}^{(0)}$  &2172.2499(2) &2657.3257(3) &2172.2499(2)  &$-$396.38743(3)  &$-$396.38743(3)  &247.43346(1)\\
\multicolumn{1}{c}{}  &$\Psi_{3,\bot}^{(0)}$  &2569.6263(2) &1862.5727(2) &2569.6263(2)  &0              &0              &$-$564.93901(4)\\
\\
\multicolumn{1}{c}{$^{4}$He}  &$\Psi_{1,\bot}^{(0)}$  &2261.3772(2) &2483.6501(2) &2261.3772(2)  &396.72600(2)     &396.72600(2)     &317.77673(2)\\
\multicolumn{1}{c}{}  &$\Psi_{2,\bot}^{(0)}$  &2173.5613(1) &2659.2815(2) &2173.5613(1)  &$-$396.72600(2)  &$-$396.72600(2)  &247.64483(2)\\
\multicolumn{1}{c}{}  &$\Psi_{3,\bot}^{(0)}$  &2571.4658(2) &1863.4728(2) &2571.4658(2) &0              &0              &$-$565.42155(3)\\
\\
\multicolumn{1}{c}{$^{3}$He}  &$\Psi_{1,\bot}^{(0)}$  &2261.8446(2) &2484.2141(2) &2261.8446(2)  &396.83685(2)     &396.83685(2)     &317.86552(2)\\
\multicolumn{1}{c}{}  &$\Psi_{2,\bot}^{(0)}$  &2173.9907(2) &2659.9219(2) &2173.9907(2)  &$-$396.83685(2)  &$-$396.83685(2)  &247.71402(2)\\
\multicolumn{1}{c}{}  &$\Psi_{3,\bot}^{(0)}$  &2572.0680(2) &1863.7672(1) &2572.0680(2)  &0              &0              &$-$565.57955(4)\\
\\
\end{tabular}
\end{ruledtabular}
%\end{table*}
\label{tab:LPer}
\end{sidewaystable}
\endgroup

\begingroup
\squeezetable
\begin{sidewaystable}
%\begin{table*}%%%% IX		
\caption{The additive and nonadditive dispersion coefficients $C_6^{(IJ)}(1,M=\pm 1)$ and $C_6^{(IJ,JK)}(1,M=\pm 1)$
of the He($n_0\,^{\lambda}S$)-He($n_0\,^{\lambda}S$)-He($n_0^{\prime}\,^{\lambda}P$) system
for three different types of the zeroth-order wave functions, where the three atoms form an isosceles right triangle, in atomic units.
The numbers in parentheses represent the computational uncertainties.
}\label{TabIX}
\centering
\begin{ruledtabular}
\begin{tabular}{lccccccc}
\multicolumn{1}{c}{Atom}  &\multicolumn{1}{c}{State}  & \multicolumn{1}{c}{$C_6^{(12)}(1,M=\pm1)$}  & \multicolumn{1}{c}{$C_6^{(23)}(1,M=\pm1)$}  & \multicolumn{1}{c}{$C_6^{(31)}(1,M=\pm1)$} & \multicolumn{1}{c}{$C_6^{(12,23)}(1,M=\pm1)$}  & \multicolumn{1}{c}{$C_6^{(23,31)}(1,M=\pm1)$}  & \multicolumn{1}{c}{$C_6^{(31,12)}(1,M=\pm1)$}\\
\hline
\\
& & & & \multicolumn{2}{c}{He($1\,^1S$)-He($1\,^1S$)-He($2\,^{1}P$)}\\
\cline{5-6}
\\
\multicolumn{1}{c}{$^{\infty}$He}  &$\Psi_{1,\bot}^{(0)}$  &33.31028394(1) &26.35677666(1) &33.31028394(1) &0.0837637587(1)     &0.0837637587(1)     &$-$0.5367568179(2)\\
\multicolumn{1}{c}{}  &$\Psi_{2,\bot}^{(0)}$  &36.05748000(1) &20.86238453(1) &36.05748000(1) &$-$0.0837637587(1)  &$-$0.0837637587(1)  &$-$0.4182969742(2)\\
\multicolumn{1}{c}{}  &$\Psi_{3,\bot}^{(0)}$  &23.60958059(1) &45.75818335(1) &23.60958059(1) &0             &0             &0.9550537923(3)\\
\\
\multicolumn{1}{c}{$^{4}$He}  &$\Psi_{1,\bot}^{(0)}$  &33.33860104(1) &26.37916140(1) &33.33860104(1) &0.0838135048(1)     &0.0838135048(1)     &$-$0.5370755900(2)\\
\multicolumn{1}{c}{}  &$\Psi_{2,\bot}^{(0)}$  &36.08814086(1) &20.88008176(1) &36.08814086(1) &$-$0.0838135048(1)  &$-$0.0838135048(1)  &$-$0.4185453947(2)\\
\multicolumn{1}{c}{}  &$\Psi_{3,\bot}^{(0)}$  &23.62962158(1) &45.79712032(1) &23.62962158(1) &0             &0             &0.9556209849(3)\\
\\
\multicolumn{1}{c}{$^{3}$He}  &$\Psi_{1,\bot}^{(0)}$  &33.34787043(2) &26.38648887(1) &33.34787043(2) &0.0838297869(1)    &0.0838297869(1)     &$-$0.5371799259(1)\\
\multicolumn{1}{c}{}  &$\Psi_{2,\bot}^{(0)}$  &36.09817745(1) &20.88587481(1) &36.09817745(1) &$-$0.0838297869(1)  &$-$0.0838297869(1)  &$-$0.4186267040(2)\\
\multicolumn{1}{c}{}  &$\Psi_{3,\bot}^{(0)}$  &23.63618184(1) &45.80986602(1) &23.63618184(1) &0             &0             &0.9558066300(3)\\
\\
& & & & \multicolumn{2}{c}{He($2\,^1S$)-He($2\,^1S$)-He($2\,^{1}P$)}\\
\cline{5-6}
\\
\multicolumn{1}{c}{$^{\infty}$He}  &$\Psi_{1,\bot}^{(0)}$  &6645.3(2) &7648.6(1) &6645.3(2) &336.19559(1)     &336.19559(1)     &$-$2154.3359(1)\\
\multicolumn{1}{c}{}  &$\Psi_{2,\bot}^{(0)}$  &6248.8(1) &8441.5(2) &6248.8(1) &$-$336.19559(1)  &$-$336.19559(1)  &$-$1678.8836(1)\\
\multicolumn{1}{c}{}  &$\Psi_{3,\bot}^{(0)}$  &8045.0(1) &4849.0(2) &8045.0(1) &0             &0             &3833.2195(1)\\
\\
\multicolumn{1}{c}{$^{4}$He}  &$\Psi_{1,\bot}^{(0)}$  &6650.1(1) &7653.9(1) &6650.1(1) &336.39760(1)     &336.39760(1)     &$-$2155.6304(1)\\
\multicolumn{1}{c}{}  &$\Psi_{2,\bot}^{(0)}$  &6253.7(2) &8447.1(1) &6253.7(2) &$-$336.39760(1)  &$-$336.39760(1)  &$-$1679.8923(1)\\
\multicolumn{1}{c}{}  &$\Psi_{3,\bot}^{(0)}$  &8050.5(1) &4853.3(2) &8050.5(1) &0             &0             &3835.5228(1)\\
\\
\multicolumn{1}{c}{$^{3}$He}  &$\Psi_{1,\bot}^{(0)}$  &6651.9(2) &7655.7(1) &6651.9(2) &336.46372(1)     &336.46372(1)     &$-$2156.0541(1)\\
\multicolumn{1}{c}{}  &$\Psi_{2,\bot}^{(0)}$  &6255.1(1) &8448.9(1) &6255.1(1) &$-$336.46372(1)  &$-$336.46372(1)  &$-$1680.2225(1)\\
\multicolumn{1}{c}{}  &$\Psi_{3,\bot}^{(0)}$  &8052.3(1) &4854.7(2) &8052.3(1) &0             &0             &3836.2766(1)\\
\\
& & & & \multicolumn{2}{c}{He($2\,^3S$)-He($2\,^3S$)-He($2\,^{3}P$)}\\
\cline{5-6}
\\
\multicolumn{1}{c}{$^{\infty}$He}  &$\Psi_{1,\bot}^{(0)}$  &2539.5150(2) &2700.4567(2) &2539.5150(2) &99.096858(7)     &99.096858(7)     &$-$635.01108(5)\\
\multicolumn{1}{c}{}  &$\Psi_{2,\bot}^{(0)}$  &2475.9300(2) &2827.6266(2) &2475.9300(2) &$-$99.096858(7)  &$-$99.096858(7)  &$-$494.86694(3)\\
\multicolumn{1}{c}{}  &$\Psi_{3,\bot}^{(0)}$  &2764.0417(2) &2251.4034(2) &2764.0417(2) &0             &0             &1129.8780(1)\\
\\
\multicolumn{1}{c}{$^{4}$He}  &$\Psi_{1,\bot}^{(0)}$  &2541.0776(2) &2702.2849(2) &2541.0776(2) &99.181504(7)     &99.181504(7)     &$-$635.55347(4)\\
\multicolumn{1}{c}{}  &$\Psi_{2,\bot}^{(0)}$  &2477.3878(2) &2829.6646(2) &2477.3878(2) &$-$99.181504(7)  &$-$99.181504(7)  &$-$495.28966(4)\\
\multicolumn{1}{c}{}  &$\Psi_{3,\bot}^{(0)}$  &2765.9749(3) &2252.4906(2) &2765.9749(3) &0             &0             &1130.8431(1)\\
\\
\multicolumn{1}{c}{$^{3}$He}  &$\Psi_{1,\bot}^{(0)}$  &2541.5891(2) &2702.8833(2) &2541.5891(2) &99.209216(7)    &99.209216(7)     &$-$635.73105(4)\\
\multicolumn{1}{c}{}  &$\Psi_{2,\bot}^{(0)}$  &2477.8649(2) &2830.3319(3) &2477.8649(2) &$-$99.209216(7)  &$-$99.209216(7)  &$-$495.42803(3)\\
\multicolumn{1}{c}{}  &$\Psi_{3,\bot}^{(0)}$  &2766.6075(2) &2252.8465(2) &2766.6075(2) &0             &0             &1131.1591(1)\\
\\
\end{tabular}
\end{ruledtabular}
%\end{table*}
\label{tab:LPer}
\end{sidewaystable}
\endgroup

\begingroup
\squeezetable
\begin{sidewaystable}
%\begin{table*}%%%% X		
\caption{The additive and nonadditive dispersion coefficients $C_8^{(IJ)}(1,M=0)$ and $C_8^{(IJ,JK)}(1,M=0)$
of the He($n_0\,^{\lambda}S$)-He($n_0\,^{\lambda}S$)-He($n_0^{\prime}\,^{\lambda}P$) system for three different types of the zeroth-order wave functions, where the three atoms form
an isosceles right triangle, in atomic units.
The numbers in parentheses represent the computational uncertainties.
}\label{TabX}
\centering
\begin{ruledtabular}
\begin{tabular}{lccccccc}
\multicolumn{1}{c}{Atom}  & \multicolumn{1}{c}{State}  & \multicolumn{1}{c}{$C_8^{(12)}(1,M=0)$}  & \multicolumn{1}{c}{$C_8^{(23)}(1,M=0)$}  & \multicolumn{1}{c}{$C_8^{(31)}(1,M=0)$} & \multicolumn{1}{c}{$C_8^{(12,23)}(1,M=0)$}  & \multicolumn{1}{c}{$C_8^{(23,31)}(1,M=0)$}  & \multicolumn{1}{c}{$C_8^{(31,12)}(1,M=0)$}\\
\hline
\\
& & & & \multicolumn{2}{c}{He($1\,^1S$)-He($1\,^1S$)-He($2\,^{1}P$)}\\
\cline{5-6}
\\
\multicolumn{1}{c}{$^{\infty}$He}  & $\Psi_{1,\bot}^{(0)}$  &288.7440(1)  &229.0510(1)    &288.7440(1)  &1.499466182(2)    &1.499466182(2)     &0\\
\multicolumn{1}{c}{}  & $\Psi_{2,\bot}^{(0)}$  &283.8354(1)  &181.6162(1)    &283.8354(1)  &$-$1.499466182(2) &$-$1.499466182(2)  &0\\
\multicolumn{1}{c}{}  & $\Psi_{3,\bot}^{(0)}$  &195.5657(1)    &357.4779(1)  &195.5657(1)    &0              &0               &0\\
\\
\multicolumn{1}{c}{$^{4}$He}  & $\Psi_{1,\bot}^{(0)}$  &288.7076(1)    &229.0245(1)    &288.7076(1)    &1.500072280(1)    &1.500072280(1)     &0\\
\multicolumn{1}{c}{}  & $\Psi_{2,\bot}^{(0)}$  &283.7814(1)  &181.5973(1)    &283.7814(1)  &$-$1.500072280(1) &$-$1.500072280(1)  &0\\
\multicolumn{1}{c}{}  & $\Psi_{3,\bot}^{(0)}$  &195.5383(1)    &357.4056(1)  &195.5383(1)    &0              &0               &0\\
\\
\multicolumn{1}{c}{$^{3}$He}  & $\Psi_{1,\bot}^{(0)}$  &288.6957(1)    &229.0157(1)    &288.6957(1)    &1.500270660(2)   &1.500270660(2)     &0\\
\multicolumn{1}{c}{}  & $\Psi_{2,\bot}^{(0)}$  &283.7637(1)    &181.5910(1)    &283.7637(1)    &$-$1.500270660(2) &$-$1.500270660(2)  &0\\
\multicolumn{1}{c}{}  & $\Psi_{3,\bot}^{(0)}$  &195.5292(1)    &357.3819(1)  &195.5292(1)    &0              &0               &0\\
\\
& & & & \multicolumn{2}{c}{He($2\,^1S$)-He($2\,^1S$)-He($2\,^{1}P$)}\\
\cline{5-6}
\\
\multicolumn{1}{c}{$^{\infty}$He}  & $\Psi_{1,\bot}^{(0)}$  &583448(2)  &634990(1)   &583448(2)  &95507.0172(1)    &95507.0172(1)     &0\\
\multicolumn{1}{c}{}  & $\Psi_{2,\bot}^{(0)}$  &509746(1)  &675213(1)   &509746(1)  &$-$95507.0172(1) &$-$95507.0172(1)  &0\\
\multicolumn{1}{c}{}  & $\Psi_{3,\bot}^{(0)}$  &636815(1)    &419806(2) &636815(1)    &0              &0               &0\\
\\
\multicolumn{1}{c}{$^{4}$He}  & $\Psi_{1,\bot}^{(0)}$  &583741(2)  &635300(2)   &583741(2)  &95561.8053(1)    &95561.8053(1)     &0\\
\multicolumn{1}{c}{}  & $\Psi_{2,\bot}^{(0)}$  &509930(2)  &675539(1)   &509930(2)  &$-$95561.8053(1) &$-$95561.8053(1)  &0\\
\multicolumn{1}{c}{}  & $\Psi_{3,\bot}^{(0)}$  &637098(2)    &419931(1) &637098(2)    &0              &0               &0\\
\\
\multicolumn{1}{c}{$^{3}$He}  & $\Psi_{1,\bot}^{(0)}$  &583838(1)    &635402(2)    &583838(1)    &95579.7342(1)    &95579.7342(1)     &0\\
\multicolumn{1}{c}{}  & $\Psi_{2,\bot}^{(0)}$  &509991(2)  &675646(1)    &509991(2)  &$-$95579.7342(1) &$-$95579.7342(1)  &0\\
\multicolumn{1}{c}{}  & $\Psi_{3,\bot}^{(0)}$  &637193(1)    &419971(3)  &637193(1)    &0              &0               &0\\
\\
& & & & \multicolumn{2}{c}{He($2\,^3S$)-He($2\,^3S$)-He($2\,^{3}P$)}\\
\cline{5-6}
\\
\multicolumn{1}{c}{$^{\infty}$He}  & $\Psi_{1,\bot}^{(0)}$  &180041.65(1)  &187152.67(1)    &180041.65(1)  &29198.2294(1)    &29198.2294(1)     &0\\
\multicolumn{1}{c}{}  & $\Psi_{2,\bot}^{(0)}$  &129184.04(1)  &192320.01(1)    &129184.04(1)  &$-$29198.2294(1) &$-$29198.2294(1)  &0\\
\multicolumn{1}{c}{}  & $\Psi_{3,\bot}^{(0)}$  &173264.08(1)    &103017.09(1)  &173264.08(1)    &0              &0               &0\\
\\
\multicolumn{1}{c}{$^{4}$He}  & $\Psi_{1,\bot}^{(0)}$  &180081.11(1)    &187205.57(1)    &180081.11(1)    &29207.5501(1)    &29207.5501(1)     &0\\
\multicolumn{1}{c}{}  & $\Psi_{2,\bot}^{(0)}$  &129223.21(1)  &192383.58(1)    &129223.21(1)  &$-$29207.5501(1) &$-$29207.5501(1)  &0\\
\multicolumn{1}{c}{}  & $\Psi_{3,\bot}^{(0)}$  &173324.04(1)    &103039.20(1)  &173324.04(1)    &0              &0               &0\\
\\
\multicolumn{1}{c}{$^{3}$He}  & $\Psi_{1,\bot}^{(0)}$  &180094.02(1)    &187222.89(1)    &180094.02(1)    &29210.6001(1)   &29210.6001(1)     &0\\
\multicolumn{1}{c}{}  & $\Psi_{2,\bot}^{(0)}$  &129236.03(1)    &192404.39(1)    &129236.03(1)    &$-$29210.6001(1) &$-$29210.6001(1)  &0\\
\multicolumn{1}{c}{}  & $\Psi_{3,\bot}^{(0)}$  &173343.67(1)    &103046.44(1)  &173343.67(1)    &0              &0               &0\\
\\
\end{tabular}
\end{ruledtabular}
%\end{table*}
\label{tab:LPer}
\end{sidewaystable}
\endgroup

\begingroup
\squeezetable
%\begin{table*}%%%% XI
\begin{sidewaystable}
\caption{The additive and nonadditive dispersion coefficients $C_8^{(IJ)}(1,M=\pm 1)$ and $C_8^{(IJ,JK)}(1,M=\pm 1)$
of the He($n_0\,^{\lambda}S$)-He($n_0\,^{\lambda}S$)-He($n_0^{\prime}\,^{\lambda}P$) system
for three different types of the zeroth-order wave functions, where the three atoms form an isosceles right, in atomic units.
The numbers in parentheses represent the computational uncertainties.
}\label{TabXI}
\centering
\begin{ruledtabular}
\begin{tabular}{lccccccc}
\multicolumn{1}{c}{Atom}  & \multicolumn{1}{c}{State}  & \multicolumn{1}{c}{$C_8^{(12)}(1,M=\pm1)$}  & \multicolumn{1}{c}{$C_8^{(23)}(1,M=\pm1)$}  & \multicolumn{1}{c}{$C_8^{(31)}(1,M=\pm1)$} & \multicolumn{1}{c}{$C_8^{(12,23)}(1,M=\pm1)$}  & \multicolumn{1}{c}{$C_8^{(23,31)}(1,M=\pm1)$}  & \multicolumn{1}{c}{$C_8^{(31,12)}(1,M=\pm1)$}\\
\hline
\\
& & & & \multicolumn{2}{c}{He($1\,^1S$)-He($1\,^1S$)-He($2\,^{1}P$)}\\
\cline{5-6}
\\
\multicolumn{1}{c}{$^{\infty}$He}  & $\Psi_{1,\bot}^{(0)}$  &3632.5979(1)  &2841.5223(1)  &3632.5979(1)  &$-$1.686899453(1) &$-$1.686899453(1) &0\\
\multicolumn{1}{c}{}  & $\Psi_{2,\bot}^{(0)}$  &4060.2445(1)  &2217.5267(1)  &4060.2445(1)  &1.686899453(1)    &1.686899453(1)    &0\\
\multicolumn{1}{c}{}  & $\Psi_{3,\bot}^{(0)}$  &2568.9867(1) &5202.7801(1)   &2568.9867(1)  &0              &0              &0\\
\\
\multicolumn{1}{c}{$^{4}$He}  & $\Psi_{1,\bot}^{(0)}$  &3635.1831(1)  &2843.5444(1) &3635.1831(1)  &$-$1.687581315(1) &$-$1.687581315(1) &0\\
\multicolumn{1}{c}{}  & $\Psi_{2,\bot}^{(0)}$  &4063.1051(1)  &2219.1043(1) &4063.1051(1)  &1.687581315(1)    &1.687581315(1)    &0\\
\multicolumn{1}{c}{}  & $\Psi_{3,\bot}^{(0)}$  &2570.8046(1)  &5206.4442(1) &2570.8046(1)  &0              &0              &0\\
\\
\multicolumn{1}{c}{$^{3}$He}  & $\Psi_{1,\bot}^{(0)}$  &3636.0293(1)  &2844.2062(1) &3636.0293(1)  &$-$1.687804492(2) &$-$1.687804492(2) &0\\
\multicolumn{1}{c}{}  & $\Psi_{2,\bot}^{(0)}$  &4064.0413(1)  &2219.6206(1) &4064.0413(1)  &1.687804492(2)    &1.687804492(2)    &0\\
\multicolumn{1}{c}{}  & $\Psi_{3,\bot}^{(0)}$  &2571.3997(1)  &5207.6434(1) &2571.3997(1)  &0              &0              &0\\
\\
& & & & \multicolumn{2}{c}{He($2\,^1S$)-He($2\,^1S$)-He($2\,^{1}P$)}\\
\cline{5-6}
\\
\multicolumn{1}{c}{$^{\infty}$He}  & $\Psi_{1,\bot}^{(0)}$  &667645(3) &687265(2) &667645(3) &$-$107445.3945(2)    &$-$107445.3945(2)     &0\\
\multicolumn{1}{c}{}  & $\Psi_{2,\bot}^{(0)}$  &2062939(3) &715953(1) &2062939(3) &107445.3945(2) &107445.3945(2)  &0\\
\multicolumn{1}{c}{}  & $\Psi_{3,\bot}^{(0)}$  &1182611(2) &2509977(5) &1182611(2) &0              &0               &0\\
\\
\multicolumn{1}{c}{$^{4}$He}  & $\Psi_{1,\bot}^{(0)}$  &668073(3) &687674(2) &668073(3) &$-$107507.0311(2)    &$-$107507.0311(2)     &0\\
\multicolumn{1}{c}{}  & $\Psi_{2,\bot}^{(0)}$  &2064160(4) &716353(2) &2064160(4) &107507.0311(2) &107507.0311(2)  &0\\
\multicolumn{1}{c}{}  & $\Psi_{3,\bot}^{(0)}$  &1183287(2) &2511492(5) &1183287(2) &0              &0               &0\\
\\
\multicolumn{1}{c}{$^{3}$He}  & $\Psi_{1,\bot}^{(0)}$  &668211(4) &687808(2) &668211(4) &$-$107527.2011(2)    &$-$107527.2011(2)     &0\\
\multicolumn{1}{c}{}  & $\Psi_{2,\bot}^{(0)}$  &2064560(4) &716485(2) &2064560(4) &107527.2011(2) &107527.2011(2)  &0\\
\multicolumn{1}{c}{}  & $\Psi_{3,\bot}^{(0)}$  &1183508(2) &2511988(5) &1183508(2) &0              &0               &0\\
\\
& & & & \multicolumn{2}{c}{He($2\,^3S$)-He($2\,^3S$)-He($2\,^{3}P$)}\\
\cline{5-6}
\\
\multicolumn{1}{c}{$^{\infty}$He}  & $\Psi_{1,\bot}^{(0)}$  &237067.32(2)  &227335.45(2)  &237067.32(2)  &$-$32848.0080(1) &$-$32848.0080(1) &0\\
\multicolumn{1}{c}{}  & $\Psi_{2,\bot}^{(0)}$  &665396.89(2)  &223634.63(2)  &665396.89(2)  &32848.0080(1)    &32848.0080(1)    &0\\
\multicolumn{1}{c}{}  & $\Psi_{3,\bot}^{(0)}$  &371010.24(1) &822504.39(2)   &371010.24(1)  &0              &0              &0\\
\\
\multicolumn{1}{c}{$^{4}$He}  & $\Psi_{1,\bot}^{(0)}$  &237122.07(2)  &227399.29(2) &237122.07(2)  &$-$32858.4939(1) &$-$32858.4939(1) &0\\
\multicolumn{1}{c}{}  & $\Psi_{2,\bot}^{(0)}$  &665562.05(2)  &223706.73(2) &665562.05(2)  &32858.4939(1)    &32858.4939(1)    &0\\
\multicolumn{1}{c}{}  & $\Psi_{3,\bot}^{(0)}$  &371117.31(2)  &822695.40(2) &371117.31(2)  &0              &0              &0\\
\\
\multicolumn{1}{c}{$^{3}$He}  & $\Psi_{1,\bot}^{(0)}$  &237139.98(2)  &227420.17(1) &237139.98(2)  &$-$32861.9252(1) &$-$32861.9252(1) &0\\
\multicolumn{1}{c}{}  & $\Psi_{2,\bot}^{(0)}$  &665616.07(1)  &223730.31(1) &665616.07(1)  &32861.9252(1)    &32861.9252(1)    &0\\
\multicolumn{1}{c}{}  & $\Psi_{3,\bot}^{(0)}$  &371152.34(2)  &822757.90(2) &371152.34(2)  &0              &0              &0\\
\\
\end{tabular}
\end{ruledtabular}
%\end{table*}
\label{tab:LPer}
\end{sidewaystable}
\endgroup

%====================== Tables for a straight line=======================================================
\begingroup
\squeezetable
%\begin{table*}%%%% XII
\begin{sidewaystable}
\caption{The additive long-range coefficients $C_3^{(IJ)}(1,M)$ of the He($n_0\,^{\lambda}S$)-He($n_0\,^{\lambda}S$)-He($n_0^{\prime}\,^{\lambda}P$) system
for three different types of the zeroth-order wave functions, where the three atoms form a straight line,
in atomic units. The numbers in parentheses represent the computational uncertainties.
}\label{TabXII}
\centering
\begin{ruledtabular}
\begin{tabular}{lccccccc}
\multicolumn{1}{c}{Atom}  & \multicolumn{1}{c}{State}  & \multicolumn{1}{c}{$C_3^{(12)}(1,M=0)$} & \multicolumn{1}{c}{$C_3^{(23)}(1,M=0)$} & \multicolumn{1}{c}{$C_3^{(31)}(1,M=0)$}  & \multicolumn{1}{c}{$C_3^{(12)}(1,M=\pm1)$}   & \multicolumn{1}{c}{$C_3^{(23)}(1,M=\pm1)$}    & \multicolumn{1}{c}{$C_3^{(31)}(1,M=\pm1)$}\\
\hline
\\
& & & & \multicolumn{2}{c}{He($1\,^1S$)-He($1\,^1S$)-He($2\,^{1}P$)}\\
\cline{5-6}
\\
\multicolumn{1}{c}{$^{\infty}$He}  & $\Psi_{1,\text{--}}^{(0)}$  &$-$0.1250751332(1) &$-$0.0924363992(1)  &$-$0.1250751332(1) &0.0625375666(1)     &0.0462181996(1)    &0.0625375666(1)\\
\multicolumn{1}{c}{}  & $\Psi_{2,\text{--}}^{(0)}$  &0.1250751332(1)    &$-$0.0846192034(1)  &0.1250751332(1)    &$-$0.0625375666(1)  &0.0423096017(1)    &$-$0.0625375666(1)\\
\multicolumn{1}{c}{}  & $\Psi_{3,\text{--}}^{(0)}$  &0               &0.1770556027(1)      &0               &0                &$-$0.0885278013(1) &0\\
\\
\multicolumn{1}{c}{$^{4}$He}  & $\Psi_{1,\text{--}}^{(0)}$  &$-$0.1250899170(1)  &$-$0.0924473251(1)  &$-$0.1250899170(1)  &0.0625449585(1)     &0.0462236625(1)     &0.0625449585(1)\\
\multicolumn{1}{c}{}  & $\Psi_{2,\text{--}}^{(0)}$  &0.1250899170(1)     &$-$0.0846292053(1)  &0.1250899170(1)     &$-$0.0625449585(1)  &0.0423146026(1)     &$-$0.0625449585(1)\\
\multicolumn{1}{c}{}  & $\Psi_{3,\text{--}}^{(0)}$  &0               &0.1770765305(1)      &0               &0                &$-$0.0885382652(1)  &0\\
\\
\multicolumn{1}{c}{$^{3}$He}  & $\Psi_{1,\text{--}}^{(0)}$  &$-$0.1250947556(1)  &$-$0.0924509011(1)  &$-$0.1250947556(1)  &0.0625473778(1)     &0.0462254505(1)     &0.0625473778(1)\\
\multicolumn{1}{c}{}  & $\Psi_{2,\text{--}}^{(0)}$  &0.1250947556(1)     &$-$0.0846324789(1)  &0.1250947556(1)     &$-$0.0625473778(1)  &0.0423162394(1)     &$-$0.0625473778(1)\\
\multicolumn{1}{c}{}  & $\Psi_{3,\text{--}}^{(0)}$  &0               &0.1770833800(1)      &0               &0                &$-$0.0885416900(1)  &0\\
\\
& & & & \multicolumn{2}{c}{He($2\,^1S$)-He($2\,^1S$)-He($2\,^{1}P$)}\\
\cline{5-6}
\\
\multicolumn{1}{c}{$^{\infty}$He}  & $\Psi_{1,\text{--}}^{(0)}$  &$-$6.0079617201(1) &$-$4.4401659550(1)  &$-$6.0079617201(1) &3.0039808600(1)    &2.2200829775(1)    &3.0039808600(1)\\
\multicolumn{1}{c}{}  & $\Psi_{2,\text{--}}^{(0)}$  &6.0079617201(1)    &$-$4.0646683475(1)  &6.0079617201(1)    &$-$3.0039808600(1)  &2.0323341737(1)    &$-$3.0039808600(1)\\
\multicolumn{1}{c}{}  & $\Psi_{3,\text{--}}^{(0)}$  &0               &8.5048343026(1)      &0               &0                &$-$4.2524171513(1) &0\\
\\
\multicolumn{1}{c}{$^{4}$He}  & $\Psi_{1,\text{--}}^{(0)}$  &$-$6.0101067462(1)  &$-$4.4417512301(1)  &$-$6.0101067462(1)  &3.0050533731(1)     &2.2208756150(1)     &3.0050533731(1)\\
\multicolumn{1}{c}{}  & $\Psi_{2,\text{--}}^{(0)}$  &6.0101067462(1)     &$-$4.0661195584(1)  &6.0101067462(1)     &$-$3.0050533731(1) &2.0330597792(1)     &$-$3.0050533731(1)\\
\multicolumn{1}{c}{}  & $\Psi_{3,\text{--}}^{(0)}$  &0               &8.5078707885(1)      &0               &0                &$-$4.2539353942(1)  &0\\
\\
\multicolumn{1}{c}{$^{3}$He}  & $\Psi_{1,\text{--}}^{(0)}$  &$-$6.0108086209(1)  &$-$4.4422699485(1)  &$-$6.0108086209(1)  &3.0054043104(1)     &2.2211349742(1)     &3.0054043104(1)\\
\multicolumn{1}{c}{}  & $\Psi_{2,\text{--}}^{(0)}$  &6.0108086209(1)     &$-$4.0665944097(1)  &6.0108086209(1)     &$-$3.0054043104(1)  &2.0332972048(1)     &$-$3.0054043104(1)\\
\multicolumn{1}{c}{}  & $\Psi_{3,\text{--}}^{(0)}$  &0               &8.5088643582(1)      &0               &0                &$-$4.2544321791(1)  &0\\
\\
& & & & \multicolumn{2}{c}{He($2\,^3S$)-He($2\,^3S$)-He($2\,^{3}P$)}\\
\cline{5-6}
\\
\multicolumn{1}{c}{$^{\infty}$He}  & $\Psi_{1,\text{--}}^{(0)}$  &$-$4.5265427447(1) &$-$3.3453277377(1)  &$-$4.5265427447(1) &2.2632713723(1)     &1.6726638688(1)    &2.2632713723(1)\\
\multicolumn{1}{c}{}  & $\Psi_{2,\text{--}}^{(0)}$  &4.5265427447(1)    &$-$3.0624188161(1)  &4.5265427447(1)    &$-$2.2632713723(1)  &1.5312094080(1)    &$-$2.2632713723(1)\\
\multicolumn{1}{c}{}  & $\Psi_{3,\text{--}}^{(0)}$  &0               &6.4077465538(1)      &0               &0                &$-$3.2038732769(1) &0\\
\\
\multicolumn{1}{c}{$^{4}$He}  & $\Psi_{1,\text{--}}^{(0)}$  &$-$4.5274900548(1)  &$-$3.3460278443(1)  &$-$4.5274900548(1)  &2.2637450274(1)     &1.6730139221(1)     &2.2637450274(1)\\
\multicolumn{1}{c}{}  & $\Psi_{2,\text{--}}^{(0)}$  &4.5274900548(1)     &$-$3.0630597158(1)  &4.5274900548(1)     &$-$2.2637450274(1)  &1.5315298579(1)     &$-$2.2637450274(1)\\
\multicolumn{1}{c}{}  & $\Psi_{3,\text{--}}^{(0)}$  &0               &6.4090875602(1)      &0               &0                &$-$3.2045437801(1)  &0\\
\\
\multicolumn{1}{c}{$^{3}$He}  & $\Psi_{1,\text{--}}^{(0)}$  &$-$4.5277999888(1)  &$-$3.3462569001(1)  &$-$4.5277999888(1)  &2.2638999944(1)     &1.6731284500(1)     &2.2638999944(1)\\
\multicolumn{1}{c}{}  & $\Psi_{2,\text{--}}^{(0)}$  &4.5277999888(1)     &$-$3.0632694008(1)  &4.5277999888(1)     &$-$2.2638999944(1)  &1.5316347004(1)     &$-$2.2638999944(1)\\
\multicolumn{1}{c}{}  & $\Psi_{3,\text{--}}^{(0)}$  &0               &6.4095263009(1)      &0               &0                &$-$3.2047631504(1)  &0\\
\\
\end{tabular}
\end{ruledtabular}
%\end{table*}
\label{tab:LPer}
\end{sidewaystable}
\endgroup

\begingroup
\squeezetable
\begin{sidewaystable}
%\begin{table*}%%%% XIII		
\caption{The additive and nonadditive dispersion coefficients $C_6^{(IJ)}(1,M=0)$ and $C_6^{(IJ,JK)}(1,M=0)$ of the He($n_0\,^{\lambda}S$)-He($n_0\,^{\lambda}S$)-He($n_0^{\prime}\,^{\lambda}P$) system for three different types of the zeroth-order wave functions, where the three atoms form a
straight line, in atomic units. The numbers in parentheses represent the computational uncertainties.
}\label{TabXIII}
\centering
\begin{ruledtabular}
\begin{tabular}{lccccccc}
\multicolumn{1}{c}{Atom}  & \multicolumn{1}{c}{State}  & \multicolumn{1}{c}{$C_6^{(12)}(1,M=0)$}  & \multicolumn{1}{c}{$C_6^{(23)}(1,M=0)$}  & \multicolumn{1}{c}{$C_6^{(31)}(1,M=0)$} & \multicolumn{1}{c}{$C_6^{(12,23)}(1,M=0)$}  & \multicolumn{1}{c}{$C_6^{(23,31)}(1,M=0)$}  & \multicolumn{1}{c}{$C_6^{(31,12)}(1,M=0)$}\\
\hline
\\
& & & & \multicolumn{2}{c}{He($1\,^1S$)-He($1\,^1S$)-He($2\,^{1}P$)}\\
\cline{5-6}
\\
\multicolumn{1}{c}{$^{\infty}$He}  & $\Psi_{1,\text{--}}^{(0)}$  &24.5033709436(3) &17.7403887642(2) &24.5033709436(3)  &0.3373332400(2)     &0.3373332400(2)     &0.249305111(1)\\
\multicolumn{1}{c}{}  & $\Psi_{2,\text{--}}^{(0)}$  &25.1917325854(2) &16.3636654802(1) &25.1917325854(2)  &$-$0.3373332400(2)  &$-$0.3373332400(2)  &0.228221784(1)\\
\multicolumn{1}{c}{}  & $\Psi_{3,\text{--}}^{(0)}$  &17.0520271221(1) &32.6430764067(3) &17.0520271221(1)  &0              &0              &$-$0.477526896(1)\\
\\
\multicolumn{1}{c}{$^{4}$He}  & $\Psi_{1,\text{--}}^{(0)}$  &24.5242915590(1) &17.7555052069(1) &24.5242915590(1)  &0.3375335775(1)     &0.3375335775(1)     &0.249453170(1)\\
\multicolumn{1}{c}{}  & $\Psi_{2,\text{--}}^{(0)}$  &25.2132439714(1) &16.3776003821(1) &25.2132439714(1)  &$-$0.3375335775(1)  &$-$0.3375335775(1)  &0.228357321(1)\\
\multicolumn{1}{c}{}  & $\Psi_{3,\text{--}}^{(0)}$  &17.0665527946(1) &32.6709827359(1) &17.0665527946(1)  &0              &0              &$-$0.477810492(1)\\
\\
\multicolumn{1}{c}{$^{3}$He}  & $\Psi_{1,\text{--}}^{(0)}$  &24.5311397688(1) &17.7604534647(1) &24.5311397688(1)  &0.3375991489(1)     &0.3375991489(1)     &0.249501629(1)\\
\multicolumn{1}{c}{}  & $\Psi_{2,\text{--}}^{(0)}$  &25.2202855655(2) &16.3821618715(1) &25.2202855655(2)  &$-$0.3375991489(1)  &$-$0.3375991489(1)  &0.228401683(1)\\
\multicolumn{1}{c}{}  & $\Psi_{3,\text{--}}^{(0)}$  &17.0713076681(1) &32.6801176662(2) &17.0713076681(1)  &0              &0              &$-$0.477903314(1)\\
\\
& & & & \multicolumn{2}{c}{He($2\,^1S$)-He($2\,^1S$)-He($2\,^{1}P$)}\\
\cline{5-6}
\\
\multicolumn{1}{c}{$^{\infty}$He}  & $\Psi_{1,\text{--}}^{(0)}$  &5940.5(1) &7496.2(1) &5940.5(1) &1353.9262(1)     &1353.9262(1)     &1000.6150(1)\\
\multicolumn{1}{c}{}  & $\Psi_{2,\text{--}}^{(0)}$  &5782.2(1) &7812.9(1) &5782.2(1)  &$-$1353.9262(1)  &$-$1353.9262(1)  &915.99470(1)\\
\multicolumn{1}{c}{}  & $\Psi_{3,\text{--}}^{(0)}$  &7654.6(1) &4068.2(1) &7654.6(1)  &0              &0              &$-$1916.6097(1)\\
\\
\multicolumn{1}{c}{$^{4}$He}  & $\Psi_{1,\text{--}}^{(0)}$  &5945.0(1) &7501.3(1) &5945.0(1)  &1354.7397(1)     &1354.7397(1)     &1001.2163(1)\\
\multicolumn{1}{c}{}  & $\Psi_{2,\text{--}}^{(0)}$  &5786.5(1) &7818.1(1) &5786.5(1)  &$-$1354.7397(1)  &$-$1354.7397(1)  &916.54509(1)\\
\multicolumn{1}{c}{}  & $\Psi_{3,\text{--}}^{(0)}$  &7659.7(1) &4071.8(1) &7659.7(1)  &0              &0              &$-$1917.7614(1)\\
\\
\multicolumn{1}{c}{$^{3}$He}  & $\Psi_{1,\text{--}}^{(0)}$  &5946.4(1) &7503.0(1) &5946.4(1)  &1355.0060(1)     &1355.0060(1)     &1001.4131(1)\\
\multicolumn{1}{c}{}  & $\Psi_{2,\text{--}}^{(0)}$  &5788.0(1) &7819.8(1) &5788.0(1)  &$-$1355.0060(1)  &$-$1355.0060(1)  &916.72523(1)\\
\multicolumn{1}{c}{}  & $\Psi_{3,\text{--}}^{(0)}$  &7661.4(1) &4073.0(1) &7661.4(1)  &0              &0              &$-$1918.1383(1)\\
\\
& & & & \multicolumn{2}{c}{He($2\,^3S$)-He($2\,^3S$)-He($2\,^{3}P$)}\\
\cline{5-6}
\\
\multicolumn{1}{c}{$^{\infty}$He}  & $\Psi_{1,\text{--}}^{(0)}$  &2231.7081(2) &2538.4092(2) &2231.7081(2)  &399.08267(3)     &399.08267(3)     &294.94085(3)\\
\multicolumn{1}{c}{}  & $\Psi_{2,\text{--}}^{(0)}$  &2200.4909(2) &2600.8435(2) &2200.4909(2)  &$-$399.08267(3)  &$-$399.08267(3)  &269.99817(2)\\
\multicolumn{1}{c}{}  & $\Psi_{3,\text{--}}^{(0)}$  &2569.6263(2) &1862.5727(2) &2569.6263(2)  &0              &0              &$-$564.93901(4)\\
\\
\multicolumn{1}{c}{$^{4}$He}  & $\Psi_{1,\text{--}}^{(0)}$  &2233.0985(1) &2540.2072(2) &2233.0985(1)  &399.42355(3)     &399.42355(3)     &295.19278(3)\\
\multicolumn{1}{c}{}  & $\Psi_{2,\text{--}}^{(0)}$  &2201.8400(2) &2602.7245(2) &2201.8400(2)  &$-$399.42355(3)  &$-$399.42355(3)  &270.22878(1)\\
\multicolumn{1}{c}{}  & $\Psi_{3,\text{--}}^{(0)}$  &2571.4658(2) &1863.4728(2) &2571.4658(2)  &0              &0              &$-$565.42155(3)\\
\\
\multicolumn{1}{c}{$^{3}$He}  & $\Psi_{1,\text{--}}^{(0)}$  &2233.5538(2) &2540.7957(2) &2233.5538(2)  &399.53514(2)    &399.53514(2)     &295.27524(2)\\
\multicolumn{1}{c}{}  & $\Psi_{2,\text{--}}^{(0)}$  &2202.2814(1) &2603.3402(2) &2202.2814(1)  &$-$399.53514(2)  &$-$399.53514(2)  &270.30430(2)\\
\multicolumn{1}{c}{}  & $\Psi_{3,\text{--}}^{(0)}$  &2572.0680(2) &1863.7672(1) &2572.0680(2)  &0              &0              &$-$565.57955(4)\\
\\
\end{tabular}
\end{ruledtabular}
%\end{table*}
\label{tab:LPer}
\end{sidewaystable}
\endgroup

\begingroup
\squeezetable
\begin{sidewaystable}
%\begin{table*}%%%% XIV		
\caption{The additive and nonadditive dispersion coefficients $C_6^{(IJ)}(1,M=\pm 1)$ and $C_6^{(IJ,JK)}(1,M=\pm 1)$
of the He($n_0\,^{\lambda}S$)-He($n_0\,^{\lambda}S$)-He($n_0^{\prime}\,^{\lambda}P$) system
for three different types of the zeroth-order wave functions, where the three atoms form a straight line, in atomic units.
The numbers in parentheses represent the computational uncertainties.
}\label{TabXIV}
\centering
\begin{ruledtabular}
\begin{tabular}{lccccccc}
\multicolumn{1}{c}{Atom}  & \multicolumn{1}{c}{State}  & \multicolumn{1}{c}{$C_6^{(12)}(1,M=\pm1)$}  & \multicolumn{1}{c}{$C_6^{(23)}(1,M=\pm1)$}  & \multicolumn{1}{c}{$C_6^{(31)}(1,M=\pm1)$} & \multicolumn{1}{c}{$C_6^{(12,23)}(1,M=\pm1)$}  & \multicolumn{1}{c}{$C_6^{(23,31)}(1,M=\pm1)$}  & \multicolumn{1}{c}{$C_6^{(31,12)}(1,M=\pm1)$}\\
\hline
\\
& & & & \multicolumn{2}{c}{He($1\,^1S$)-He($1\,^1S$)-He($2\,^{1}P$)}\\
\cline{5-6}
\\
\multicolumn{1}{c}{$^{\infty}$He}  & $\Psi_{1,\text{--}}^{(0)}$  &34.1949396258(1)  &24.5874652972(1) &34.1949396258(1)  &0.8433331004(3)    &0.8433331004(3)     &0.623262779(1)\\
\multicolumn{1}{c}{}  & $\Psi_{2,\text{--}}^{(0)}$  &35.1728243264(1)  &22.6316958960(1) &35.1728243264(1)  &$-$0.8433331004(3) &$-$0.8433331004(3)  &0.570554460(1)\\
\multicolumn{1}{c}{}  & $\Psi_{3,\text{--}}^{(0)}$  &23.6095805966(1)  &45.758183355(1) &23.6095805966(1)  &0            &0             &$-$1.193817240(1)\\
\\
\multicolumn{1}{c}{$^{4}$He}  & $\Psi_{1,\text{--}}^{(0)}$  &34.2240114649(2) &24.6083405653(2) &34.2240114649(2)  &0.8438339437(3)     &0.8438339437(3)     &0.623632926(1)\\
\multicolumn{1}{c}{}  & $\Psi_{2,\text{--}}^{(0)}$  &35.2027304431(3) &22.6509026088(1) &35.2027304431(3)  &$-$0.8438339437(3)  &$-$0.8438339437(3)  &0.570893304(1)\\
\multicolumn{1}{c}{}  & $\Psi_{3,\text{--}}^{(0)}$  &23.6296215870(2) &45.797120320(1) &23.6296215870(2)  &0             &0             &$-$1.194526231(1)\\
\\
\multicolumn{1}{c}{$^{3}$He}  & $\Psi_{1,\text{--}}^{(0)}$  &34.2335278990(2) &24.6151739190(1) &34.2335278990(2)  &0.8439978723(3)     &0.8439978723(3)     &0.623754077(1)\\
\multicolumn{1}{c}{}  & $\Psi_{2,\text{--}}^{(0)}$  &35.2125199713(2) &22.6571897745(1) &35.2125199713(2)  &$-$0.8439978723(3)  &$-$0.8439978723(3)  &0.571004208(2)\\
\multicolumn{1}{c}{}  & $\Psi_{3,\text{--}}^{(0)}$  &23.6361818467(2) &45.809866023(1) &23.6361818467(2)  &0             &0             &$-$1.194758287(1)\\
\\
& & & & \multicolumn{2}{c}{He($2\,^1S$)-He($2\,^1S$)-He($2\,^{1}P$)}\\
\cline{5-6}
\\
\multicolumn{1}{c}{$^{\infty}$He}  & $\Psi_{1,\text{--}}^{(0)}$  &6517.6(2)  &7903.9(1) &6517.6(2)  &3384.8155(1)    &3384.8155(1)     &2501.5377(1)\\
\multicolumn{1}{c}{}  & $\Psi_{2,\text{--}}^{(0)}$  &6376.5(2)  &8186.1(1) &6376.5(2)  &$-$3384.8155(1) &$-$3384.8155(1)  &2289.9867(1)\\
\multicolumn{1}{c}{}  & $\Psi_{3,\text{--}}^{(0)}$  &8045.0(1)  &4849.0(1) &8045.0(1)  &0            &0             &$-$4791.5244(1)\\
\\
\multicolumn{1}{c}{$^{4}$He}  & $\Psi_{1,\text{--}}^{(0)}$  &6522.5(2) &7909.3(1) &6522.5(2)  &3386.8493(1)     &3386.8493(1)     &2503.0408(1)\\
\multicolumn{1}{c}{}  & $\Psi_{2,\text{--}}^{(0)}$  &6381.4(2) &8191.6(1) &6381.4(2)  &$-$3386.8493(1)  &$-$3386.8493(1) &2291.3627(1)\\
\multicolumn{1}{c}{}  & $\Psi_{3,\text{--}}^{(0)}$  &8050.5(1) &4853.3(2) &8050.5(1)  &0             &0             &$-$4794.4035(1)\\
\\
\multicolumn{1}{c}{$^{3}$He}  & $\Psi_{1,\text{--}}^{(0)}$  &6524.0(1) &7911.1(1) &6524.0(1)  &3387.5150(1)     &3387.5150(1)     &2503.5327(1)\\
\multicolumn{1}{c}{}  & $\Psi_{2,\text{--}}^{(0)}$  &6382.9(2) &8193.5(1) &6382.9(2) &$-$3387.5150(1)  &$-$3387.5150(1)  &2291.8130(1)\\
\multicolumn{1}{c}{}  & $\Psi_{3,\text{--}}^{(0)}$  &8052.3(1) &4854.7(2) &8052.3(1)  &0             &0             &$-$4795.3458(1)\\
\\
& & & & \multicolumn{2}{c}{He($2\,^3S$)-He($2\,^3S$)-He($2\,^{3}P$)}\\
\cline{5-6}
\\
\multicolumn{1}{c}{$^{\infty}$He}  & $\Psi_{1,\text{--}}^{(0)}$  &2519.0393(2)  &2741.4083(3) &2519.0393(2)  &997.7066(1)    &997.7066(1)     &737.3521(1)\\
\multicolumn{1}{c}{}  & $\Psi_{2,\text{--}}^{(0)}$  &2496.4058(2)  &2786.6752(2) &2496.4058(2)  &$-$997.7066(1) &$-$997.7066(1)  &674.9954(1)\\
\multicolumn{1}{c}{}  & $\Psi_{3,\text{--}}^{(0)}$  &2764.0417(2)  &2251.4034(2) &2764.0417(2)  &0            &0             &$-$1412.3475(1)\\
\\
\multicolumn{1}{c}{$^{4}$He}  & $\Psi_{1,\text{--}}^{(0)}$  &2520.5681(2) &2743.3039(2) &2520.5681(2)  &998.5588(1)     &998.5588(1)     &737.9819(1)\\
\multicolumn{1}{c}{}  & $\Psi_{2,\text{--}}^{(0)}$  &2497.8973(2) &2788.6456(2) &2497.8973(2)  &$-$998.5588(1)  &$-$998.5588(1)  &675.5719(1)\\
\multicolumn{1}{c}{}  & $\Psi_{3,\text{--}}^{(0)}$  &2765.9749(3) &2252.4905(1) &2765.9749(3)  &0             &0             &$-$1413.5539(1)\\
\\
\multicolumn{1}{c}{$^{3}$He}  & $\Psi_{1,\text{--}}^{(0)}$  &2521.0686(2) &2743.9244(2) &2521.0686(2)  &998.8379(1)     &998.8379(1)     &738.1881(1)\\
\multicolumn{1}{c}{}  & $\Psi_{2,\text{--}}^{(0)}$  &2498.3853(1) &2789.2908(3) &2498.3853(1)  &$-$998.8379(1)  &$-$998.8379(1) &675.7607(1)\\
\multicolumn{1}{c}{}  & $\Psi_{3,\text{--}}^{(0)}$  &2766.6075(2) &2252.8465(2) &2766.6075(2)  &0             &0             &$-$1413.9488(1)\\
\\
\end{tabular}
\end{ruledtabular}
%\end{table*}
\label{tab:LPer}
\end{sidewaystable}
\endgroup

\begingroup
\squeezetable
\begin{sidewaystable}
%\begin{table*}%%%% XV		
\caption{The additive and nonadditive dispersion coefficients $C_8^{(IJ)}(1,M=0)$ and $C_8^{(IJ,JK)}(1,M=0)$
of the He($n_0\,^{\lambda}S$)-He($n_0\,^{\lambda}S$)-He($n_0^{\prime}\,^{\lambda}P$) system for three different types of the zeroth-order wave functions, where the three atoms form
a straight line, in atomic units.
The numbers in parentheses represent the computational uncertainties.
}\label{TabXV}
\centering
\begin{ruledtabular}
\begin{tabular}{lccccccc}
\multicolumn{1}{c}{Atom}  & \multicolumn{1}{c}{State}  & \multicolumn{1}{c}{$C_8^{(12)}(1,M=0)$}  & \multicolumn{1}{c}{$C_8^{(23)}(1,M=0)$}  & \multicolumn{1}{c}{$C_8^{(31)}(1,M=0)$} & \multicolumn{1}{c}{$C_8^{(12,23)}(1,M=0)$}  & \multicolumn{1}{c}{$C_8^{(23,31)}(1,M=0)$}  & \multicolumn{1}{c}{$C_8^{(31,12)}(1,M=0)$}\\
\hline
\\
& & & & \multicolumn{2}{c}{He($1\,^1S$)-He($1\,^1S$)-He($2\,^{1}P$)}\\
\cline{5-6}
\\
\multicolumn{1}{c}{$^{\infty}$He}  & $\Psi_{1,\text{--}}^{(0)}$  &296.0844(1)  &213.7760(1)   &296.0844(1)  &2.134984185(2)    &2.134984185(2)     &$-$1.577853610(1)\\
\multicolumn{1}{c}{}  & $\Psi_{2,\text{--}}^{(0)}$  &276.4949(1)  &196.8912(1)   &276.4949(1)  &$-$2.134984185(2) &$-$2.134984185(2)  &$-$1.444417100(2)\\
\multicolumn{1}{c}{}  & $\Psi_{3,\text{--}}^{(0)}$  &195.5657(1)    &357.4779(1) &195.5657(1)    &0              &0               &3.022270710(3)\\
\\
\multicolumn{1}{c}{$^{4}$He}  & $\Psi_{1,\text{--}}^{(0)}$  &296.0468(1)  &213.7519(1)   &296.0468(1)  &2.135847167(2)    &2.135847167(2)     &$-$1.578491396(2)\\
\multicolumn{1}{c}{}  & $\Psi_{2,\text{--}}^{(0)}$  &276.4423(1)  &196.8698(1)   &276.4423(1)  &$-$2.135847167(2) &$-$2.135847167(2)  &$-$1.445000948(2)\\
\multicolumn{1}{c}{}  & $\Psi_{3,\text{--}}^{(0)}$  &195.5383(1)    &357.4056(1) &195.5383(1)    &0              &0               &3.023492343(3)\\
\\
\multicolumn{1}{c}{$^{3}$He}  & $\Psi_{1,\text{--}}^{(0)}$  &296.0344(1)    &213.7440(1)    &296.0344(1)    &2.136129624(2)    &2.136129624(2)     &$-$1.578700145(2)\\
\multicolumn{1}{c}{}  & $\Psi_{2,\text{--}}^{(0)}$  &276.4250(1)  &196.8628(1)    &276.4250(1)  &$-$2.136129624(2) &$-$2.136129624(2)  &$-$1.445192042(1)\\
\multicolumn{1}{c}{}  & $\Psi_{3,\text{--}}^{(0)}$  &195.5292(1)    &357.3819(1)  &195.5292(1)    &0              &0               &3.023892187(3)\\
\\
& & & & \multicolumn{2}{c}{He($2\,^1S$)-He($2\,^1S$)-He($2\,^{1}P$)}\\
\cline{5-6}
\\
\multicolumn{1}{c}{$^{\infty}$He}  & $\Psi_{1,\text{--}}^{(0)}$  &576416(2)  &647941(2)   &576416(2)  &135985.7088(2)    &135985.7088(2)     &$-$100499.8272(2)\\
\multicolumn{1}{c}{}  & $\Psi_{2,\text{--}}^{(0)}$  &516777(2)  &662260(1)  &516777(2)  &$-$135985.7088(2) &$-$135985.7088(2)  &$-$92000.72(1)\\
\multicolumn{1}{c}{}  & $\Psi_{3,\text{--}}^{(0)}$  &636815(1)    &419806(2) &636815(1)    &0              &0               &192500.5476(3)\\
\\
\multicolumn{1}{c}{$^{4}$He}  & $\Psi_{1,\text{--}}^{(0)}$  &576705(2)  &648259(1)   &576705(2)  &136063.7177(2)    &136063.7177(2)     &$-$100557.4795(2)\\
\multicolumn{1}{c}{}  & $\Psi_{2,\text{--}}^{(0)}$  &516966(2)  &662582(1)   &516966(2)  &$-$136063.7177(2) &$-$136063.7177(2)  &$-$92053.50(1)\\
\multicolumn{1}{c}{}  & $\Psi_{3,\text{--}}^{(0)}$  &637098(2)    &419931(2) &637098(2)    &0              &0               &192610.9765(3)\\
\\
\multicolumn{1}{c}{$^{3}$He}  & $\Psi_{1,\text{--}}^{(0)}$  &576800(2)    &648361(2)    &576800(2)    &136089.2454(2)    &136089.2454(2)     &$-$100576.3455(1)\\
\multicolumn{1}{c}{}  & $\Psi_{2,\text{--}}^{(0)}$  &517028(2)  &662687(1)    &517028(2)  &$-$136089.2454(2) &$-$136089.2454(2)  &$-$92070.76(1)\\
\multicolumn{1}{c}{}  & $\Psi_{3,\text{--}}^{(0)}$  &637193(1)    &419971(3)  &637193(1)    &0              &0               &192647.1134(3)\\
\\
& & & & \multicolumn{2}{c}{He($2\,^3S$)-He($2\,^3S$)-He($2\,^{3}P$)}\\
\cline{5-6}
\\
\multicolumn{1}{c}{$^{\infty}$He}  & $\Psi_{1,\text{--}}^{(0)}$  &178708.89(1)  &188816.66(1)   &178708.89(1)  &41573.3004(1)    &41573.3004(1)     &$-$30724.6220(1)\\
\multicolumn{1}{c}{}  & $\Psi_{2,\text{--}}^{(0)}$  &130516.79(1)  &190656.01(1)   &130516.79(1)  &$-$41573.3004(1) &$-$41573.3004(1)  &$-$28126.2907(1)\\
\multicolumn{1}{c}{}  & $\Psi_{3,\text{--}}^{(0)}$  &173264.08(1)    &103017.09(1) &173264.08(1)    &0              &0               &58850.9128(1)\\
\\
\multicolumn{1}{c}{$^{4}$He}  & $\Psi_{1,\text{--}}^{(0)}$  &178746.69(1)  &188873.00(1)   &178746.69(1)  &41586.5716(1)    &41586.5716(1)     &$-$30734.4300(1)\\
\multicolumn{1}{c}{}  & $\Psi_{2,\text{--}}^{(0)}$  &130557.63(1)  &190716.15(1)   &130557.63(1)  &$-$41586.5716(1) &$-$41586.5716(1)  &$-$28135.2693(1)\\
\multicolumn{1}{c}{}  & $\Psi_{3,\text{--}}^{(0)}$  &173324.04(1)    &103039.20(1) &173324.04(1)    &0              &0               &58869.6994(1)\\
\\
\multicolumn{1}{c}{$^{3}$He}  & $\Psi_{1,\text{--}}^{(0)}$  &178759.05(1)    &188891.44(1)    &178759.05(1)    &41590.9142(1)   &41590.9142(1)     &$-$30737.6395(1)\\
\multicolumn{1}{c}{}  & $\Psi_{2,\text{--}}^{(0)}$  &130571.00(1)  &190735.83(1)    &130571.00(1)  &$-$41590.9142(1) &$-$41590.9142(1)  &$-$28138.2073(1)\\
\multicolumn{1}{c}{}  & $\Psi_{3,\text{--}}^{(0)}$  &173343.66(1)    &103046.44(1)  &173343.66(1)    &0              &0               &58875.8469(1)\\
\\
\end{tabular}
\end{ruledtabular}
%\end{table*}
\label{tab:LPer}
\end{sidewaystable}
\endgroup

\clearpage

\begingroup
\squeezetable
\begin{sidewaystable}
%\begin{table*}%%%% XVI		
\caption{The additive and nonadditive dispersion coefficients $C_8^{(IJ)}(1,M=\pm 1)$ and $C_8^{(IJ,JK)}(1,M=\pm 1)$
of the He($1\,^1S$)-He($1\,^1S$)-He($2\,^{1}P$) system
for three different types of the zeroth-order wave functions, where the three atoms form a straight line, in atomic units.
The numbers in parentheses represent the computational uncertainties.
}\label{TabXVI}
\centering
\begin{ruledtabular}
\begin{tabular}{lccccccc}
\multicolumn{1}{c}{Atom}  & \multicolumn{1}{c}{State}  & \multicolumn{1}{c}{$C_8^{(12)}(1,M=\pm1)$}  & \multicolumn{1}{c}{$C_8^{(23)}(1,M=\pm1)$}  & \multicolumn{1}{c}{$C_8^{(31)}(1,M=\pm1)$} & \multicolumn{1}{c}{$C_8^{(12,23)}(1,M=\pm1)$}  & \multicolumn{1}{c}{$C_8^{(23,31)}(1,M=\pm1)$}  & \multicolumn{1}{c}{$C_8^{(31,12)}(1,M=\pm1)$}\\
\hline
\\
& & & & \multicolumn{2}{c}{He($1\,^1S$)-He($1\,^1S$)-He($2\,^{1}P$)}\\
\cline{5-6}
\\
\multicolumn{1}{c}{$^{\infty}$He}  & $\Psi_{1,\text{--}}^{(0)}$  &3734.2675(1) &2640.5824(1) &3734.2675(1) &4.269968371(4)    &4.269968371(4)     &$-$3.155707222(3)\\
\multicolumn{1}{c}{}  & $\Psi_{2,\text{--}}^{(0)}$  &3958.5749(1) &2418.4666(1) &3958.5749(1) &$-$4.269968371(4) &$-$4.269968371(4)  &$-$2.888834199(3)\\
\multicolumn{1}{c}{}  & $\Psi_{3,\text{--}}^{(0)}$  &2568.9867(1) &5202.7801(1) &2568.9867(1) &0              &0               &6.044541421(6)\\
\\
\multicolumn{1}{c}{$^{4}$He}  & $\Psi_{1,\text{--}}^{(0)}$  &3736.9249(1) &2642.4613(1) &3736.9249(1) &4.271694335(4)    &4.271694335(4)     &$-$3.156982791(3)\\
\multicolumn{1}{c}{}  & $\Psi_{2,\text{--}}^{(0)}$  &3961.3633(1) &2420.1873(1) &3961.3633(1) &$-$4.271694335(4) &$-$4.271694335(4)  &$-$2.890001895(3)\\
\multicolumn{1}{c}{}  & $\Psi_{3,\text{--}}^{(0)}$  &2570.8046(1) &5206.4442(1) &2570.8046(1) &0              &0               &6.046984686(6)\\
\\
\multicolumn{1}{c}{$^{3}$He}  & $\Psi_{1,\text{--}}^{(0)}$  &3737.7946(1) &2643.0763(1) &3737.7946(1) &4.272259249(4)    &4.272259249(4)     &$-$3.157400290(4)\\
\multicolumn{1}{c}{}  & $\Psi_{2,\text{--}}^{(0)}$  &3962.2760(1) &2420.7505(1) &3962.2760(1) &$-$4.272259249(4) &$-$4.272259249(4)  &$-$2.890384086(3)\\
\multicolumn{1}{c}{}  & $\Psi_{3,\text{--}}^{(0)}$  &2571.3997(1) &5207.6434(1) &2571.3997(1) &0              &0               &6.047784375(6)\\
\\
& & & & \multicolumn{2}{c}{He($2\,^1S$)-He($2\,^1S$)-He($2\,^{1}P$)}\\
\cline{5-6}
\\
\multicolumn{1}{c}{$^{\infty}$He}  & $\Psi_{1,\text{--}}^{(0)}$  &677649(3) &696503(2) &677649(3) &271971.4176(4)    &271971.4176(4)     &$-$200999.6545(4)\\
\multicolumn{1}{c}{}  & $\Psi_{2,\text{--}}^{(0)}$  &2052935(3) &706714(2) &2052935(3) &$-$271971.4176(4) &$-$271971.4176(4)  &$-$184001.46(2)\\
\multicolumn{1}{c}{}  & $\Psi_{3,\text{--}}^{(0)}$  &1182611(2) &2509977(5) &1182611(2) &0              &0               &385001.0952(6)\\
\\
\multicolumn{1}{c}{$^{4}$He}  & $\Psi_{1,\text{--}}^{(0)}$  &678086(3) &696908(3) &678086(3) &272127.4354(4)    &272127.4354(4)     &$-$201114.9590(4)\\
\multicolumn{1}{c}{}  & $\Psi_{2,\text{--}}^{(0)}$  &2054147(4) &707118(2) &2054147(4) &$-$272127.4354(4) &$-$272127.4354(4)  &$-$184106.99(1)\\
\multicolumn{1}{c}{}  & $\Psi_{3,\text{--}}^{(0)}$  &1183287(2) &2511492(5) &1183287(2) &0              &0               &385221.9532(7)\\
\\
\multicolumn{1}{c}{$^{3}$He}  & $\Psi_{1,\text{--}}^{(0)}$  &678229(3) &697041(3) &678229(3) &272178.4910(5)    &272178.4910(5)     &$-$201152.6912(3)\\
\multicolumn{1}{c}{}  & $\Psi_{2,\text{--}}^{(0)}$  &2054544(4) &707250(2) &2054544(4) &$-$272178.4910(5) &$-$272178.4910(5)  &$-$184141.53(1)\\
\multicolumn{1}{c}{}  & $\Psi_{3,\text{--}}^{(0)}$  &1183508(2) &2511988(5) &1183508(2) &0              &0               &385294.2270(7)\\
\\
& & & & \multicolumn{2}{c}{He($2\,^3S$)-He($2\,^3S$)-He($2\,^{3}P$)}\\
\cline{5-6}
\\
\multicolumn{1}{c}{$^{\infty}$He}  & $\Psi_{1,\text{--}}^{(0)}$  &242087.18(2) &226143.71(2) &242087.18(2) &83146.60083(2)    &83146.60083(2)     &$-$61449.24409(1)\\
\multicolumn{1}{c}{}  & $\Psi_{2,\text{--}}^{(0)}$  &660377.03(2) &224826.36(1) &660377.03(2) &$-$83146.60083(2) &$-$83146.60083(2)  &$-$56252.58154(1)\\
\multicolumn{1}{c}{}  & $\Psi_{3,\text{--}}^{(0)}$  &371010.24(1) &822504.39(2) &371010.24(1) &0              &0               &117701.82562(2)\\
\\
\multicolumn{1}{c}{$^{4}$He}  & $\Psi_{1,\text{--}}^{(0)}$  &242141.77(1) &226210.21(2) &242141.77(1) &83173.14325(2)    &83173.14325(2)     &$-$61468.86023(1)\\
\multicolumn{1}{c}{}  & $\Psi_{2,\text{--}}^{(0)}$  &660542.33(2) &224895.80(1) &660542.33(2) &$-$83173.14325(2) &$-$83173.14325(2)  &$-$56270.53874(1)\\
\multicolumn{1}{c}{}  & $\Psi_{3,\text{--}}^{(0)}$  &371117.31(2) &822695.40(2) &371117.31(2) &0              &0               &117739.39893(2)\\
\\
\multicolumn{1}{c}{$^{3}$He}  & $\Psi_{1,\text{--}}^{(0)}$  &242159.65(2) &226231.97(2) &242159.65(2) &83181.82859(2)    &83181.82859(2)     &$-$61475.27906(1)\\
\multicolumn{1}{c}{}  & $\Psi_{2,\text{--}}^{(0)}$  &660596.40(1) &224918.54(2) &660596.40(1) &$-$83181.82859(2) &$-$83181.82859(2)  &$-$56276.41479(2)\\
\multicolumn{1}{c}{}  & $\Psi_{3,\text{--}}^{(0)}$  &371152.34(2) &822757.90(2) &371152.34(2) &0              &0               &117751.69384(2)\\
\\
\end{tabular}
\end{ruledtabular}
%\end{table*}
\label{tab:LPer}
\end{sidewaystable}
\endgroup

\clearpage



\end{document}